\documentclass[onecolumn]{IEEEtran}
\usepackage[T1]{fontenc}
\usepackage[utf8]{inputenc}
\usepackage{array}
\usepackage{multirow}
\usepackage{amsmath}
\usepackage{amsthm}
\usepackage{amssymb}
\usepackage{graphicx}

\makeatletter

\providecommand{\tabularnewline}{\\}

\theoremstyle{plain}
\newtheorem{thm}{\protect\theoremname}
\theoremstyle{definition}
\newtheorem{defn}[thm]{\protect\definitionname}
\theoremstyle{plain}
\newtheorem{prop}[thm]{\protect\propositionname}
\theoremstyle{plain}
\newtheorem{lem}[thm]{\protect\lemmaname}

\makeatother

\usepackage{babel}
\providecommand{\definitionname}{Definition}
\providecommand{\lemmaname}{Lemma}
\providecommand{\propositionname}{Proposition}
\providecommand{\theoremname}{Theorem}

\begin{document}
\title{Coding-Logic Correspondence: Turning Information and Communication Networks into Logical Formulae via Hypergraph Heyting Algebra}
\author{Cheuk Ting Li \\
Department of Information Engineering, The Chinese University of Hong Kong, Hong Kong, China \\
 (Email: ctli@ie.cuhk.edu.hk)}
\maketitle
\begin{abstract}
We propose using confusion hypergraph (hyperconfusion) as a model of information. In contrast to the conventional approach using random variables, we can now perform conjunction, disjunction and implication of information, forming a Heyting algebra. Confusion hypergraphs have an interpretation similar to propositions in generalized inquisitive logic. Using the connection between Heyting algebra and intuitionistic logic, we can express the requirements of a communication network (e.g., network coding, index coding, Slepian-Wolf coding) as a logical formula, allowing us to use the hypergraph Heyting algebra to directly compute the optimal coding scheme. The optimal communication cost is simply given by the entropy of the hypergraph (within a logarithmic gap). This gives a surprising correspondence between coding settings and logical formulae, similar to the Curry-Howard correspondence between computer programs and proofs.
\end{abstract}


\section{Introduction}

Information is conventionally modelled as random variables, or equivalently, $\sigma$-algebras or partitions of the sample space. The operations we can perform on information defined in this manner are limited. For example, given two pieces of information $X,Y$ as random variables, we can take their conjunction, i.e., the information that corresponds to knowing both $X$ and $Y$, which is given as the joint random variable $(X,Y)$. This unambiguously gives the smallest information that contains $X$ and $Y$, as any other $M$ that can deduce $X$ and $Y$ must be larger than $(X,Y)$ in the functional partial order (i.e., $(X,Y)$ is a function of $M$). 

Nevertheless, the disjunction ``$X\vee Y$'', corresponding to knowing $X$ or $Y$, does not have an unambiguous definition, that is, there is no smallest information $U$ (with respect to functional partial order) that can deduce $X$ or $Y$.\footnote{``$U$ can deduce $X$ or $Y$'' means there is a decoder that reads $U$, declares which of $X$ or $Y$ to decode, and outputs its value. The decoding function $f:\mathcal{U}\to(\{1\}\times\mathcal{X})\cup(\{2\}\times\mathcal{Y})$ satisfies ($K=1$ and $Z=X$) or ($K=2$ and $Z=Y$), where $(K,Z)=f(U)$. There is no smallest $U$ with respect to functional ordering  as $U=X$ and $U=Y$ are incomparable. There are examples where neither $U=X$ nor $U=Y$ gives the smallest entropy; see Section \ref{subsec:disjunction}.} The difference or implication ``$X\rightarrow Y$'', corresponding to the minimal additional information a terminal needs to decode $Y$ if it already knows $X$, also has no unambiguous definition, but instead has several incomparable definitions. See Section \ref{subsec:prev_other}.

Following the success of the analogy between entropy and set measures (information diagram and the I-measure) \cite{ting1962amount,cover2006elements,yeung1991new}, concrete correspondences have been constructed by Ellerman \cite{ellerman2014introduction,ellerman2017logical,ellerman2018logical}, Down and Mediano \cite{down2023logarithmic} and Li \cite{li2025poisson}, which embed information into a Boolean algebra, allowing us to take conjunction, disjunction and difference of information. While these results are theoretically interesting,  operational implications are limited.  The logarithmic decomposition \cite{down2023logarithmic} sometimes gives negative measure,\footnote{The intersection of three pieces of information $X,Y,Z$, i.e., multivariate mutual information $I(X;Y;Z)$, can be negative.} which prevents it from having a clear meaning. This suggests that Boolean algebra is perhaps not the correct algebraic structure for information. More discussions  will be included in Sections \ref{subsec:prev_partition} and \ref{subsec:prev_log}.

Events in a probability space are also often regarded as information. This is related but different from the information of a random variable. The information of an event $E$ is the knowledge that $E$ occurs (e.g., knowing that tomorrow is going to rain), whereas the information of a random variable $X$ is the knowledge of the value of $X$ (e.g., knowing whether tomorrow is going to rain, i.e., the indicator random variable of whether it will rain).  The amount of information in an event is measured by the self-information $-\log\mathbb{P}(E)$ \cite{shannon1948mathematical}. Although events naturally form a Boolean algebra, they are not as useful as random variables for coding tasks, where sources and messages are random variables.

In this paper, which is the full version of \cite{li2026coding},\footnote{The conference version \cite{li2026coding} does not contain most properties of the entropy of hyperconfusions in Section \ref{sec:hyperconfusion}, hyperconfusable set-valued functions and joint-source-channel coding in Sections \ref{sec:setvalued} and \ref{sec:joint}, and the algorithms in Section \ref{sec:algorithm}. It only contains a short version of the discussion of the coding-logic correspondence in Section \ref{sec:formulae}, with many examples omitted.} we use a more general model of information that encompasses both random variables and events. Consider the following illustrative example. Suppose you see Alice driving a car. What information can you gain about Alice's car ownership status: whether she has a convertible car (``$\mathrm{C}$'', can switch between having and not having a roof), has a car with a fixed roof (``$\mathrm{R}$''), has an open-air car without a roof (``$\mathrm{O}$''), or has no car (``$\mathrm{N}$'')? If you see her driving a car with a roof, then the possibilities are $\mathrm{C}$ and $\mathrm{R}$ (you cannot distinguish these two), so $\{\mathrm{C},\mathrm{R}\}$ is the \emph{confusion set}. If you see a car without a roof, the confusion set is $\{\mathrm{C},\mathrm{O}\}$. Hence, the information of seeing Alice's car is captured by the \emph{confusion hypergraph} (or \emph{hyperconfusion}) $\{\{\mathrm{C},\mathrm{R}\},\{\mathrm{C},\mathrm{O}\}\}$.\footnote{Technically, we have to include the subsets of $\{\mathrm{C},\mathrm{R}\}$ and $\{\mathrm{C},\mathrm{O}\}$ as well. The precise definition is given in Definition \ref{def:hyperconfusion}.} Confusion hypergraphs are different from partitions (or random variables) in two aspects. First, confusion sets may overlap (e.g., $\mathrm{C}$ is in $\{\mathrm{C},\mathrm{R}\}$ and $\{\mathrm{C},\mathrm{O}\}$). Second, there may be outcomes that are in none of the confusion sets (e.g., $\mathrm{N}$), meaning that this information implies a certain event (e.g., you can see Alice's car only if she has a car, which excludes $\mathrm{N}$).

Confusion hypergraphs were originally studied in a different context of zero-error channel capacity \cite{korner1990capacity,adei2023entanglement}. They are mathematically equivalent to, and have similar interpretations as the concept of propositions in generalized inquisitive logic \cite{ciardelli2009generalized,ciardelli2011inquisitive,roelofsen2013algebraic} (see Section \ref{subsec:inquisitive}). They have a richer structure compared to partitions.  For example, as shown in \cite{roelofsen2013algebraic}, hyperconfusions form a Heyting algebra \cite{heyting1930formalen,esakia2019heyting} (not a Boolean algebra),\footnote{The partition logic \cite{ellerman2014introduction} also uses Heyting algebra. See Section \ref{subsec:prev_partition}.} allowing us to take conjunction (meet) $\cap$, disjunction (join) $\cup$, difference (implication) $\rightarrow$ and negation $\lnot$ of information and events. This connection between information and Heyting algebra reveals a new correspondence between coding and intuitionistic logic (more specifically, Medvedev logic \cite{medvedev1962finiteEN,sorbi2008intermediate}, as shown in \cite{ciardelli2011inquisitive}), which we call \emph{coding-logic correspondence}. 

\begin{figure}
\begin{centering}
\includegraphics{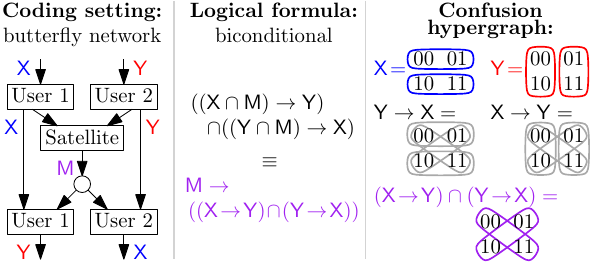}
\par\end{centering}
\caption{Correspondence between coding setting (e.g., butterfly network) and logical formula. The optimal code can be computed via confusion hypergraphs.}\label{fig:butterfly}

\end{figure}

For example, in the butterfly network \cite{ahlswede2000network,yeung2008information}, a satellite which knows the hyperconfusions $\mathsf{X},\mathsf{Y}$ wants to broadcast a message $\mathsf{M}$ (also a hyperconfusion) to allow User 1 who posesses $\mathsf{X}$ to decode $\mathsf{Y}$, and allow User 2 who posesses $\mathsf{Y}$ to decode $\mathsf{X}$. The requirement ``$\mathsf{X},\mathsf{M}$ can be used to decode $\mathsf{Y}$, and $\mathsf{Y},\mathsf{M}$ can be used to decode $\mathsf{X}$'' is written as the following logical formula
\begin{equation}
((\mathsf{X}\cap\mathsf{M})\rightarrow\mathsf{Y})\cap((\mathsf{Y}\cap\mathsf{M})\rightarrow\mathsf{X}).\label{eq:butterfly_intro}
\end{equation}
Using the rules of intuitionistic logic (e.g., \emph{currying} \cite{schonfinkel1924bausteine,curry1958combinatory} or \emph{exportation} \cite{copi2016introduction}), we can simplify it to
\begin{equation}
\mathsf{M}\rightarrow((\mathsf{X}\rightarrow\mathsf{Y})\cap(\mathsf{Y}\rightarrow\mathsf{X})),\label{eq:butterfly_intro_2}
\end{equation}
so the optimal $\mathsf{M}$ (with the smallest entropy) is $(\mathsf{X}\rightarrow\mathsf{Y})\cap(\mathsf{Y}\rightarrow\mathsf{X})$, the ``biconditional'' between $\mathsf{X},\mathsf{Y}$. Interestingly, we also have $(\mathsf{X}\rightarrow\mathsf{Y})\cap(\mathsf{Y}\rightarrow\mathsf{X})=(\mathsf{X}\cup\mathsf{Y})\to(\mathsf{X}\cap\mathsf{Y})$, so $\mathsf{M}$ is the information needed to turn ``$\mathsf{X}$ or $\mathsf{Y}$'' into ``$\mathsf{X}$ and $\mathsf{Y}$''. See Figure \ref{fig:butterfly}.  Other settings such as general network coding \cite{ahlswede2000network,li2003linear}, index coding \cite{birk1998informed,bar2011index,el2010index}, Slepian-Wolf coding \cite{slepian1973noiseless} and Gray-Wyner network \cite{gray1974source} can also be expressed as logical formulae. The coding-logic correspondence is reminiscent of the celebrated Curry-Howard correspondence between proofs and computer programs \cite{curry1958combinatory,howard1980formulae}, the Brouwer-Heyting interpretation of proofs  \cite{troelstra1988constructivism}, the Kolmogorov-Kleene logic of problems \cite{kolmogorov1932deutung,kleene1945interpretation} and its applications to Kolmogorov complexity \cite{shen2002logical,shen2017kolmogorov}. See Tables \ref{tab:correspondence_intro}, \ref{tab:correspondence} for the correspondence between logic and coding tasks, and Sections \ref{subsec:CH}, \ref{subsec:problems} for discussions.

\begin{table*}
\centering
\caption{Comparing coding-logic correspondence (with confusion hypergraphs) with Curry-Howard correspondence and  information theory with random variables.}\label{tab:correspondence_intro}

{\renewcommand*{\arraystretch}{1.5}%
\begin{tabular}{cccc}
\hline 
\textbf{Logic} & $\begin{array}{c}
\textbf{Curry-Howard}\\
\textbf{correspondence}
\end{array}$ & $\begin{array}{c}
\textbf{Information theory}\\
\textbf{with random variables}
\end{array}$ & $\begin{array}{c}
\textbf{Coding-logic correspondence}\\
\textbf{with confusion hypergraphs}
\end{array}$\tabularnewline
\hline 
Formula & Type & Random variable & Hypergraph / coding task\tabularnewline
\hline 
Proof & Program & --- & Coding scheme\tabularnewline
\hline 
Truth $\top$ & Unit type & Constant RV & Fully-confused source \tabularnewline
\hline 
Falsehood $\bot$ & Empty type & --- & Impossibility / omniscience\tabularnewline
\hline 
$\begin{array}{c}
\text{Stable proposition}\\
\phi=\lnot\lnot\phi
\end{array}$ & --- & --- & Events\tabularnewline
\hline 
Conjunction $\phi\wedge\psi$ & Product type & Joint RV & Decode both sources\tabularnewline
\hline 
Disjunction $\phi\vee\psi$ & Sum type & \emph{(ambiguous)} & Decode one of two sources\tabularnewline
\hline 
Implication $\phi\rightarrow\psi$ & Function type & \emph{(ambiguous)} & Decode with side information\tabularnewline
\hline 
Modus ponens & Apply function & --- & Black box\tabularnewline
\hline 
Exportation & Currying & --- & Currying\tabularnewline
\hline 
--- & --- & Shannon entropy & Average encoding length\tabularnewline
\hline 
--- & --- & Min-entropy & Negative log success probability\tabularnewline
\hline 
\end{tabular}}
\end{table*}

The coding-logic correspondence is not only a theoretical curiosity, but also gives an algorithm that directly \emph{computes} the optimal coding scheme of various settings, simply by evaluating the corresponding logical formula over the Heyting algebra of hypergraphs. We define a generalized notion of entropy (slightly generalizing the entropy of hypergraphs \cite{korner1990capacity,korner1988new,simonyi1995graph}, which generalizes the K\"{o}rner graph entropy \cite{korner1971coding}, which  generalizes Shannon entropy) such that the communication cost of a setting is the entropy of the corresponding formula. For example, the communication cost of the butterfly network for general discrete joint distributions of $\mathsf{X},\mathsf{Y}$ can be computed as the entropy 
\[
H((\mathsf{X}\rightarrow\mathsf{Y})\cap(\mathsf{Y}\rightarrow\mathsf{X})),
\]
within a logarithmic gap (Theorem \ref{thm:butterfly}). For the special case where $\mathsf{X},\mathsf{Y}$ are two i.i.d. bits, $(\mathsf{X}\rightarrow\mathsf{Y})\cap(\mathsf{Y}\rightarrow\mathsf{X})$ is the XOR of $\mathsf{X}$ and $\mathsf{Y}$, and the above entropy is $1$, correctly identifying that the XOR is the optimal code. The correspondence applies to coding with error as well. For example, if we allow the users in the butterfly network to fail to decode with a small probability, then the negative log success probability is the generalized min-entropy $H_{\infty}(\mathsf{M}\rightarrow((\mathsf{X}\rightarrow\mathsf{Y})\cap(\mathsf{Y}\rightarrow\mathsf{X})))$ of the formula in (\ref{eq:butterfly_intro_2}), giving a new meaning to min-entropy. 

These computations are carried out with $\mathsf{X},\mathsf{Y},\mathsf{M}$ being hyperconfusions, not  random variables. This allows the operations $\cup,\cap,\rightarrow$ to be computed unambiguously. Nevertheless, if an ordinary random variable is desired (e.g., if we use Huffman coding \cite{huffman1952method} to compress the message $\mathsf{M}$), we can convert a hyperconfusion back to a random variable (within a logarithmic gap) via a result we call the \emph{unconfusing lemma}, proved using the strong functional representation lemma \cite{sfrl_trans,li2025discrete}. This gives a standardized workflow for constructing a coding scheme for general coding settings: express the requirements as a logical formula, use intuitionistic logic to simplify the formula, evaluate the formula using hyperconfusions, and finally invoke the unconfusing lemma to convert the hyperconfusions to random variables.

This paper is organized as follows. In Section \ref{sec:previous}, we review some relevant works. Section \ref{sec:hyperconfusion} gives the definition of hyperconfusion and entropy. Section \ref{sec:operations} defines the logical operations on hyperconfusions. Section \ref{sec:formulae} explains the coding-logic correspondence. The remaining sections describe applications to zero-error joint source-channel coding, and algorithms for hyperconfusions. 

\smallskip{}

\subsection*{Notations}

Entropy is in bits, and logarithms are to the base $2$. Write $[n]:=\{1,\ldots,n\}$. The power set of a set $\Omega$ (the set of subsets of $\Omega$) is denoted as $2^{\Omega}$. We focus on finite probability spaces $(\Omega,p)$ unless otherwise stated, where $\Omega$ is a finite sample space and $p:\Omega\to[0,1]$ is a probability mass function. Write $\mathrm{supp}(p):=\{\omega\in\Omega:\,p(\omega)>0\}$. We use capital serif letters $X,Y,M$ for random variables, $A,B,S$ for sets, and capital sans-serif letters $\mathsf{X},\mathsf{Y},\mathsf{M}$ for hyperconfusions.

\smallskip{}

\section{Previous Works}\label{sec:previous}

\subsection{Zero-Error Channel Coding and Source Coding}\label{subsec:zero_error}

In zero-error channel coding \cite{shannon1956zero}, the encoder transmits a message $M$ through a noisy channel $p_{Y|X}$, and the decoder must recover $M$ with probability $1$. The maximum number of values of $M$ that can be transmitted through one use of the channel is given by the size of the largest independent set of the confusion graph, where two vertices $x_{1},x_{2}\in\mathcal{X}$ are adjacent (confusable) if there exists $y\in\mathcal{Y}$ where $p_{Y|X}(y|x_{1}),p_{Y|X}(y|x_{2})>0$ \cite{shannon1956zero,alon2006shannon}. This has been generalized to list decoding \cite{elias1988zero}, and hypergraphs are used in \cite{korner1990capacity} to analyze the list decoding capacity.  Applications of our results to zero-error channel coding are discussed in Section \ref{sec:joint}.

A related source coding setting was studied in \cite{korner1971coding}, where a source $X\sim p_{X}$ is compressed into a message $M$, but we only require that non-confusable values of $X$ (according to a confusion graph) are mapped to different values of $M$. Equivalently, we are allowed to decode to a set of values of $X$ as long as every pair of values is confusable. The asymptotic rate (with vanishing error probability) is given by the K\"{o}rner graph entropy \cite{korner1971coding}. The K\"{o}rner graph entropy can be generalized to hypergraphs \cite{korner1990capacity,korner1988new,simonyi1995graph}.

\smallskip{}

\subsection{Generalized Inquisitive Logic}\label{subsec:inquisitive}

A closely related line of research in logic is the generalized inquisitive logic \cite{ciardelli2009generalized,ciardelli2011inquisitive,roelofsen2013algebraic}, where a \emph{possibility} (or a piece of information) is a set of possible worlds $\alpha\subseteq W$ (where $W$ is the set of all possible worlds), and a \emph{proposition} $A$ is defined as a non-empty set of possibilities. Alternatively, a proposition is a downward closed set of sets of possible worlds \cite{roelofsen2013algebraic}. This can be shown to be equivalent to the confusion hypergraph (hyperconfusion) in this paper. The connection between generalized inquisitive logic and Medvedev logic \cite{medvedev1962finiteEN} was shown in \cite{ciardelli2011inquisitive}.

The interpretations of the generalized inquisitive logic and the hyperconfusions in this paper are similar. In generalized inquisitive logic, a piece of information $\alpha$ settles $A$ if $\alpha\in A$, whereas in hyperconfusions, we can ``know'' a hyperconfusion $\mathsf{X}$ by giving an event $B\in\mathsf{X}$ which occurs (i.e., the outcome $\omega\in B$). ``Possible world'', ``possibility/piece of information'' and ``proposition'' in generalized inquisitive logic correspond to ``outcome'', ``event'' and ``hyperconfusion/information'' in this paper, respectively. 

The main novelty of this paper is that we consider a probabilistic setting where we can study the entropy of hypergraphs, and apply it to find the communication cost of coding problems. Hence, we adopt the terminology in probability and information theory (outcome, event, confusion hypergraph, etc.) instead of the terminology in generalized inquisitive logic. 

Another notable difference is the interpretation of ``information''. In generalized inquisitive logic, a piece of information is a set of possible worlds. In this paper, a piece of information is modelled as a hyperconfusion, which generalizes random variables, and is not merely a set of outcomes. This is in line with information theory which models information using random variables.

\smallskip{}

\subsection{Curry-Howard Correspondence}\label{subsec:CH}

In the Curry-Howard correspondence between proofs and computer programs \cite{curry1958combinatory,howard1980formulae}, logical formulae correspond to types. Consider the formula $(\mathsf{X}\cap\mathsf{M})\rightarrow\mathsf{Y}$ in (\ref{eq:butterfly_intro}). In Curry-Howard, this is the type of a function from a product type $\mathsf{X}\cap\mathsf{M}$ (type of pairs $(x,m)$) to the type $\mathsf{Y}$. To inhabit this type is to implement a function $f$ that maps a pair $(x,m)$ to an output $y=f(x,m)$ of type $\mathsf{Y}$. The equivalence between $(\mathsf{X}\cap\mathsf{M})\rightarrow\mathsf{Y}$ and $\mathsf{M}\rightarrow(\mathsf{X}\rightarrow\mathsf{Y})$ is known as \emph{currying} \cite{schonfinkel1924bausteine,curry1958combinatory}.  Truth $\top$ is represented by the unit type (can always be inhabited), whereas falsehood $\bot$ is represented by the empty type (impossible to inhabit).

In the coding-logic correspondence in this paper, we interpret $(\mathsf{X}\cap\mathsf{M})\rightarrow\mathsf{Y}$ as an encoder/decoder that takes the information in $\mathsf{X}$ and $\mathsf{M}$ as input, and output the information $\mathsf{Y}$. To inhabit this coder is to give a concrete encoding/decoding function. The formula $\mathsf{M}\rightarrow(\mathsf{X}\rightarrow\mathsf{Y})$ means that a coder takes $\mathsf{M}$ as input, and outputs a coder that takes $\mathsf{X}$ as input and outputs $\mathsf{Y}$. Truth $\top$ is represented by ``no information'' (can always be achieved), whereas falsehood $\bot$ is represented by ``omniscience'' (being able to decode every present and future information, impossible to achieve).

The Curry-Howard correspondence and the coding-logic correspondence have some similarities, as both concern functions, though one is about the \emph{types} of the input and output of the functions, and the other is about the \emph{information} of the input and output. See Table \ref{tab:correspondence_intro}. The contribution of this paper is a systematic method and algorithm to keep track of \emph{all} possible functions and finding the smallest output size of these functions, via the entropy of the hyperconfusion of the formula. We focus on coding settings where minimizing the communication size is the main goal. The theory developed in this paper might have applications to time complexity or communication complexity, though we defer this to future studies.

Both coding-logic and Curry-Howard correspondence (without control operators \cite{griffin1990formulae}) concern intuitionistic logic, so double negation elimination $\lnot\lnot\mathsf{X}=\mathsf{X}$ is not guaranteed to hold. One interesting feature of the coding-logic correspondence that has no counterpart in Curry-Howard is the notion of \emph{events}, which are information $\mathsf{E}$ where $\lnot\lnot\mathsf{E}=\mathsf{E}$ (it is \emph{regular}). Such an $\mathsf{E}$ represents an event in the probability space. This allows a unified treatment of information and events.\footnote{Information and events are also treated uniformly in \cite{li2025poisson}, though \cite{li2025poisson} uses Boolean algebra, and cannot express the union of events in terms of operations on information.} 

 Curry-Howard correspondence shows that we can prove theorems using computer programs. We show that it is also possible to prove statements using coding, in the sense that a logical formula holds in classical logic if a code for the corresponding setting always exists. More precisely, the formulae that can be proven this way are the theorems in Medvedev logic \cite{medvedev1962finiteEN} for finite spaces, or the logic of infinite problems \cite{skvortsov1979logic} for infinite spaces. See Section \ref{subsec:logic_coding} and Appendix \ref{subsec:pf_medvedev}. Unfortunately, this is less useful than Curry-Howard since proving a code always exists is usually harder than proving the formula directly. The usefulness of our correspondence lies in the other direction: using logic to reason about information and coding.

\smallskip{}

\subsection{The Calculus of Problems}\label{subsec:problems}

In the calculus of problems by Kolmogorov \cite{kolmogorov1932deutung} (also see Kleene's notion of realizability \cite{kleene1945interpretation}), logical formulae correspond to the abstract notion of problems. Conjunction $x\wedge y$ means to solve both problems $x$ and $y$, disjunction $x\vee y$ means to solve any one problem, and implication $x\rightarrow y$ means to solve $y$ using a solution to $x$. A concrete meaning of problems was studied in the context of Kolmogorov complexity \cite{shen2002logical,shen2017kolmogorov}, where a problem is a set of binary strings, and a solution is an algorithm to generate one string in the set. A related notion is the mass problems studied by Medvedev \cite{medvedev1955degrees}, where a problem is a set of functions from natural numbers to natural numbers. Refer to \cite{skvortsova1988faithful,terwijn2006constructive} for extensions.

In a sense, the coding-logic correspondence can be regarded as another concrete meaning of problems. Nevertheless, unlike most notions of problems (e.g., \cite{medvedev1955degrees,shen2002logical,shen2017kolmogorov}) where the validity of a solution does not depend on any randomness or uncertainty, whether a coding task is successfully performed is a random event. The dependence on the random outcome opens up new interpretations of logical operations. More specifically, negations can be interpreted as events: for any coding task $\mathsf{X}$ (as a hypergraph), $\lnot\mathsf{X}$ is the event where the task is impossible, and $\lnot\lnot\mathsf{X}$ is the event where the task is possible. In comparison, for Kolmogorov complexity, negation is undefined \cite{shen2017kolmogorov}.\footnote{One can take $\lnot x=\emptyset$ if $x\neq\emptyset$, or take $\lnot x$ to be the set of all sequences if $x=\emptyset$. Nevertheless, this only gives two possible values for $\lnot x$, which is the same limitation as \cite{medvedev1955degrees}.} For mass problems, the weak law of excluded middle $\lnot x\vee\lnot\lnot x$ holds, so $\lnot x$ only has two possible values: $\top$ or $\bot$ \cite{medvedev1955degrees}. We can see that negation in the coding-logic correspondence has a richer structure, as it can model events, and by extension, many concepts in probability theory that are built upon events.

The notion of problems that is closest to the coding-logic correspondence is perhaps Medvedev's notion of finite problems \cite{medvedev1962finiteEN}, where a problem is a finite set $F$ (the possibilities) together with a subset $X\subseteq F$ (the solutions). Finite problems also corresponds to the Medvedev logic \cite{medvedev1962finiteEN}. The difference is that, since this paper aims at capturing the concept of information, we cannot have a fixed subset as the set of solutions, but rather the ``solutions'' should depend on the random source or message. Therefore, we have to use hypergraphs (families of subsets) instead to model codes in communication settings.

\smallskip{}

\subsection{Partition Logic}\label{subsec:prev_partition}

The partition logic, which concerns the operations that can be performed over partitions of the sample space, was studied by Ellerman \cite{ellerman2014introduction,ellerman2017logical,ellerman2018logical}. Given a partition $\pi$ of the sample space $\Omega$, the \emph{ditset} $\mathrm{dit}(\pi)$ is the set of pairs $(x,y)\in\Omega^{2}$ where $x,y$ belong to different parts. An elegant property of $\mathrm{dit}(\pi)$ is that the ditset of the common refinement of two partitions is simply the union of the two ditsets. Nevertheless, other Boolean operations on ditsets often result in sets that are not ditsets, and we have to take the interior to convert the sets to ditsets. A Heyting algebra of ditsets can  be defined via Boolean operations followed by taking the interior \cite{ellerman2014introduction}.

Nevertheless, partition is ``too coarse'' to be useful for coding settings, and the interior operation  often results in a significant gap (and may even give the trivial partition).Consider $\Omega=\{0,1\}^{2}$ consisting of two bits, $\pi_{1}=\{\{00,01\},\{10,11\}\}$ is the partition induced by the first bit, and $\pi_{2}=\{\{00,10\},\{01,11\}\}$ is the second bit. The ``disjunction'' $\mathrm{int}(\mathrm{dit}(\pi_{1})\cap\mathrm{dit}(\pi_{2}))$ gives the common part \cite{gacs1973common} of $\pi_{1}$ and $\pi_{2}$ which is the trivial partition $\{\Omega\}$. The Heyting implication $\mathrm{int}(\mathrm{dit}(\pi_{1})^{\mathrm{c}}\cap\mathrm{dit}(\pi_{2}))$ is also the trivial partition. A lattice of random variables has also been studied in \cite{li2007information}, though it is subject to the same limitations.

In comparison, the hyperconfusion considered in this paper allows us to define nontrivial disjunction and implication in this example. For example, hyperconfusion implication ``$\pi_{1}\rightarrow\pi_{2}$'' correctly identifies that knowing the XOR of the two bits will allow someone who has the first bit to deduce the second bit (Section \ref{subsec:butterfly}). Moreover, the entropy of the resultant hyperconfusion  gives the smallest communication cost within a logarithmic gap.\footnote{We also remark that the logical entropy \cite{ellerman2017logical,ellerman2018logical} recovers the order-2 Tsallis entropy \cite{tsallis1988possible} $1-\sum_{\omega}p(\omega)^{2}$, which is usually not the operational limit of coding settings.} 

\smallskip{}

\subsection{Logarithmic Decomposition}\label{subsec:prev_log}

Down and Mediano \cite{down2023logarithmic} considered embedding a partition $\pi$ into a subset of $\Delta\Omega:=2^{\Omega}\backslash(\{\{\omega\}:\omega\in\Omega\}\cup\{\emptyset\})$, where a set $S\in\Delta\Omega$ is included in the embedding if and only if $S$ is being split by $\pi$ (i.e., $S$ is not a subset of a single part of the partition $\pi$). This is the complement of the confusion hypergraph. This was generalized by Li \cite{li2025poisson} to general probability spaces. These works regard the space of embedded ``generalized information'' as a field of sets, or a Boolean algebra, so the analogy between Shannon entropy and set measures \cite{ting1962amount,campbell1965entropy,yeung1991new} can be recovered. A downside is that this generalized information may be negative, and does not appear to have operational meaning except when it coincides with existing notions.\footnote{If we regard ``$X\cap Y$'' as the common part of $X$ and $Y$, then if $X,Y$ are the information of random variables, the measure of ``$X\cap Y$'' is indeed the mutual information. However, if $X,Y$ are general sets (e.g., if they are formed by intersecting other random variables), then ``$X\cap Y$'' has no clear meaning. Even ``$X\cap(Y\cap Z)$'' can have a negative measure (multivariate mutual information $I(X;Y;Z)$), so it cannot be regarded as a measure of the amount of information.}

Since the logarithmic decomposition is the unique measure that recovers the I-measure \cite{down2023logarithmic}, negative measures are unavoidable as long as we use Boolean algebra. In this paper, we use Heyting algebra instead, where the logical operations are defined differently. The entropy in this paper is nonnegative and has a clear operational meaning as communication cost. 

\smallskip{}

\subsection{Other Previous Works}\label{subsec:prev_other}

Regarding logic operations on information, there are several notions of common information \cite{gacs1973common,wyner1975common,kumar2014exact}, though none of them attempts to capture the size of the disjunction ``knowing $X$ or $Y$''. There are attempts to define the difference between information \cite{sfrl_trans,li2017extended,arikan1996inequality,bunte2014encoding,li2025discrete}. Nevertheless, the definitions in \cite{sfrl_trans,li2017extended} are approximate, whereas the constructions in \cite{arikan1996inequality,bunte2014encoding,li2025discrete} are dependent on the distribution, not only the partitions that the information induces. None of them are Heyting implications.

 A formal concept \cite{wille2005formal} is a set of attributes that correspond to the set of objects that all possess those attributes, and can be regarded as a knowledge unit. For example, ``red sports car'' is a concept that corresponds to the set of cars that are red and are sports cars. Formal concepts form a lattice \cite{wille2005formal}, which is different from the Heyting algebra of hyperconfusions.\footnote{The meet of two concepts is given by the intersection of their sets of attributes, e.g., the meet of the concepts ``red sports car'' and ``blue sports car'' is ``sports car''. For hyperconfusions, the disjunction of the events ``red sports car'' and ``blue sports car'' is the information of not only knowing that the car is a sports car, but also that it is red or blue, and knowing which color it is (knowing the ``color'' random variable of the car). The ability to capture random variables make hyperconfusions useful for coding tasks.}  Refer to \cite{harremoes2025probability} for connections between the concept lattice and probability.

Logical probability \cite{bar1953semantic} provides a framework for probability and information via logic. Another connection between logic and information theory has been proposed in \cite{lastras2023towards}. These lines of research mostly focus on classical logic, and regard the amount of information as a function of the (logical) probability that a logical sentence holds. In contrast, the hyperconfusion entropy in this paper (Definition \ref{def:entropy}) is not a simple function of the probability of a certain event. We will see that our definition correctly captures the communication cost of various settings. There are other works that relate logic and information flow \cite{eijck1994logic,barwise1995logic}, though the meaning of ``information'' in these works is closer to the general notion of human knowledge, rather than sources and messages in communication settings as in this paper.

\smallskip{}

\section{Hyperconfusion and Entropy}\label{sec:hyperconfusion}

\subsection{Definition of Hyperconfusion}

Conventionally, information is represented as a partition of the sample space. For example, assume that a student has obtained a score $X\in\{0,\ldots,7\}$ in an exam. The student's grade in the course is computed as $Y=\lfloor X/3\rfloor\in\{0,1,2\}$. If we only concern the grade $Y$, then this information induces a partition $\{\{0,1,2\},\{3,4,5\},\{6,7\}\}$ of the set of scores. Whether the score is $X=3$ or $4$ would not matter to the grade. Hence, $3$ and $4$ are \emph{confusable}, and $\{3,4,5\}$ is a \emph{confusable set} since every pair of scores in it is confusable. The confusable sets are $\{0,1,2\}$, $\{3,4,5\}$, $\{6,7\}$ and their subsets.

We may also have confusable sets that are not induced by a partition. For example, consider the scenario where we only concern the approximate score, and we can store the score as $\hat{X}$ as long as $|X-\hat{X}|\le1$. In this case, $\{3,4,5\}$ is still confusable since we can store the score as $\hat{X}=4$ for any $X\in\{3,4,5\}$. The confusable sets are $\{0,1,2\}$, $\{1,2,3\}$, ..., $\{5,6,7\}$ and their subsets. These confusable sets are not induced by a single partition. More generally, if we are to compress $X\in\Omega$ into $\hat{X}\in\hat{\Omega}$, so that $(X,\hat{X})\in\mathcal{S}$ where $\mathcal{S}\subseteq\Omega\times\hat{\Omega}$ is the set of allowed source-reconstruction pairs, then a set $A\subseteq\Omega$ is confusable if there exists $\hat{x}\in\hat{\Omega}$ such that $(x,\hat{x})\in\mathcal{S}$ for every $x\in A$. This is because even if we do not know the precise value of $X$, if we know $X\in A$, then we are still able to choose a reconstruction $\hat{X}$ that is guaranteed to be allowed. 

In this paper, we use the following general definition of a family of confusable sets, which we call a \emph{confusion hypergraph}, or \emph{hyperconfusion} for the sake of brevity.\footnote{Simply calling it ``hypergraph'' is prone to confusion with \cite{korner1990capacity}, which studies the hypergraph of distinguished elements (the ``complement'' of the confusion hypergraph) instead.} The only requirement on this family is that if a set is confusable, then all its subsets are confusable, i.e., it is an abstract simplicial complex \cite{munkres2018elements}. Similar (but slightly different) definitions have appeared in \cite{korner1990capacity,adei2023entanglement}.

\smallskip{}

\begin{defn}
\label{def:hyperconfusion}Given a set $\Omega$ (the \emph{sample space}), a \emph{confusion hypergraph} (or \emph{hyperconfusion} in short) over $\Omega$ is a nonempty set $\mathsf{X}\subseteq2^{\Omega}$ that is downward closed, i.e., if $A\in\mathsf{X}$, then for every $B\subseteq A$, we also have $B\in\mathsf{X}$. The set of hyperconfusions over $\Omega$ is\footnote{We remark that $\mathrm{Hyps}(\Omega)\cup\{\emptyset\}$ gives an Alexandrov topology over $2^{\Omega}$ where the closed sets are $\mathrm{Hyps}(\Omega)\cup\{\emptyset\}$, and the closure operation is the downward closure $\mathrm{cl}(\mathsf{X}):=\bigcup_{A\in\mathsf{X}}2^{A}$. We will see in Section \ref{subsec:logic_coding} that $\mathrm{Hyps}(\Omega)$ is isomorphic to the dual Heyting algebra of the Medvedev frame \cite{medvedev1962finiteEN}. We also remark that \cite{stell2015symmetric} studied a Heyting algebra of sub-hypergraphs under a different definition.}
\[
\mathrm{Hyps}(\Omega):=\Big\{\mathsf{X}\in2^{2^{\Omega}}\backslash\{\emptyset\}:\,\mathsf{X}=\bigcup_{A\in\mathsf{X}}2^{A}\Big\}.
\]
For a hyperconfusion $\mathsf{X}$, its \emph{support} is
\[
\mathrm{supp}(\mathsf{X}):=\{\omega\in\Omega:\,\{\omega\}\in\mathsf{X}\}\subseteq\Omega.
\]
\end{defn}
\smallskip{}

The conventional notion of information is a partition of $\Omega$, which can be regarded as a hyperconfusion where the confusable sets are sets that are contained within one part in the partition. This is captured by the following definition of ordinary information.

\smallskip{}

\begin{defn}
\label{def:ordinary}Given a hyperconfusion $\mathsf{X}$ over $\Omega$, we say that $\mathsf{X}$ is an \emph{ordinary information} if  it can be written as $\mathsf{X}=\bigcup_{A\in\mathcal{S}}2^{A}$, where $\mathcal{S}\subseteq2^{\Omega}$ is a nonempty collection of disjoint subsets of $\Omega$. Write $\mathrm{OIs}(\Omega)$ for the set of all ordinary informations over $\Omega$. Given a function $f:\Omega\to\Omega'$, the \emph{ordinary information induced by} $f$ is 
\[
\mathrm{oi}(f):=\bigcup_{y\in\Omega'}2^{f^{-1}(\{y\})}=\{A\subseteq\Omega:\,|f(A)|\le1\},
\]
which is in $\mathrm{OIs}(\Omega)$.\footnote{Not all ordinary informations can be written as $\mathrm{oi}(f)$ for some $f$. We have $\mathrm{supp}(\mathrm{oi}(f))=\Omega$, though we generally do not require an ordinary information to have full support.} In particular, the \emph{singleton hyperconfusion} is
\[
\mathrm{sing}(\Omega):=\bigcup_{\omega\in\Omega}2^{\{\omega\}}=\mathrm{oi}(\omega\mapsto\omega).
\]
\end{defn}
\smallskip{}

The meaning of \emph{knowing} a hyperconfusion $\mathsf{X}$ is that, when the outcome $\omega$ is unknown (or random),  we can know a set $A\in\mathsf{X}$ (which may depend on $\omega$) satisfying $\omega\in A$. If $\mathsf{X}=\mathrm{oi}(f)$, then knowing $\mathsf{X}$ is equivalent to knowing the value of the random variable $Y=f(\omega)$, since knowing $\mathsf{X}$ (i.e., knowing $A\in\mathsf{X}$, $A\ni\omega$) allows us to obtain $Y$ as the unique element of the image $f(A)$, and knowing $Y$ allows us to know $A=f^{-1}(\{Y\})\in\mathsf{X}$ with $A\ni\omega$.

If $\mathsf{X}\subseteq\mathsf{Y}$, then we say that $\mathsf{X}$ is\emph{ less ambiguous} (or \emph{more informative}) than $\mathsf{Y}$. Knowing $\mathsf{X}$ implies knowing $\mathsf{Y}$, since if we know $A\in\mathsf{X}$, $A\ni\omega$, then we also know $A\in\mathsf{Y},A\ni\omega$.   We prefer using ``less ambiguous'' over ``more informative'' since the ambiguity ordering coincides with the inclusion ordering, i.e., a smaller hyperconfusion confuses fewer sets and contains less ambiguity.

The most ambiguous hyperconfusion is $2^{\Omega}$, called the \emph{full hyperconfusion}, corresponding to knowing nothing (every outcome is confusable). It is the ordinary information of a constant function, i.e., $2^{\Omega}=\mathrm{oi}(\omega\mapsto c)$ for fixed constant $c$. The least ambiguous hyperconfusion is $2^{\emptyset}=\{\emptyset\}$, called the \emph{null hyperconfusion}, corresponding to \emph{impossibility} or \emph{omniscience} (knowing everything in the present and the future; see Section \ref{subsec:conjunction}). We have $2^{\emptyset}\subseteq\mathsf{X}\subseteq2^{\Omega}$ for every hyperconfusion $\mathsf{X}$, i.e., knowing $2^{\emptyset}$ implies knowing $\mathsf{X}$, and knowing $\mathsf{X}$ implies knowing $2^{\Omega}$.

Note that $2^{\emptyset}$ is even less ambiguous than the singleton hyperconfusion $\mathrm{sing}(\Omega)=\mathrm{oi}(\omega\mapsto\omega)$, corresponding to knowing $\omega$ precisely (only the empty set and singleton sets are confusable). The singleton hyperconfusion represents knowing everything about the current sample space $\Omega$ (which is possible), whereas the null hyperconfusion represents knowing every information about the current space as well as the future (which is impossible).  The impossibility of knowing $2^{\emptyset}$ is due to the fact that it is impossible to have $A\in2^{\emptyset}$ satisfying $A\ni\omega$, regardless of $\omega$. Generally, a hyperconfusion $\mathsf{X}$ is not required to confuse singleton sets, i.e., $\mathrm{supp}(\mathsf{X})$ does not need to be $\Omega$. This allows us to consider events as hyperconfusions (Section \ref{subsec:negation}). A hyperconfusion that does not confuse all singleton sets corresponds to a piece of knowledge that has an assumption, such as knowing whether a person's car has a roof (which assumes the person has a car); see Section \ref{sec:gen}.

The goal of this paper is to show that hyperconfusions have a richer structure and support more operations compared to ordinary informations. In case if the situation requires ordinary information, we can still convert a hyperconfusion to an ordinary information with a small penalty, which will be shown in Section \ref{subsec:convert}.

\smallskip{}

\subsection{Entropy of Hyperconfusion}\label{subsec:entropy}

We define the entropy of a hyperconfusion using a generalization of the source coding setting in \cite{korner1971coding} (also see \cite{posner1971epsilon}). Recall that knowing $\mathsf{X}$ is to know a set $A\in\mathsf{X}$ satisfying $\omega\in A$. One can regard $A$ as a lossy compression of $\omega$, with a distortion function $d:\Omega\times\mathsf{X}\to[0,\infty]$ where $d(\omega,A)=0$ if $\omega\in A$ and $d(\omega,A)=\infty$ otherwise. By rate-distortion theory \cite{berger1971rate}, the optimal asymptotic compression rate is given by the following quantity, which we call the \emph{entropy} of $\mathsf{X}$. An operational meaning via one-shot compression will also be given in Section \ref{subsec:convert}.

\smallskip{}

\begin{defn}
\label{def:entropy}For a hyperconfusion $\mathsf{X}$ over a probability space $(\Omega,p)$, its \emph{entropy} is 
\[
H(\mathsf{X}):=\min_{p_{A|Z}:\,Z\in A\in\mathsf{X}\;\mathrm{a.s.}}I(Z;A),
\]
where $Z\in\Omega$ follows the distribution $p$, and we minimize over conditional probability mass functions $p_{A|Z}$ from $\Omega$ to $\mathsf{X}$ which satisfies that $Z\in A$ almost surely where $A\in\mathsf{X}$ is a random set following $p_{A|Z}$ given $Z$. If no such $p_{A|Z}$ exists, we take $H(\mathsf{X})=\infty$.
\end{defn}
\smallskip{}

This quantity generalizes K\"{o}rner graph entropy \cite{korner1971coding} from confusion graphs to hypergraphs. A similar quantity has been studied in \cite{korner1990capacity,korner1988new}, which concerns uniform hypergraphs of distinguishable elements instead of confusable elements, and is more restrictive since it concerns only zero-error channel coding with list decoding of a fixed list length. Definition \ref{def:entropy} can also be expressed as the entropy of a convex corner \cite{grotschel1986relaxations,csiszar1990entropy,simonyi1995graph}, as shown below. The proof is in Appendix \ref{subsec:pf_convex}.

\smallskip{}

\begin{prop}
\label{prop:convex}Let $\mathbf{1}_{A}\in\{0,1\}^{\Omega}$ be the indicator vector of $A\in\mathsf{X}$, i.e., $\mathbf{1}_{A}(\omega)=1$ if and only if $\omega\in A$. Let $\mathcal{C}\subseteq\mathbb{R}^{\Omega}$ be the convex hull of the vectors $\mathbf{1}_{A}$ for $A\in\mathsf{X}$. We have
\[
H(\mathsf{X})=\min_{v\in\mathcal{C}}\sum_{\omega\in\Omega}p(\omega)\log\frac{1}{v(\omega)}.
\]
We treat $0\log(1/0)=0$ and $p(\omega)\log(1/0)=\infty$ if $p(\omega)>0$.
\end{prop}
\smallskip{}

We also define the min-entropy and the max-entropy which generalizes the conventional min-entropy \cite{renyi1961measures} $H_{\infty}(X)=-\log\max_{x}p_{X}(x)$ and max-entropy (Hartley entropy) \cite{hartley1928transmission} $H_{0}(X)=\log|\{x:p_{X}(x)>0\}|$ of ordinary random variables. There are two generalizations of max-entropy. One is based on integral covering number, and the other is based on fractional covering number.

\smallskip{}

\begin{defn}
\label{def:entropy-1}For a hyperconfusion $\mathsf{X}$ over a probability space $(\Omega,p)$, its \emph{min-entropy} is 
\[
H_{\infty}(\mathsf{X}):=-\log\max_{A\in\mathsf{X}}p(A),
\]
its \emph{integral max-entropy} is
\[
H_{0}(\mathsf{X}):=\log\min_{\mathcal{S}\subseteq\mathsf{X}:\,\bigcup_{A\in\mathcal{S}}A=\mathrm{supp}(p)}|\mathcal{S}|,
\]
where $\mathrm{supp}(p):=\{\omega\in\Omega:\,p(\omega)>0\}$ (take $H_{0}(\mathsf{X})=\infty$ if no such $\mathcal{S}$ exists), and its \emph{fractional max-entropy} is
\[
H_{\epsilon}(\mathsf{X}):=\log\min_{\mu:\mathsf{X}\to[0,\infty)}\sum_{A\in\mathsf{X}}\mu(A),
\]
where the minimum is over functions $\mu:\mathsf{X}\to[0,\infty)$ satisfying $\sum_{A\in\mathsf{X},A\ni\omega}\mu(A)\ge1$ for all $\omega\in\mathrm{supp}(p)$ (i.e., $\mu$ is a fractional covering of $\mathrm{supp}(p)$). If no such $\mu$ exists, we take $H_{\epsilon}(\mathsf{X})=\infty$.
\end{defn}
\smallskip{}

If the probability $p$ is not given, we define $H_{0}$ and $H_{\epsilon}$ by assuming $\mathrm{supp}(p)=\Omega$. Intuitively, $H(\mathsf{X})$ measures the compression size of $\mathsf{X}$ in either variable-length encoding or asymptotic settings, whereas $H_{0}(\mathsf{X})$ measures its size in fixed-length encoding. The meaning of $H_{\infty}(\mathsf{X})$ is less clear, but we will see in Section \ref{subsec:source_error} that it has an unexpected connection to the probability of decoding success. We show some properties of the entropies. The proof is in Appendix \ref{subsec:pf_properties}.

\smallskip{}

\begin{prop}
\label{prop:properties}The following holds for every hyperconfusion $\mathsf{X}$ over $(\Omega,p)$ and every choice of $a\in\{0,\epsilon,1,\infty\}$ (where $H_{1}(\mathsf{X})=H(\mathsf{X})$):
\begin{itemize}
\item (Comparison)
\[
H_{\infty}(\mathsf{X})\le H(\mathsf{X})\le H_{\epsilon}(\mathsf{X})\le H_{0}(\mathsf{X}).
\]
\item (Full and null) $H_{a}(2^{\Omega})=0$ and $H_{a}(2^{\emptyset})=\infty$.
\item (Non-negativity) $H_{a}(\mathsf{X})\ge0$, with equality if and only if $2^{\mathrm{supp}(p)}\subseteq\mathsf{X}$.
\item (Finiteness) $H_{a}(\mathsf{X})<\infty$ if and only if $\mathrm{supp}(p)\subseteq\mathrm{supp}(\mathsf{X})$ for $a\in\{0,\epsilon,1\}$. $H_{\infty}(\mathsf{X})<\infty$ if and only if $\mathrm{supp}(p)\cap\mathrm{supp}(\mathsf{X})\neq\emptyset$.
\item (Monotonicity) If $\mathsf{X}\subseteq\mathsf{Y}$, then $H_{a}(\mathsf{X})\ge H_{a}(\mathsf{Y})$.
\item (Generalizing ordinary entropy) If $\mathsf{X}=\mathrm{oi}(\tilde{X})$ for an ordinary random variable $\tilde{X}$, then $H_{a}(\mathsf{X})=H_{a}(\tilde{X})$ (take $H_{\epsilon}(\tilde{X})=H_{0}(\tilde{X})$). 
\end{itemize}
\end{prop}
\smallskip{}

\subsection{Converting Hyperconfusion to Ordinary Information}\label{subsec:convert}

One concern about hyperconfusion is that it might not be applicable to situations where ordinary information is expected. For example, we can use Huffman coding \cite{huffman1952method} to compress a random variable $X$ to a prefix-free codeword with expected length less than $H(X)+1$, but it is unclear whether we can do so for a hyperconfusion $\mathsf{X}$.  The following result, called the \emph{unconfusing lemma}, is a simple yet central result to this paper which shows that a hyperconfusion can always be converted into ordinary information with only a small loss. More specifically, for any hyperconfusion $\mathsf{X}$, we can always find an ordinary information $\mathsf{Y}\in\mathrm{OIs}(\Omega)$ that is less ambiguous than $\mathsf{X}$ (i.e., $\mathsf{Y}\subseteq\mathsf{X}$), with $H(\mathsf{Y})\approx H(\mathsf{X})$ within a logarithmic gap. As a result, $\mathsf{X}$ can be compressed using prefix-free code with expected length within a logarithmic gap from $H(\mathsf{X})$, justifying the use of $H(\mathsf{X})$ as a measure of information. It is proved in Appendix \ref{subsec:pf_unconfuse} using the strong functional representation lemma \cite{sfrl_trans,li2025discrete}. Alternatively, we may also use greedy rejection sampling \cite{harsha2010communication} or the results on epsilon entropy in \cite{posner1971epsilon,kostina2026single} to give similar (but slightly weaker) bounds.

\smallskip{}

\begin{lem}
[Unconfusing lemma]\label{lem:unconfuse}For any hyperconfusion $\mathsf{X}$ over a probability space $(\Omega,p)$, 
\[
H(\mathsf{X})\le\min_{\mathsf{Y}\in\mathrm{OIs}(\Omega):\,\mathsf{Y}\subseteq\mathsf{X}}H(\mathsf{Y})\le H(\mathsf{X})+\log(H(\mathsf{X})+3.4)+1.
\]
\end{lem}
\smallskip{}

If we consider min-entropy or integral max-entropy instead, then the logarithmic gap disappears, as shown in the following lemma. The proof is in Appendix \ref{subsec:pf_unconfuse}.

\smallskip{}

\begin{lem}
[Unconfusing lemma for $H_{0}$ and $H_{\infty}$]\label{lem:unconfuse2}For any hyperconfusion $\mathsf{X}$ over a probability space $(\Omega,p)$, 
\[
\min_{\mathsf{Y}\in\mathrm{OIs}(\Omega):\,\mathsf{Y}\subseteq\mathsf{X}}H_{0}(\mathsf{Y})=H_{0}(\mathsf{X}).
\]
The same holds with $H_{\infty}$ in place of $H_{0}$.
\end{lem}
\smallskip{}

\section{The Operations of Hyperconfusion}\label{sec:operations}

In this section, we study various operations on hyperconfusions, namely conjunction $\mathsf{X}\cap\mathsf{Y}$, disjunction $\mathsf{X}\cup\mathsf{Y}$, implication $\mathsf{X}\rightarrow\mathsf{Y}$ and negation $\lnot\mathsf{X}$. We will see in Section \ref{subsec:Heyting} that these operations turn $\mathrm{Hyps}(\Omega)$ into a Heyting algebra. This can be shown to be equivalent to the Heyting algebra of propositions in generalized inquisitive logic \cite{roelofsen2013algebraic}.

\smallskip{}

\subsection{Conjunction and Product of Information}\label{subsec:conjunction}

Given two hyperconfusions $\mathsf{X},\mathsf{Y}$, their \emph{conjunction} is $\mathsf{X}\cap\mathsf{Y}$, where a set $A$ is confusable if and only if it is confusable in both $\mathsf{X}$ and $\mathsf{Y}$. Intuitively, $\mathsf{X}\cap\mathsf{Y}$ is the knowledge of both $\mathsf{X}$ and $\mathsf{Y}$, in the sense that \emph{knowing both $\mathsf{X}$ and $\mathsf{Y}$ is equivalent to knowing $\mathsf{X}\cap\mathsf{Y}$}. To show the forward direction, knowing $\mathsf{X}$ and $\mathsf{Y}$ means knowing $A\in\mathsf{X}$, $A\ni\omega$ and $B\in\mathsf{Y}$, $B\ni\omega$, so we can form $A\cap B\in\mathsf{X}\cap\mathsf{Y}$ and $A\cap B\ni\omega$ ($A\cap B\in\mathsf{X}$ since $A\in\mathsf{X}$ and $\mathsf{X}$ is downward closed), and hence we know $\mathsf{X}\cap\mathsf{Y}$. For the reverse direction, knowing $C\in\mathsf{X}\cap\mathsf{Y}$ and $C\ni\omega$ gives us $\omega\in C\in\mathsf{X}$ and $\omega\in C\in\mathsf{Y}$, so we know $\mathsf{X}$ and $\mathsf{Y}$.

\begin{figure}
\begin{centering}
\includegraphics{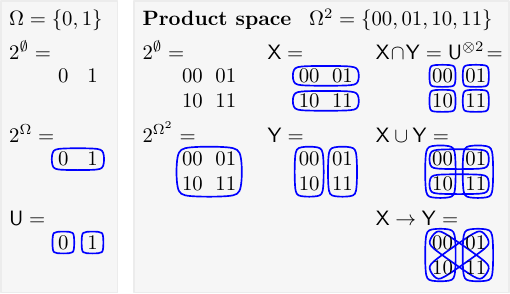}
\par\end{centering}
\caption{The hyperconfusion $\mathsf{U}=\mathrm{sing}(\Omega)$ over $\Omega=\{0,1\}$, its embeddings $\mathsf{X},\mathsf{Y}$ in $\Omega^{2}$, and the operations $\mathsf{X}\cap\mathsf{Y}=\mathsf{U}\otimes\mathsf{U}$, $\mathsf{X}\cup\mathsf{Y}$ and $\mathsf{X}\rightarrow\mathsf{Y}$. The blue circles are maximal confusable sets (no strict supersets are confusable).}\label{fig:op}
\end{figure}

In probability theory, when new randomness is generated, we extend the probability space, for example, by taking the product probability space, and embed the existing random variables into the product probability space. For two hyperconfusions $\mathsf{X}_{1},\mathsf{X}_{2}$ over $\Omega_{1},\Omega_{2}$, respectively, we can also embed them as $\tilde{\mathsf{X}}_{1},\tilde{\mathsf{X}}_{2}$ in the product space $\Omega_{1}\times\Omega_{2}$, so that they correspond to $\mathsf{X}_{1},\mathsf{X}_{2}$ generated independently.

\smallskip{}

\begin{defn}
\label{def:product}Given a hyperconfusion $\mathsf{X}_{i}$ over the sample space $\Omega_{i}$ for $i=1,\ldots,n$, consider the product sample space $\Omega:=\prod_{i=1}^{n}\Omega_{i}$. The \emph{embeddings} of $\mathsf{X}_{1},\ldots,\mathsf{X}_{n}$ onto the product space $\Omega$ are given by $\tilde{\mathsf{X}}_{1},\ldots,\tilde{\mathsf{X}}_{n}$, where 
\[
\tilde{\mathsf{X}}_{i}:=\bigcup_{A\in\mathsf{X}_{i}}2^{\Omega_{1}\times\cdots\times\Omega_{i-1}\times A\times\Omega_{i+1}\times\cdots\times\Omega_{n}}
\]
is a hyperconfusion over $\Omega$. The \emph{product hyperconfusion} is defined as
\[
\bigotimes_{i=1}^{n}\mathsf{X}_{i}:=\bigcap_{i=1}^{n}\tilde{\mathsf{X}}_{i}=\bigcup_{A_{1}\in\mathsf{X}_{1},\ldots,A_{n}\in\mathsf{X}_{n}}\!\!2^{\prod_{i=1}^{n}A_{i}}.
\]
In particular, given a hyperconfusion $\mathsf{X}$ over the probability space $(\Omega,p)$, we can define an \emph{i.i.d. hyperconfusion sequence} $\tilde{\mathsf{X}}_{1},\ldots,\tilde{\mathsf{X}}_{n}$ over the product probability space $(\Omega^{n},p^{\otimes n})$, by taking $\tilde{\mathsf{X}}_{1},\ldots,\tilde{\mathsf{X}}_{n}$ to be the embeddings of $\mathsf{X},\ldots,\mathsf{X}$ ($n$ times) onto the product space $\Omega^{n}$.
\end{defn}
\smallskip{}

For example, consider the sample space of one bit $\Omega=\{0,1\}$ and the hyperconfusion $\mathsf{U}=\mathrm{sing}(\Omega)=\{\emptyset,\{0\},\{1\}\}$. The embeddings of $\mathsf{U},\mathsf{U}$ onto the product $\Omega^{2}=\{00,01,10,11\}$ are 
\begin{align}
\mathsf{X} & =\{\emptyset,\{00\},\{01\},\{10\},\{11\},\{00,01\},\{10,11\}\},\nonumber \\
\mathsf{Y} & =\{\emptyset,\{00\},\{01\},\{10\},\{11\},\{00,10\},\{01,11\}\},\label{eq:twobits}
\end{align}
corresponding to the information of the first and the second bit, respectively. The product hyperconfusion is $\mathsf{U}^{\otimes2}=\mathsf{U}\otimes\mathsf{U}=\mathsf{X}\cap\mathsf{Y}=\{\emptyset,\{00\},\{01\},\{10\},\{11\}\}$. Refer to Figure \ref{fig:op} for an illustration of $\mathsf{X}$, $\mathsf{Y}$ and $\mathsf{X}\cup\mathsf{Y}$ (only maximal confusable sets are drawn). 

If $\mathsf{X}_{1}=2^{\emptyset}$ is the null hyperconfusion over $\Omega_{1}$, then its embedding $\tilde{\mathsf{X}}_{1}$ over $\Omega_{1}\times\Omega_{2}$ is also the null hyperconfusion. Hence, the null hyperconfusion not only has all information in the current space, but also in all future extensions of the space (e.g., knowing all future coin flips). For this reason, null hyperconfusion represents omniscience, which is impossible.

The embeddings $\tilde{\mathsf{X}}_{1},\tilde{\mathsf{X}}_{2}$ over a product probability space are intuitively independent of each other. We now define the general notion of probabilistic independence between two hyperconfusions $\mathsf{X},\mathsf{Y}$. If we require all pairs of confusable sets $A\in\mathsf{X}$, $B\in\mathsf{Y}$ to be independent as events, this is almost never satisfied since $A,B$ can be singleton sets. Therefore, we only require that all pairs of maximal confusable sets (and unions of maximal confusable sets) to be independent.

\smallskip{}

\begin{defn}
\label{def:ind}Given hyperconfusions $\mathsf{X},\mathsf{Y}$ over $(\Omega,p)$, they are \emph{probabilistically independent} if the two events $\bigcup_{A\in\mathcal{S}}A$ and $\bigcup_{B\in\mathcal{T}}B$ are independent for every $\mathcal{S}\subseteq\mathrm{maxs}(\mathsf{X})$ and $\mathcal{T}\subseteq\mathrm{maxs}(\mathsf{Y})$, where $\mathrm{maxs}(\mathsf{X}):=\{A\in\mathsf{X}:\,\forall B\in\mathsf{X}.\,B\supseteq A\,\Rightarrow\,B=A\}$ are the maximal sets in $\mathsf{X}$.
\end{defn}
\smallskip{}

For ordinary random variables, their dependence can be measured by mutual information. We generalize the notion of mutual information to hyperconfusions using the same inclusion-exclusion formula.

\smallskip{}

\begin{defn}
\label{def:mutual}Given hyperconfusions $\mathsf{X},\mathsf{Y}$ over $(\Omega,p)$, their \emph{mutual information} is
\[
I(\mathsf{X};\mathsf{Y}):=H(\mathsf{X})+H(\mathsf{Y})-H(\mathsf{X}\cap\mathsf{Y}).
\]
\end{defn}
\smallskip{}

The following proposition shows that $I(\mathsf{X};\mathsf{Y})\ge0$. A similar result about uniform hypergraphs of distinguished elements was given in \cite{korner1988new,korner1990capacity}. The proof is in Appendix \ref{subsec:pf_conjunction}.

\smallskip{}

\begin{prop}
\label{prop:conjunction}For hyperconfusions $\mathsf{X},\mathsf{Y}$ over $(\Omega,p)$ with $H(\mathsf{X}),H(\mathsf{Y})<\infty$, we have $I(\mathsf{X};\mathsf{Y})\ge0$, or equivalently, $H(\mathsf{X}\cap\mathsf{Y})\le H(\mathsf{X})+H(\mathsf{Y})$. Equality holds if $\mathsf{X}$ and $\mathsf{Y}$ are probabilistically independent.
\end{prop}
\smallskip{}

We now describe the \emph{conjunctive source coding} setting for hyperconfusions, which generalizes the setting in \cite{korner1971coding}. Consider a hyperconfusion $\mathsf{X}$ over the probability space $(\Omega,p)$, which induces an i.i.d. hyperconfusion sequence $\mathsf{X}_{1},\ldots,\mathsf{X}_{n}$ over the product probability space $(\Omega^{n},p^{\otimes n})$. Proposition \ref{prop:conjunction} tells us that $H(\bigcap_{i=1}^{n}\mathsf{X}_{i})=nH(\mathsf{X})$. The unconfusing lemma (Lemma \ref{lem:unconfuse}) tells us that there exists an ordinary information $\mathsf{M}=\mathrm{oi}(M)$ with $H(M)\le nH(\mathsf{X})+\log(nH(\mathsf{X})+3.4)+1$ and $\mathsf{M}\subseteq\bigcap_{i=1}^{n}\mathsf{X}_{i}$, i.e., we can know $\mathsf{X}_{1},\ldots,\mathsf{X}_{n}$ using $\mathsf{M}$. In other words, there exists an encoding function $f:\Omega^{n}\to\mathcal{M}$ and a decoding function $g:\mathcal{M}\to\mathsf{X}^{n}$ such that when $Z_{1},\ldots,Z_{n}\in\Omega$ are i.i.d. following $p$, $M=f(Z_{1},\ldots,Z_{n})$, $g(M)=(A_{1},\ldots,A_{n})$, then we have $Z_{i}\in A_{i}$ for $i=1,\ldots,n$ almost surely. The encoding rate is $n^{-1}H(M)=H(\mathsf{X})+O(n^{-1}\log n)$ which approaches $H(\mathsf{X})$. This gives the operational meaning of $H(\mathsf{X})$. The following theorem follows immediately.

\smallskip{}

\begin{thm}
[Conjunctive source coding theorem]Consider a hyperconfusion $\mathsf{X}$ over the probability space $(\Omega,p)$, and its $n$-fold i.i.d. hyperconfusion sequence $\mathsf{X}_{1},\ldots,\mathsf{X}_{n}$ over the product space. We have 
\[
\lim_{n}\frac{H(\bigcap_{i=1}^{n}\mathsf{X}_{i})}{n}=\lim_{n}\min_{\mathsf{M}\in\mathrm{OIs}(\Omega^{n}):\,\mathsf{M}\subseteq\bigcap_{i=1}^{n}\mathsf{X}_{i}}\frac{H(\mathsf{M})}{n}=H(\mathsf{X}).
\]
\end{thm}
\smallskip{}

\subsection{Disjunction of Information}\label{subsec:disjunction}

Given two hyperconfusions $\mathsf{X},\mathsf{Y}$, their \emph{disjunction} is $\mathsf{X}\cup\mathsf{Y}$, where a set $C$ is confusable if and only if it is confusable in $\mathsf{X}$ or in $\mathsf{Y}$. Intuitively, $\mathsf{X}\cup\mathsf{Y}$ is the knowledge of knowing at least one of $\mathsf{X}$, $\mathsf{Y}$ at a time, in the sense that \emph{knowing $\mathsf{X}$ or $\mathsf{Y}$ is equivalent to knowing $\mathsf{X}\cup\mathsf{Y}$}. Here, knowing $\mathsf{X}$ or $\mathsf{Y}$ means knowing $C\ni\omega$ where $C\in\mathsf{X}$ or $C\in\mathsf{Y}$,  which is the same as knowing $C\in\mathsf{X}\cup\mathsf{Y}$, $C\ni\omega$. 

Knowing $\mathsf{X}\cup\mathsf{Y}$ does not mean that we must always know $\mathsf{X}$ or always know $\mathsf{Y}$. If $\mathsf{X}=\mathrm{oi}(X)$, $\mathsf{Y}=\mathrm{oi}(Y)$, $X\in\mathcal{X}$, $Y\in\mathcal{Y}$ are ordinary random variables, knowing $\mathsf{X}\cup\mathsf{Y}$ means outputting a pair $(K,Z)\in(\{1\}\times\mathcal{X})\cup(\{2\}\times\mathcal{Y})$, where ``($K=1$ and $X=Z$) or ($K=2$ and $Y=Z$)'' always holds, i.e., we declare which one of $X,Y$ we know, and output the value we know. This is weaker than always knowing $X$ or always knowing $Y$, since $K$ can depend on the values of $(X,Y)$. For example, consider two runners in a race, where $X,Y$ are their completion times behind the winner (i.e., how many seconds they are slower than the winner). We can convey $\mathsf{X}\cup\mathsf{Y}$ using $(K,0)$, where $K\in\{1,2\}$ indicates the winner, and the winner's time behind the winner is always $0$. This only takes $1$ bits, much easier than conveying $X$ or conveying $Y$.

By monotonicity (Proposition \ref{prop:properties}), for $a\in\{0,\epsilon,1,\infty\}$,
\[
H_{a}(\mathsf{X}\cup\mathsf{Y})\le\min\{H_{a}(\mathsf{X}),H_{a}(\mathsf{Y})\}.
\]
For $H_{\infty}$, equality always holds by definition. For $H$, this inequality can be strict, i.e., $\mathsf{X}\cup\mathsf{Y}$ can carry strictly less information than both $\mathsf{X}$ and $\mathsf{Y}$, even when $\mathsf{X}$, $\mathsf{Y}$ are independent ordinary informations. 

To demonstrate this phenomenon, consider the following setting which we call \emph{asymptotic disjunctive source coding}, which is to compress an i.i.d. sequence such that \emph{one of the entries} can be decoded, unlike the usual conjunctive source coding theorem where all entries should be decoded. Consider a hyperconfusion $\mathsf{X}$ over the probability space $(\Omega,p)$, and its $n$-fold i.i.d. hyperconfusion sequence $\mathsf{X}_{1},\ldots,\mathsf{X}_{n}$ over the product space (Definition \ref{def:product}). We are interested in the limit $H^{*}=\lim_{n}H(\bigcup_{i=1}^{n}\mathsf{X}_{i})$. For the special case where $\mathsf{X}=\mathrm{oi}(X)$ is an ordinary information, the unconfusing lemma (Lemma \ref{lem:unconfuse}) tells us that there exists $n$ and an ordinary information $\mathsf{M}=\mathrm{oi}(M)$ with $\mathsf{M}\subseteq\bigcup_{i=1}^{n}\mathsf{X}_{i}$ and $H(M)\le H^{*}+\log(H^{*}+3.4)+1$, so there exists an encoding function $f:\mathcal{X}^{n}\to\mathcal{M}$ and decoding functions $k:\mathcal{M}\to[n]$, $g:\mathcal{M}\to\mathcal{X}$ such that $X_{k(M)}=g(M)$, i.e., the decoder declares that the $k(M)$-th entry is decoded, and output its value as $g(M)$. The compression size $H(M)$ is approximately $H^{*}$ within a logarithmic gap. This $H^{*}$ can be shown to coincide with the min-entropy $H_{\infty}(\mathsf{X})$, which can be strictly less than $H(\mathsf{X})$, so merely compressing $\mathsf{X}_{1}$ can be suboptimal. The proof is in Appendix \ref{subsec:pf_disjunctive}.

\smallskip{}

\begin{thm}
[Disjunctive source coding theorem]\label{thm:disjunctive}Consider a hyperconfusion $\mathsf{X}$ over the probability space $(\Omega,p)$ with $\mathrm{supp}(\mathsf{X})\supseteq\mathrm{supp}(p)$, and its $n$-fold i.i.d. hyperconfusion sequence $\mathsf{X}_{1},\ldots,\mathsf{X}_{n}$ over the product space. We have $\lim_{n}H(\bigcup_{i=1}^{n}\mathsf{X}_{i})=H_{\infty}(\mathsf{X})$.
\end{thm}
\smallskip{}

\subsection{Conditioning of Information}

Hyperconfusion can capture the concept of events as well. The event $E\subseteq\Omega$ corresponds to the hyperconfusion $2^{E}$, which we call the \emph{event hyperconfusion} of $E$. Knowing $2^{E}$ means knowing that the event $E$ occurs.  Note that $\mathrm{supp}(2^{E})=E$,
\[
H(2^{E})=H_{0}(2^{E})=H_{\epsilon}(2^{E})=\begin{cases}
0 & \mathrm{if}\;p(E)=1,\\
\infty & \mathrm{if}\;p(E)<1,
\end{cases}
\]
and
\begin{equation}
H_{\infty}(2^{E})=-\log p(E)\label{eq:Hinfty_event}
\end{equation}
is the self-information \cite{shannon1948mathematical}. In probability theory, a basic operation is conditioning, where we restrict a probability space $(\Omega,p)$ to the conditional probability space $(E,p(\cdot|E))$, where $p(\omega|E)=p(\omega)/p(E)$ for $\omega\in E$. Conditioning for hyperconfusion can also be defined.\smallskip{}

\begin{defn}
\label{def:condition}Given hyperconfusion $\mathsf{X}$ over the probability space $(\Omega,p)$, and an event $E\subseteq\Omega$, the \emph{conditional hyperconfusion} is defined as
\[
\mathsf{X}\upharpoonright E:=\mathsf{X}\cap2^{E},
\]
which is a hyperconfusion over the conditional probability space $(E,p(\cdot|E))$. The \emph{conditional entropy} of $\mathsf{X}$ given $E$ is
\[
H(\mathsf{X}|E):=H(\mathsf{X}\upharpoonright E),
\]
where $H(\mathsf{X}\upharpoonright E)$ is evaluated over $(E,p(\cdot|E))$.
\end{defn}
\smallskip{}

The reason we introduce the notation $\mathsf{X}\upharpoonright E$ instead of simply writing $\mathsf{X}\cap2^{E}$ is that $\mathsf{X}\cap2^{E}$ is interpreted as a hyperconfusion over the original sample space $\Omega$, whereas $\mathsf{X}\upharpoonright E$ is interpreted as a hyperconfusion over the conditional space $E$. Therefore, $H(\mathsf{X}\upharpoonright E)$ is different from $H(\mathsf{X}\cap2^{E})$ (which is often infinite).

We can also define the conditional entropy of a hyperconfusion given another hyperconfusion, which is given using the same inclusion-exclusion formula as Shannon entropy. Nevertheless, since defining $H(\mathsf{X}|\mathsf{Y})$ as $H(\mathsf{X}\cap\mathsf{Y})-H(\mathsf{Y})$ may result in an indeterminate form ``$\infty-\infty$'', we have to condition on the support of $\mathsf{Y}$ to make the negative term finite.

\smallskip{}

\begin{defn}
\label{def:condition_HXY}Given hyperconfusions $\mathsf{X},\mathsf{Y}$ over $(\Omega,p)$, the \emph{conditional entropy} of $\mathsf{X}$ given $\mathsf{Y}$ is
\[
H(\mathsf{X}|\mathsf{Y}):=p(E)\big(H(\mathsf{X}\cap\mathsf{Y}|E)-H(\mathsf{Y}|E)\big),
\]
where $E:=\mathrm{supp}(\mathsf{Y})$. The above is taken to be $0$ if $p(E)=0$.\footnote{Alternatively, we can take $H(\mathsf{X}|\mathsf{Y})=H(\lnot\lnot\mathsf{Y}\rightarrow(\mathsf{X}\cap\mathsf{Y}))-H(\lnot\lnot\mathsf{Y}\rightarrow\mathsf{Y})$, where ``$\rightarrow$'' and ``$\lnot$'' are defined in later sections.} 
\end{defn}
\smallskip{}

We give some properties of the conditional entropy. The proof is in Appendix \ref{subsec:pf_condition}.

\smallskip{}

\begin{prop}
\label{prop:condition}For hyperconfusions $\mathsf{X},\mathsf{Y},\mathsf{Z}$, we have the following properties:
\begin{itemize}
\item $0\le H(\mathsf{X}|\mathsf{Y})\le H(\mathsf{X})$.
\item $H(\mathsf{X}|2^{\Omega})=H(\mathsf{X})$ and $H(\mathsf{X}|2^{\emptyset})=0$.
\item $H(\mathsf{X}|\mathsf{Y})+H(\mathsf{Y})=H(\mathsf{X}\cap\mathsf{Y})$.
\item If $\mathsf{Y}\cap\mathsf{Z}=2^{\emptyset}$, then $H(\mathsf{X}|\mathsf{Y}\cup\mathsf{Z})=H(\mathsf{X}|\mathsf{Y})+H(\mathsf{X}|\mathsf{Z})$.
\item If $\mathsf{Y}\in\mathrm{OIs}(\Omega)$, then
\[
H(\mathsf{X}|\mathsf{Y})=\sum_{E\in\mathrm{maxs}(\mathsf{Y})}p(E)H(\mathsf{X}|E),
\]
where $\mathrm{maxs}(\mathsf{Y})$ are the maximal confusable sets in $\mathsf{Y}$ (Definition \ref{def:ind}).  In particular, if $\mathsf{Y}=\mathrm{oi}(Y)$ for an ordinary random variable $Y\in\mathcal{Y}$, then $H(\mathsf{X}|\mathsf{Y})=\sum_{y\in\mathcal{Y}}p_{Y}(y)H(\mathsf{X}|Y=y)$. Also, $H(\mathsf{X}|2^{E})=p(E)H(\mathsf{X}|E)$ for $E\subseteq\Omega$.
\end{itemize}
\end{prop}
\smallskip{}

Note that $H(\mathsf{X}|\mathsf{Y})=0$ does not imply $\mathsf{X}\supseteq\mathsf{Y}$, and hence $H(\mathsf{X}|\mathsf{Y})$ does not exactly have the interpretation ``the amount of information in $\mathsf{X}$ that is not in $\mathsf{Y}$''. For example, consider the two bits example in (\ref{eq:twobits}), we have  $H(\mathsf{X}|\mathsf{X}\cup\mathsf{Y})=H(\mathsf{X})-H(\mathsf{X}\cup\mathsf{Y})=1-1=0$. In the next section, we will see how we can take the difference between two pieces of information.

We also remark that we can define conditional mutual information by combining Definitions \ref{def:mutual} and \ref{def:condition_HXY}:
\[
I(\mathsf{X};\mathsf{Y}|\mathsf{Z}):=H(\mathsf{X}|\mathsf{Z})+H(\mathsf{Y}|\mathsf{Z})-H(\mathsf{X}\cap\mathsf{Y}|\mathsf{Z}).
\]
If $\mathsf{Z}\in\mathrm{OIs}(\Omega)$ and $H(\mathsf{X}|\mathsf{Z}),H(\mathsf{Y}|\mathsf{Z})<\infty$, then Propositions \ref{prop:conjunction} and \ref{prop:condition} imply $I(\mathsf{X};\mathsf{Y}|\mathsf{Z})\ge0$, with equality if $\mathsf{X}\upharpoonright E$ is independent of $\mathsf{Y}\upharpoonright E$ for every $E\in\mathrm{maxs}(\mathsf{Z})$ with $p(E)>0$ (i.e., $\mathsf{X},\mathsf{Y}$ are conditionally independent given $\mathsf{Z}$). Nevertheless, for general $\mathsf{Z}$, it is possible for $I(\mathsf{X};\mathsf{Y}|\mathsf{Z})$ to be negative. For example, for the $\mathsf{X},\mathsf{Y}$ in (\ref{eq:twobits}), we have $I(\mathsf{X};\mathsf{Y}|\mathsf{X}\cup\mathsf{Y})=H(\mathsf{X})+H(\mathsf{Y})-H(\mathsf{X}\cap\mathsf{Y})-H(\mathsf{X}\cup\mathsf{Y})=-1$. We can see that entropy is not a submodular function.

\smallskip{}

\subsection{Implication and Difference of Information}\label{subsec:implication}

Given two pieces of information $X,Y$, we are often interested in finding ``$Y\backslash X$'', i.e., the part of $Y$ that is not in $X$. In other words, we want to find the smallest information $M$ such that $Y$ can be deduced from $X$ and $M$. For ordinary information, this problem is often ill-defined. For example, consider the situation where $X,Y\in\{0,1\}$ are binary random variables. Then the part of $Y$ that is not in $X$ can simply be taken to be $M=Y$, or it can also be taken to be $M=(X+Y)\;\mathrm{mod}\;2$ (the XOR of $X$ and $Y$). Both satisfy that $Y$ can be deduced from $(X,M)$. In terms of refinement of partition (or inclusion of $\sigma$-algebra), $M=Y$ and $M=(X+Y)\;\mathrm{mod}\;2$ are incomparable, so we cannot find a single choice that is the ``least informative'' in the sense of having the most coarse partition. In terms of entropy, whether $M=Y$ or $M=(X+Y)\;\mathrm{mod}\;2$ has a smaller entropy depends on the joint distribution of $X,Y$. Refer to \cite{sfrl_trans,li2017extended,arikan1996inequality,bunte2014encoding,li2025discrete} for attempts to define the difference between two pieces of (ordinary) information.

Interestingly, this problem can be eliminated using hyperconfusion. We want to find the most ambiguous (least informative) hyperconfusion $\mathsf{M}$ such that $\mathsf{X}\cap\mathsf{M}\subseteq\mathsf{Y}$, i.e., we can know $\mathsf{Y}$ if we know $\mathsf{X}$ and $\mathsf{M}$. Since $\mathsf{M}$ is downward closed, we can include $A\subseteq\Omega$ (and all subsets of $A$) in $\mathsf{M}$ if $\mathsf{X}\cap2^{A}\subseteq\mathsf{Y}$. Hence, the most ambiguous $\mathsf{M}$ is as follows.

\smallskip{}

\begin{defn}
For hyperconfusions $\mathsf{X},\mathsf{Y}$ over $\Omega$, the \emph{Heyting implication} is defined as\footnote{This is equivalent to the relative pseudo-complement in \cite{roelofsen2013algebraic}.} 
\begin{align*}
\mathsf{X}\rightarrow\mathsf{Y} & :=\left\{ A\subseteq\Omega:\,\mathsf{X}\cap2^{A}\subseteq\mathsf{Y}\right\} \\
 & =\bigcup_{\mathsf{Z}\in\mathrm{Hyps}(\Omega):\,\mathsf{X}\cap\mathsf{Z}\subseteq\mathsf{Y}}\mathsf{Z}.
\end{align*}
\end{defn}
\smallskip{}

It is straightforward to check that $\mathsf{X}\rightarrow\mathsf{Y}$ is the most ambiguous hyperconfusion $\mathsf{M}$ such that $\mathsf{X}\cap\mathsf{M}\subseteq\mathsf{Y}$ with respect to inclusion, that is, for a hyperconfusion $\mathsf{M}$, we have $\mathsf{X}\cap\mathsf{M}\subseteq\mathsf{Y}$ if and only if $\mathsf{M}\subseteq\mathsf{X}\rightarrow\mathsf{Y}$. In particular, $\mathsf{X}\subseteq\mathsf{Y}$ if and only if $\mathsf{X}\rightarrow\mathsf{Y}\,=\,2^{\Omega}$ is the full hyperconfusion.

The notation $\mathsf{X}\rightarrow\mathsf{Y}$ might be confused with the notation for functions, which is intentional since $\mathsf{X}\rightarrow\mathsf{Y}$ is indeed the information of the functions from $\mathsf{X}$ to $\mathsf{Y}$. More precisely, knowing the hyperconfusion $\mathsf{X}\rightarrow\mathsf{Y}$ is equivalent to having black box access to a function $f_{\omega}:\mathsf{X}\rightarrow\mathsf{Y}$ (which may depend on the outcome $\omega\in\Omega$) satisfying $f_{\omega}(A)\ni\omega$ whenever $\omega\in\Omega$, $A\in\mathsf{X}$ satisfy $A\ni\omega$. For the forward direction, if we know $\mathsf{X}\rightarrow\mathsf{Y}$, i.e., we have $S\in(\mathsf{X}\rightarrow\mathsf{Y})$, $S\ni\omega$, then we can construct $f_{\omega}(A)=A\cap S\in\mathsf{Y}$ since $\mathsf{X}\cap2^{S}\subseteq\mathsf{Y}$. For the reverse direction, we show how the black box access of $f_{\omega}$ allows us to know $\mathsf{X}\rightarrow\mathsf{Y}$. We know $\omega\notin A\backslash f_{\omega}(A)$ for every $A\in\mathsf{X}$. Let $S_{\omega}:=\Omega\backslash\bigcup_{A\in\mathsf{X}}(A\backslash f_{\omega}(A))=\bigcap_{A\in\mathsf{X}}((\Omega\backslash A)\cup f_{\omega}(A))$. We have $S_{\omega}\ni\omega$ since $\omega\notin A\backslash f_{\omega}(A)$. If $C\in\mathsf{X}$ satisfies $C\subseteq S_{\omega}$, then $C\subseteq(\Omega\backslash C)\cup f_{\omega}(C)$, so $C\subseteq f_{\omega}(C)\in\mathsf{Y}$. Hence, $\mathsf{X}\cap2^{S_{\omega}}\subseteq\mathsf{Y}$, and $S_{\omega}\in(\mathsf{X}\rightarrow\mathsf{Y})$ by definition. Therefore, having black box access to $f_{\omega}$ can give us a set $S_{\omega}\in(\mathsf{X}\rightarrow\mathsf{Y})$ with $S_{\omega}\ni\omega$, so we know $\mathsf{X}\rightarrow\mathsf{Y}$.

For the aforementioned example of two binary random variables, where $\mathsf{X},\mathsf{Y}$ represents the information of the first and second bits as in (\ref{eq:twobits}), we have
\begin{align*}
\mathsf{X}\rightarrow\mathsf{Y} & =\{\emptyset,\{00\},\{01\},\{10\},\{11\},\\
 & \quad\;\{00,10\},\{00,11\},\{01,10\},\{01,11\}\}.
\end{align*}
We can see that $\mathsf{X}\rightarrow\mathsf{Y}$ contains both $\mathsf{Y}$ and $\mathrm{oi}(Z)=\{\emptyset,\{00\},\{01\},\{10\},\{11\},\{01,10\},\{01,11\}\}$ where $Z=(X+Y)\;\mathrm{mod}\;2$. See Figure \ref{fig:op}. This allows us to define the most ambiguous $\mathsf{M}^{*}$ for the coding task which encompasses \emph{all} possible codes for this task, in the sense that any successful code $\mathsf{M}$ must satisfy $\mathsf{M}\subseteq\mathsf{M}^{*}$, unlike ordinary information where there is no single ``least informative'' choice of the encoding $M$. We will see in Section \ref{sec:formulae} that having a well-defined implication/difference allows us to directly compute the optimal code for more complex tasks.

The following proposition shows that $H(\mathsf{X}\rightarrow\mathsf{Y})$ is lower-bounded by the conditional entropy (Definition \ref{def:condition_HXY}), and equality holds if $\mathsf{X}$ is an ordinary information. The proof is in Appendix \ref{subsec:pf_implication}. Note that Proposition \ref{prop:implication} and the unconfusing lemma (Lemma \ref{lem:unconfuse}) imply the conditional prefix-free compression result in \cite{li2025discrete} (with slightly worse constants).

\smallskip{}

\begin{prop}
\label{prop:implication}For hyperconfusions $\mathsf{X},\mathsf{Y}$ over $(\Omega,p)$, we have
\[
H(\mathsf{X}\rightarrow\mathsf{Y})\ge H(\mathsf{Y}|\mathsf{X}).
\]
Equality holds if $\mathsf{X}\in\mathrm{OIs}(\Omega)$ is an ordinary information. In particular, for an event $E\subseteq\Omega$, $H(2^{E}\rightarrow\mathsf{Y})=p(E)H(\mathsf{Y}|E)$. 
\end{prop}

\smallskip{}

\subsection{Events, Negation and Double Negation}\label{subsec:negation}

The definition of negation in intuitionistic logic is $\lnot x:=(x\to\bot)$. We define negation for hyperconfusion analogously.

\smallskip{}

\begin{defn}
For a hyperconfusion $\mathsf{X}$ over $\Omega$, its \emph{negation} is\footnote{This is equivalent to the relative complement in \cite{roelofsen2013algebraic}.}
\begin{align*}
\lnot\mathsf{X} & :=(\mathsf{X}\rightarrow2^{\emptyset})=2^{\Omega\backslash\mathrm{supp}(\mathsf{X})}.
\end{align*}
Its \emph{double negation} is
\begin{align*}
\lnot\lnot\mathsf{X} & =\lnot(\lnot\mathsf{X})=2^{\mathrm{supp}(\mathsf{X})}.
\end{align*}
\end{defn}
\smallskip{}

In intuitionistic logic, $\lnot\lnot x$ is generally not equivalent to $x$. For hyperconfusion, $\lnot\lnot\mathsf{X}$ is also generally not equivalent to $\mathsf{X}$. An event $E\subseteq\Omega$ corresponds to the event hyperconfusion $2^{E}$, which has a negation $\lnot2^{E}=2^{E^{\mathrm{c}}}$ corresponding to the complement event $E^{\mathrm{c}}$. We can observe that $\mathsf{F}\in\mathrm{Hyps}(\Omega)$ is an event hyperconfusion if and only if $\lnot\lnot\mathsf{F}=\mathsf{F}$ ($\mathsf{F}$ is a \emph{regular element} \cite{grilletti2024esakia}), if and only if $\mathsf{F}=\lnot\mathsf{G}$ for some $\mathsf{G}\in\mathrm{Hyps}(\Omega)$. Operations on events $E,F\subseteq\Omega$ can be expressed as operations on the corresponding hyperconfusions $\mathsf{E}=2^{E}$, $\mathsf{F}=2^{F}$:
\begin{itemize}
\item Intersection $E\cap F$ corresponds to conjunction $\mathsf{E}\cap\mathsf{F}=2^{E\cap F}$.
\item Complement $E^{\mathrm{c}}=\Omega\backslash E$ corresponds to negation $\lnot\mathsf{E}=2^{E^{\mathrm{c}}}$.
\item Implication $E\rightarrow F=E^{\mathrm{c}}\cup F$ corresponds to implication $\mathsf{E}\rightarrow\mathsf{F}=2^{E\rightarrow F}$.
\item Union $E\cup F$ corresponds to $\lnot(\lnot\mathsf{E}\cap\lnot\mathsf{F})=2^{E\cup F}$ (the \emph{De Morgan translation}).\footnote{This aspect is similar to \cite{li2025poisson}, which studies information and events in a unified manner, but cannot handle union of events. In comparison, hyperconfusions can handle union of events, though it does not directly correspond to union of hyperconfusions.} It does not correspond to $\mathsf{E}\cup\mathsf{F}$, which represents a different piece of information. For example, for $E$ being the event ``tomorrow is going to rain'', and $F=E^{\mathrm{c}}$, knowing $\lnot(\lnot\mathsf{E}\cap\lnot\mathsf{F})$ is knowing that the event $E\cup F$ occurs, i.e., knowing the fact that ``tomorrow is either going to rain or not'', which is trivial; whereas knowing $\mathsf{E}\cup\mathsf{F}$ is knowing whether tomorrow is going to rain, which is a valuable piece of information.\footnote{This aspect is similar to generalized inquisitive logic \cite{ciardelli2009generalized,ciardelli2011inquisitive,roelofsen2013algebraic}, where disjunction behaves differently compared to classical logic.}
\end{itemize}

Intuitively, for a general hyperconfusion $\mathsf{X}$, $\lnot\mathsf{X}$ (which is always an event hyperconfusion) is the ``event of impossible knowledge''. Knowing $\lnot\mathsf{X}$ is knowing that it is impossible to know $\mathsf{X}$, i.e., knowing $\mathsf{X}$ will lead to contradiction (or omniscience). The double negation $\lnot\lnot\mathsf{X}$ (the ``event of possible knowledge'') is knowing that it is possible to know $\mathsf{X}$, i.e., knowing $\mathsf{X}$ will not lead to contradiction. Also, $\lnot\mathsf{X}\cup\lnot\lnot\mathsf{X}$ is the ordinary information of the binary random variable which is the indicator of whether it is possible to know $\mathsf{X}$. Refer to Section \ref{sec:gen} for details.

By Definition \ref{def:hyperconfusion}, every hyperconfusion is a union of event hyperconfusions. An ordinary information is a hyperconfusion formed by disjoint events. By Definition \ref{def:ordinary}, for finite $\Omega$, $\mathsf{X}\in\mathrm{OIs}(\Omega)$ if and only if for every $\mathsf{F},\mathsf{G}\in\mathrm{Hyps}(\Omega)$, if $\lnot\mathsf{F}\cup\lnot\mathsf{G}\subseteq\mathsf{X}$, then
\[
\lnot\mathsf{F}\cap\lnot\mathsf{G}=2^{\emptyset}\;\mathrm{or}\;\lnot(\mathsf{F}\cap\mathsf{G})\subseteq\mathsf{X}.
\]

\smallskip{}

\subsection{Hyperconfusion as a Heyting Algebra}\label{subsec:Heyting}

Recall that a Heyting algebra \cite{heyting1930formalen,esakia2019heyting} is a bounded lattice $\mathcal{H}$ with partial order $\le$, join $\vee$, meet $\wedge$,\footnote{A lattice $\mathcal{H}$ is a partially ordered set with a join operation $x\vee y$ satisfying $\{z\in\mathcal{H}:\,z\ge x,\,z\ge y\}=\{z:\,z\ge x\vee y\}$, and a meet operation $x\wedge y$ satisfying $\{z:\,z\le x,\,z\le y\}=\{z:\,z\le x\wedge y\}$.} least element $\bot$ and greatest element $\top$ (i.e., $\bot\le x\le\top$ for all $x\in\mathcal{H}$), with an additional implication operation $\rightarrow:\mathcal{H}\times\mathcal{H}\to\mathcal{H}$ such that $(z\wedge x)\le y$ if and only if $z\le(x\rightarrow y)$. From the previous sections, we can see that $\mathrm{Hyps}(\Omega)$ (the set of hyperconfusions for a fixed $\Omega$) is a Heyting algebra with partial order $\subseteq$, join $\cup$, meet $\cap$, least element $2^{\emptyset}$, greatest element $2^{\Omega}$, and implication $\rightarrow$.  This is equivalent to the Heyting algebra of propositions in generalized inquisitive logic \cite{roelofsen2013algebraic}.

One may raise the reverse question: can every finite Heyting algebra be regarded as a collection of hyperconfusions? Can we embed any finite Heyting algebra into $\mathrm{Hyps}(\Omega)$ for some $\Omega$? We require some additional conditions on the Heyting algebra for this to be possible. See Section \ref{sec:Heyting_to}.

\smallskip{}

\section{Coding-Logic Correspondence}\label{sec:formulae}

In information theory, the optimal code for a coding setting is often given as the solution to an optimization problem, which might not be straightforward to solve. Nevertheless, if we relax the code construction to allow hyperconfusions instead of ordinary information, we can use the operations in the previous sections to directly \emph{compute} the optimal code construction. Although the resultant code might require messages that are hyperconfusions (which may or may not be possible in practice), we can convert the hyperconfusion back to ordinary information via the unconfusing lemma (Lemma \ref{lem:unconfuse}), incurring only a logarithmic loss. Hence, this procedure allows us to directly compute the exact optimal code with hyperconfusion, or the almost optimal code with ordinary information (up to a logarithmic gap), in a mechanical manner that can be automated by an algorithm. Moreover, it reveals an elegant correspondence between coding tasks and logical formulae, providing new semantics to logical formulae.

\smallskip{}

\subsection{Intuitionistic Logic and Coding Tasks}\label{subsec:intuitionistic_task}

A formula corresponds a coding task, or an encoder or decoder which accomplishes this task. For example, the formula ``$\mathsf{X}$'' with only one term corresponds to the task: ``output $\mathsf{X}$''. If $\mathsf{X}=\mathrm{oi}(X)$ is the ordinary information of a random variable $X$, the task is to output the value of $X$. If $\mathsf{X}$ is a general hyperconfusion, the task is to output a set $A\in\mathsf{X}$ so $\omega\in A$ always holds, i.e., the coder declares that an event $A$ occurs (i.e., the outcome $\omega$ is in a set $A$), where we allow $A$ to be any confusable set in $\mathsf{X}$. A coder cannot accomplish this task without prior information. The minimal prior information needed by the coder is precisely $\mathsf{X}$. Similarly, the formula $\mathsf{X}\cap\mathsf{Y}$ is the task ``output $\mathsf{X}$ and $\mathsf{Y}$'', and the formula $\mathsf{X}\cup\mathsf{Y}$ is the task ``output $\mathsf{X}$ or $\mathsf{Y}$'' (i.e., output $A\in\mathsf{X}$, $A\ni\omega$, or output $B\in\mathsf{Y}$, $B\ni\omega$).

The formula $\mathsf{X}\rightarrow\mathsf{Y}$ is the task ``given $\mathsf{X}$ as input, output $\mathsf{Y}$''. If $\mathsf{X}=\mathrm{oi}(X)$, $\mathsf{Y}=\mathrm{oi}(Y)$, the task is to output the value of $Y$ given the value of $X$. For general hyperconfusions, the task is: given $A\in\mathsf{X}$ satisfying $A\ni\omega$, output $B\in\mathsf{Y}$ which must satisfy $B\ni\omega$. This task is possible without prior information for every $\omega$ if and only if $\mathsf{X}\subseteq\mathsf{Y}$ (where we simply output $B=A$). Generally, the coder requires a prior information $\mathsf{M}$ to accomplish this task. We need $(\mathsf{M}\cap\mathsf{X})\subseteq\mathsf{Y}$ since now the input can be regarded as both $\mathsf{M}$ and $\mathsf{X}$. This is equivalent to $\mathsf{M}\subseteq(\mathsf{X}\rightarrow\mathsf{Y})$ by definition of implication (Section \ref{subsec:implication}), so the most ambiguous $\mathsf{M}$ is $\mathsf{X}\rightarrow\mathsf{Y}$. In general, evaluating a formula gives the most ambiguous prior information needed by a coder that accomplishs the corresponding task.

The formula
\[
(\mathsf{X}\cap\mathsf{Y})\rightarrow\mathsf{X}
\]
is the task ``given $\mathsf{X}$, $\mathsf{Y}$ as inputs, output $\mathsf{Y}$''. This corresponds to the logical formula $(x\wedge y)\rightarrow x$ (if $x$ and $y$, then $x$), known as \emph{conjunction elimination}, that is true is intuitionistic logic. Due to the fact that any formula deducible from intuitionistic logic must evaluate to the greatest element in any Heyting algebra \cite{heyting1930formalen}, we have $((\mathsf{X}\cap\mathsf{Y})\rightarrow\mathsf{X})=2^{\Omega}$ being the full hyperconfusion. This provides a quick way to evaluate the hyperconfusion $(\mathsf{X}\cap\mathsf{Y})\rightarrow\mathsf{X}$ using intuitionistic logic. Intuitively, this means that the coder only needs the full hyperconfusion $\mathsf{M}=2^{\Omega}$ as prior information (i.e., it does not need any prior information), because the coder can simply output $\mathsf{X}$ from the input $\mathsf{X},\mathsf{Y}$. We call a task that always requires no prior information a \emph{trivial task}.

The formula $\mathsf{X}\rightarrow(\mathsf{X}\cup\mathsf{Y})$ (\emph{disjunction elimination}) corresponds to the task: ``given $\mathsf{X}$ as input, output $\mathsf{X}$ or $\mathsf{Y}$''. This formula also evaluates to $2^{\Omega}$ since it is deducible from intuitionistic logic. The task is also trivial.

Refer to Table \ref{tab:correspondence} for the correspondence between logic operations, hyperconfusion and coding tasks.

\begin{table*}
\centering
\caption{Correspondence between intuitionistic logic, hyperconfusion, and coding task. Hyperconfusions $\mathsf{X},\mathsf{Y}$ represent information, and $\mathsf{A},\mathsf{B}$ are compound formulae representing tasks.}\label{tab:correspondence}

{\renewcommand*{\arraystretch}{1.5}%
\begin{tabular}{ccc}
\hline 
\textbf{Intuitionistic logic} & \textbf{Hyperconfusion} & \textbf{Coding task}\tabularnewline
\hline 
Atom $x$ & Hyperconfusion $\mathsf{X}$ & Output $\mathsf{X}$\tabularnewline
\hline 
Truth $\top$ & $2^{\Omega}$ & No requirement (trivial)\tabularnewline
\hline 
Falsehood $\bot$ & $2^{\emptyset}$ & Achieve omniscience (impossible)\tabularnewline
\hline 
$\begin{array}{c}
\text{Stable proposition}\\
\phi=\lnot\lnot\phi
\end{array}$ & $\begin{array}{c}
\text{Event hyperconfusion}\\
\mathsf{E}=\lnot\lnot\mathsf{E}
\end{array}$ & $\begin{array}{c}
\text{No requirement if event occurs,}\\
\text{achieve omniscience if event does not occur}
\end{array}$\tabularnewline
\hline 
Conjunction $\phi\wedge\psi$ & $\mathsf{A}\cap\mathsf{B}$ & Do both tasks $\mathsf{A}$ and $\mathsf{B}$\tabularnewline
\hline 
Disjunction $\phi\vee\psi$ & $\mathsf{A}\cup\mathsf{B}$ & Do at least one task $\mathsf{A}$ or $\mathsf{B}$ at a time\tabularnewline
\hline 
\multirow{3}{*}{Implication $\phi\rightarrow\psi$} & $\mathsf{X}\rightarrow\mathsf{B}$ & Do task $\mathsf{B}$ using input $\mathsf{X}$\tabularnewline
\cline{2-3}
 & $\mathsf{E}\rightarrow\mathsf{B}$ for $\mathsf{E}=\lnot\lnot\mathsf{E}$ & Do task $\mathsf{B}$ when event $\mathsf{E}$ occurs\tabularnewline
\cline{2-3}
 & $\mathsf{A}\rightarrow\mathsf{B}$ & Do task $\mathsf{B}$ using coder for task $\mathsf{A}$ as black box\tabularnewline
\hline 
Negation $\lnot\phi$ & $\begin{array}{c}
\lnot\mathsf{A}=(\mathsf{A}\rightarrow2^{\emptyset})\\
\text{(event hyperconfusion)}
\end{array}$ & $\begin{array}{c}
\text{Achieve omniscience using input \ensuremath{\mathsf{A}}}\\
\text{(No requirement if \ensuremath{\mathsf{A}} gives omniscience,}\\
\text{otherwise need omniscience)}
\end{array}$\tabularnewline
\hline 
Biconditional $\phi\leftrightarrow\psi$ & $(\mathsf{X}\rightarrow\mathsf{Y})\cap(\mathsf{Y}\rightarrow\mathsf{X})$ & Butterfly network\tabularnewline
\hline 
$\begin{array}{c}
\text{Weak excluded middle}\\
\lnot\phi\vee\lnot\lnot\phi
\end{array}$ & $\lnot\mathsf{E}\cup\lnot\lnot\mathsf{E}$ & Determine whether event $\mathsf{E}$ occurs\tabularnewline
\hline 
\end{tabular}}
\end{table*}

\smallskip{}

\subsection{Modus Ponens and Black Box Access}\label{subsec:modus_ponens}

The formula 
\begin{equation}
(\mathsf{X}\cap(\mathsf{X}\rightarrow\mathsf{Y}))\rightarrow\mathsf{Y}\label{eq:modusponens}
\end{equation}
is known as \emph{modus ponens}. In Section \ref{subsec:implication}, we explained that the hyperconfusion $\mathsf{X}\rightarrow\mathsf{Y}$ is equivalent to the black box access to an inner coder, i.e., access to a function $f_{\omega}:\mathsf{X}\to\mathsf{Y}$ (which may depend on $\omega$) satisfying $f_{\omega}(A)\ni\omega$ whenever $A\ni\omega$. Hence, (\ref{eq:modusponens}) corresponds to the setting where we construct an outer coder for the task: ``given $\mathsf{X}$ and an inner coder accomplishing the task ``given $\mathsf{X}$ as input, output $\mathsf{Y}$'' as inputs, output $\mathsf{Y}$''. There are two layers of coders, where the outer coder can invoke the inner coder as a black box. Intuitively, the outer coder can invoke the inner coder with the input $\mathsf{X}$ to obtain $\mathsf{Y}$. Hence, the task is trivial and requires no prior information. This matches the fact that modus ponens holds in intuitionistic logic, so $(\mathsf{X}\cap(\mathsf{X}\rightarrow\mathsf{Y}))\rightarrow\mathsf{Y}=2^{\Omega}$. 

\smallskip{}

\subsection{Butterfly Network}\label{subsec:butterfly}

Any formula deducible from intuitionistic logic  must evaluate to the full hyperconfusion, corresponding to trivial coding tasks where no prior information is required. Nontrivial coding settings concerns formulae not deducible from intuitionistic logic. In this sense, the entropy of a formula (Definition \ref{def:entropy}) is a ``measure of non-intuitionistic-ness'' of a formula. For example, consider the biconditional formula
\[
(\mathsf{X}\rightarrow\mathsf{Y})\cap(\mathsf{Y}\rightarrow\mathsf{X}),
\]
which is not deducible from intuitionistic logic. Interestingly, this formula corresponds to the well-known butterfly network \cite{ahlswede2000network} described as follows. There are two pieces of information $\mathsf{X},\mathsf{Y}$ (for full generality, we consider them as hyperconfusions that may not be independent), which are held by User 1 and User 2, respectively. The users send their information to a satellite, which will then broadcast the information $\mathsf{M}$ to the users. The goal is to allow both users to recover both $\mathsf{X}$ and $\mathsf{Y}$. For User 1 to decode $\mathsf{Y}$ using $\mathsf{X}$ and $\mathsf{M}$, we require $\mathsf{X}\cap\mathsf{M}\subseteq\mathsf{Y}$, equivalently $\mathsf{M}\subseteq\mathsf{X}\rightarrow\mathsf{Y}$. Similarly, User 2 requires $\mathsf{M}\subseteq\mathsf{Y}\rightarrow\mathsf{X}$. Hence, the requirement is $\mathsf{M}\subseteq(\mathsf{X}\rightarrow\mathsf{Y})\cap(\mathsf{Y}\rightarrow\mathsf{X})$. The most confusing hyperconfusion for $\mathsf{M}$ is therefore $\mathsf{M}^{*}:=(\mathsf{X}\rightarrow\mathsf{Y})\cap(\mathsf{Y}\rightarrow\mathsf{X})$. See Figure \ref{fig:butterfly}.

We can also arrive at this conclusion by manipulating the logical formula using rules in intuitionistic logic. The decoding requirement is $((\mathsf{X}\cap\mathsf{M})\rightarrow\mathsf{Y})\cap((\mathsf{Y}\cap\mathsf{M})\rightarrow\mathsf{X})$, which must evaluate to $2^{\Omega}$ for decoding to succeed. Simplifying this formula,
\begin{align}
 & ((\mathsf{X}\cap\mathsf{M})\rightarrow\mathsf{Y})\cap((\mathsf{Y}\cap\mathsf{M})\rightarrow\mathsf{X})\nonumber \\
 & =\,(\mathsf{M}\rightarrow(\mathsf{X}\rightarrow\mathsf{Y}))\cap(\mathsf{M}\rightarrow(\mathsf{Y}\rightarrow\mathsf{X}))\label{eq:butterfly_deduce}\\
 & =\,\mathsf{M}\rightarrow((\mathsf{X}\rightarrow\mathsf{Y})\cap(\mathsf{Y}\rightarrow\mathsf{X}))\label{eq:butterfly_biconditional}\\
 & =\,\mathsf{M}\rightarrow((\mathsf{X}\cup\mathsf{Y})\rightarrow(\mathsf{X}\cap\mathsf{Y})).\label{eq:butterfly_or_and}
\end{align}
From (\ref{eq:butterfly_biconditional}), we arrive at the same conclusion that $\mathsf{M}\rightarrow((\mathsf{X}\rightarrow\mathsf{Y})\cap(\mathsf{Y}\rightarrow\mathsf{X}))$ must be $2^{\Omega}$, that is, we must have $\mathsf{M}\subseteq\mathsf{M}^{*}=(\mathsf{X}\rightarrow\mathsf{Y})\cap(\mathsf{Y}\rightarrow\mathsf{X})$. An alternative expression is $\mathsf{M}^{*}=(\mathsf{X}\cup\mathsf{Y})\rightarrow(\mathsf{X}\cap\mathsf{Y})$ in (\ref{eq:butterfly_or_and}), corresponding to the task ``given $\mathsf{X}$ or $\mathsf{Y}$, output $\mathsf{X}$ and $\mathsf{Y}$'', which can be seen to be equivalent to the task of the butterfly network.

If the goal is to minimize the entropy of $\mathsf{M}$, the answer is simply given by $H(\mathsf{M}^{*})$. Nevertheless, $\mathsf{M}^{*}$ may not be an ordinary information. If ordinary information is necessary, i.e., we require $\mathsf{M}\in\mathrm{OIs}(\Omega)$ (the satellite can only broadcast an ordinary random variable), we invoke the unconfusing lemma (Lemma \ref{lem:unconfuse}) to convert the hyperconfusion to an ordinary random variable. The smallest communication cost is close to $H(\mathsf{M}^{*})$ within a logarithmic gap. The following result is immediate.

\smallskip{}

\begin{thm}
\label{thm:butterfly}Let $H^{*}:=\inf_{\mathsf{M}\in\mathrm{OIs}(\Omega):\,\mathsf{X}\cap\mathsf{M}\subseteq\mathsf{Y},\,\mathsf{Y}\cap\mathsf{M}\subseteq\mathsf{X}}H(\mathsf{M})$ be the smallest entropy of an ordinary random variable that satisfies the requirement of the butterfly network. We have
\[
H(\mathsf{M}^{*})\le H^{*}\le H(\mathsf{M}^{*})+\log(H(\mathsf{M}^{*})+3.4)+1,
\]
where $\mathsf{M}^{*}:=(\mathsf{X}\rightarrow\mathsf{Y})\cap(\mathsf{Y}\rightarrow\mathsf{X})$. Moreover, letting $H_{0}^{*}:=\inf_{\mathsf{M}\in\mathrm{OIs}(\Omega):\,\mathsf{X}\cap\mathsf{M}\subseteq\mathsf{Y},\,\mathsf{Y}\cap\mathsf{M}\subseteq\mathsf{X}}H_{0}(\mathsf{M})$, we have 
\[
H^{*}\le H_{0}^{*}=H_{0}(\mathsf{M}^{*}).
\]
\end{thm}
\smallskip{}

We now demonstrate how we can compute the optimal $\mathsf{M}^{*}$ for an example, where $\mathsf{X},\mathsf{Y}$ are two i.i.d. fair bits, shown in Figure \ref{fig:butterfly}. Consider the singleton hyperconfusion $\mathsf{U}=\{\emptyset,\{0\},\{1\}\}$ over the uniform probability space over $\{0,1\}$, and its 2-fold product (Definition \ref{def:product}), giving an i.i.d. hyperconfusion sequence $\mathsf{X},\mathsf{Y}$ over the sample space $\{0,1\}^{2}=\{00,01,10,11\}$, as in (\ref{eq:twobits}). $\mathsf{X}$ and $\mathsf{Y}$ correspond to the information of the first and second bit, respectively. Direct computation yields
\begin{align*}
\mathsf{X}\rightarrow\mathsf{Y} & =\{\emptyset,\{00\},\{01\},\{10\},\{11\},\\
 & \quad\;\{00,10\},\{00,11\},\{01,10\},\{01,11\}\},\\
\mathsf{Y}\rightarrow\mathsf{X} & =\{\emptyset,\{00\},\{01\},\{10\},\{11\},\\
 & \quad\;\{00,01\},\{00,11\},\{10,01\},\{10,11\}\}.
\end{align*}
Hence,
\begin{align*}
\mathsf{M}^{*} & =(\mathsf{X}\rightarrow\mathsf{Y})\cap(\mathsf{Y}\rightarrow\mathsf{X})\\
 & =\{\emptyset,\{00\},\{01\},\{10\},\{11\},\{00,11\},\{01,10\}\},
\end{align*}
which is precisely the hyperconfusion induced by the XOR of the two bits (which is an ordinary information). We have $H(\mathsf{M}^{*})=1\le H^{*}\le H_{0}(\mathsf{M}^{*})=1$, so we need precisely 1 bit.  This gives a complete proof that the XOR of the two bits is the optimal code for the butterfly network based on direct computation.  More generally, if $\mathsf{X}$ and $\mathsf{Y}$ are i.i.d. and uniform over $k$ choices, then $H(\mathsf{M}^{*})=\log k$, and the hyperconfusion induced by the code $M=(X+Y)\;\mathrm{mod}\;k$ is included in $\mathsf{M}^{*}$ (together with other possible codes).

In case if $\mathsf{X}=\bigcap_{i=1}^{n}\mathsf{X}_{i}$ and $\mathsf{Y}=\bigcap_{i=1}^{n}\mathsf{Y}_{i}$ consist of sequences $\mathsf{X}_{1},\ldots,\mathsf{X}_{n}$ and $\mathsf{Y}_{1},\ldots,\mathsf{Y}_{n}$ of (not necessarily i.i.d.) hyperconfusions, while we can still use the formula $\mathsf{M}^{*}=(\mathsf{X}\rightarrow\mathsf{Y})\cap(\mathsf{Y}\rightarrow\mathsf{X})$, it might be preferrable to code each pair $\mathsf{X}_{i},\mathsf{Y}_{i}$ separately for the sake of computational tractability. Using the fact that 
\begin{align*}
 & (\mathsf{X}_{1}\cap\mathsf{X}_{2})\rightarrow(\mathsf{Y}_{1}\cap\mathsf{Y}_{2})\\
 & =((\mathsf{X}_{1}\cap\mathsf{X}_{2})\rightarrow\mathsf{Y}_{1})\cap((\mathsf{X}_{1}\cap\mathsf{X}_{2})\rightarrow\mathsf{Y}_{2})\\
 & \supseteq(\mathsf{X}_{1}\rightarrow\mathsf{Y}_{1})\cap(\mathsf{X}_{2}\rightarrow\mathsf{Y}_{2}),
\end{align*}
we know that $\mathsf{M}=\bigcap_{i=1}^{n}(\mathsf{X}_{i}\rightarrow\mathsf{Y}_{i})\cap(\mathsf{Y}_{i}\rightarrow\mathsf{X}_{i})$ satisfies the requirement (but is not necessarily optimal). By Proposition \ref{prop:conjunction}, $H(\mathsf{M})\le\sum_{i}H((\mathsf{X}_{i}\rightarrow\mathsf{Y}_{i})\cap(\mathsf{Y}_{i}\rightarrow\mathsf{X}_{i}))$. Applying the unconfusing lemma, as long as each pair $(\mathsf{X}_{i},\mathsf{Y}_{i})$ has the same distribution, the optimal asymptotic rate is bounded by 
\begin{align*}
 & \underset{n\to\infty}{\mathrm{limsup}}\,n^{-1}\inf_{\mathsf{M}\in\mathrm{OIs}(\Omega):\,\mathsf{X}\cap\mathsf{M}\subseteq\mathsf{Y},\,\mathsf{Y}\cap\mathsf{M}\subseteq\mathsf{X}}H(\mathsf{M})\\
 & \le H((\mathsf{X}_{1}\rightarrow\mathsf{Y}_{1})\cap(\mathsf{Y}_{1}\rightarrow\mathsf{X}_{1})).
\end{align*}

The intuitive reason of the usefulness of hyperconfusion is that conjunction, disjunction and implication can be unambiguously defined. In comparison, for ordinary information (random variables), the only naturally defined operation is taking the joint random variable. If one is given a task ``either decode $X$ or decode $Y$'' or ``decode $Y$ with the help of $X$'', then there is no single best choice of ordinary information for the task, and choosing any ordinary information will incur an irreversible loss of information. For example, if we choose $M=Y$ for the task ``design $M$ which can be used to decode $Y$ with the help of $X$'', then we will lose the choice $M=(X+Y)\;\mathrm{mod}\;2$ which might turn out to be better. Hyperconfusion allows us to keep track of all coding schemes that are possibly optimal, and only commit to one particular scheme (i.e., convert the hyperconfusion to ordinary information) at the end when we invoke the unconfusing lemma, so we only pay the loss once at the end. 

This is slightly reminiscent of quantum computing, where the deferred measurement principle \cite{nielsen2010quantum} allows us to defer all measurements to the end. A very loose analogy would be: ordinary information $\leftrightarrow$ classical information, hyperconfusion $\leftrightarrow$ superposition, and unconfusing lemma $\leftrightarrow$ measurement. Nevertheless, unlike quantum computing, invoking the unconfusing lemma later rather than earlier can affect the result and reduce the compression size.

\smallskip{}

\subsection{Disjunctive Butterfly Network}\label{subsec:dis_butterfly}

We relax the setting to the \emph{disjunctive butterfly network} so that the satellite only has to ensure that \emph{at least one} of the two users can recover $\mathsf{X}$ and $\mathsf{Y}$, that is, after receiving $\mathsf{M}$, User 1 outputs $A\subseteq\Omega$ and User 2 outputs $B\subseteq\Omega$, so that $\omega\in A$ and $\omega\in B$ and ($A\in\mathsf{X}$ or $B\in\mathsf{Y}$) must hold. If $\mathsf{X},\mathsf{Y},\mathsf{M}$ are ordinary information induced by random variables $X,Y,M$, this means User 1 outputs $\hat{Y}$ as a function of $(X,M)$, and User 2 outputs $\hat{X}$ as a function of $(Y,M)$, so ($X=\hat{X}$ or $Y=\hat{Y}$) always holds. In this case, $\mathsf{M}$ should deduce the disjunction of $\mathsf{X}\rightarrow\mathsf{Y}$ and $\mathsf{Y}\rightarrow\mathsf{X}$, i.e., $\mathsf{M}\subseteq(\mathsf{X}\rightarrow\mathsf{Y})\cup(\mathsf{Y}\rightarrow\mathsf{X})$. Hence, the most ambiguous hyperconfusion would be 
\[
\mathsf{M}^{*}=(\mathsf{X}\rightarrow\mathsf{Y})\cup(\mathsf{Y}\rightarrow\mathsf{X}).
\]
This formula is known as \emph{Dummett's axiom} in logic \cite{dummett1959propositional}, which holds in classical and G\"{o}del logic, but not provable in intuitionistic logic.

For the two bits example, we can check that $\mathsf{M}^{*}$ is not an ordinary information, and $H(\mathsf{M}^{*})=1$, which shows that the relaxed setting still requires $1$ bit of broadcast, regardless of whether the broadcast message is a hyperconfusion or an ordinary information (Lemma \ref{lem:unconfuse}). This gives a proof that the relaxed setting does not lower the communication requirement for this example. 

One may attempt to modify (\ref{eq:butterfly_deduce}) to argue that the requirement of the disjunctive butterfly network is written as 
\begin{equation}
(\mathsf{M}\rightarrow(\mathsf{X}\rightarrow\mathsf{Y}))\cup(\mathsf{M}\rightarrow(\mathsf{Y}\rightarrow\mathsf{X})).\label{eq:dis_butterfly_2}
\end{equation}
This formula is equivalent to the correct formula 
\begin{equation}
\mathsf{M}\rightarrow((\mathsf{X}\rightarrow\mathsf{Y})\cup(\mathsf{Y}\rightarrow\mathsf{X})).\label{eq:dis_butterfly_1}
\end{equation}
in classical logic, but (\ref{eq:dis_butterfly_2}) is strictly more stringent than (\ref{eq:dis_butterfly_1}) in intuitionistic logic. Requiring $(\mathsf{M}\rightarrow(\mathsf{X}\rightarrow\mathsf{Y}))\cup(\mathsf{M}\rightarrow(\mathsf{Y}\rightarrow\mathsf{X}))=2^{\Omega}$ implies $\mathsf{M}\rightarrow(\mathsf{X}\rightarrow\mathsf{Y})=2^{\Omega}$ or $\mathsf{M}\rightarrow(\mathsf{Y}\rightarrow\mathsf{X})=2^{\Omega}$, meaning that we have to guarantee that User 1 can always decode, or guarantee that User 2 can always decode. This is more stringent than the disjunctive butterfly network (\ref{eq:dis_butterfly_1}), which only requires that one of the users can successfully decode, but which user is successful can depend on the values of the sources.  This mirrors the fact in intuitionistic logic that if we have a proof for ``$\mathsf{M}$ implies ($\mathsf{A}$ or $\mathsf{B}$)'', this does not mean that we will have a proof for ``$\mathsf{M}$ implies $\mathsf{A}$'' or a proof for ``$\mathsf{M}$ implies $\mathsf{B}$''. The minimum entropy for (\ref{eq:dis_butterfly_2}) is $\min\{H(\mathsf{X}\rightarrow\mathsf{Y}),H(\mathsf{Y}\rightarrow\mathsf{X})\}$, generally greater than the minimum entropy for (\ref{eq:dis_butterfly_1}) which is $H((\mathsf{X}\rightarrow\mathsf{Y})\cup(\mathsf{Y}\rightarrow\mathsf{X}))$.

\smallskip{}

\subsection{Information Retrieval}\label{subsec:retrieve}

Suppose there are two pieces of information $\mathsf{X},\mathsf{Y}$ at a server. The user sends a request for either one of $\mathsf{X},\mathsf{Y}$, and the server responds a message $\mathsf{M}$ from which the user can recover the desired information. While the server can simply send $\mathsf{X}$ or $\mathsf{Y}$ depending on the request, if the choice of the user depends on $\mathsf{X},\mathsf{Y}$, then the communication can sometimes be reduced. For example, if $\mathsf{X},\mathsf{Y}$ are two random books, and we know a priori that the user will not request $\mathsf{X}$ if $\mathsf{X}$ is a fiction, then if the user requests $\mathsf{X}$, the server does not have to transmit the bit that indicates whether $\mathsf{X}$ is a fiction.

Let $E\subseteq\Omega$ be the event that the user requests $\mathsf{X}$. The event is represented by the event hyperconfusion $\mathsf{E}=2^{E}$ (Section \ref{subsec:negation}), corresponding to knowing nothing within $E$ (other than the fact $\omega\in E$), but omniscience outside of $E$. The requirement ``decode $\mathsf{X}$ using $\mathsf{M}$ if event $E$ occurs'' can be written as $\mathsf{E}\rightarrow(\mathsf{M}\rightarrow\mathsf{X})$, equivalent to $(\mathsf{E}\cap\mathsf{M})\rightarrow\mathsf{X}$ and $\mathsf{M}\rightarrow(\mathsf{E}\rightarrow\mathsf{X})$ due to currying. Intuitively, $\mathsf{E}\rightarrow(\mathsf{M}\rightarrow\mathsf{X})$ means that a coder with $\mathsf{E}$ as prior information will be able to accomplish the task ``given the input $\mathsf{M}$, output $\mathsf{X}$''. The event hyperconfusion $\mathsf{E}$ provides omniscience if $E$ does not occur, so providing $\mathsf{E}$ to a coder allows it to perform any future coding task if $E$ does not occur, so the requirement is only active if $E$ occurs. 

Recall that the complement $E^{\mathrm{c}}$ correspond to $\lnot\mathsf{E}$. The overall requirement is $(\mathsf{E}\rightarrow(\mathsf{M}\rightarrow\mathsf{X}))\cap(\lnot\mathsf{E}\rightarrow(\mathsf{M}\rightarrow\mathsf{Y}))$. Recall that $\mathsf{E}$ is an event hyperconfusion if and only if $\mathsf{E}=\lnot\lnot\mathsf{E}$, so we can also write the requirement as
\begin{align*}
 & (\lnot\lnot\mathsf{E}\rightarrow(\mathsf{M}\rightarrow\mathsf{X}))\cap(\lnot\mathsf{E}\rightarrow(\mathsf{M}\rightarrow\mathsf{Y}))\\
 & =\;((\lnot\lnot\mathsf{E}\cap\mathsf{M})\rightarrow\mathsf{X}))\cap((\lnot\mathsf{E}\cap\mathsf{M})\rightarrow\mathsf{Y}))\\
 & =\;\mathsf{M}\rightarrow((\lnot\lnot\mathsf{E}\rightarrow\mathsf{X})\cap(\lnot\mathsf{E}\rightarrow\mathsf{Y})).
\end{align*}
The optimal $\mathsf{M}$ is $(\lnot\lnot\mathsf{E}\rightarrow\mathsf{X})\cap(\lnot\mathsf{E}\rightarrow\mathsf{Y})$. We use $\lnot\lnot\mathsf{E}$ instead of $\mathsf{E}$, because both $\lnot\mathsf{E}$ and $\lnot\lnot\mathsf{E}$ are guaranteed to be event hyperconfusions regardless of whether $\mathsf{E}$ is an event hyperconfusion, so we do not need to specify that $\mathsf{E}$ is an event hyperconfusion for the formula to work.

\smallskip{}

\subsection{Index Coding}\label{subsec:index_coding}

In index coding \cite{birk1998informed,bar2011index,el2010index}, a server knows the sources $\mathsf{X}_{1},\ldots,\mathsf{X}_{n}$, and broadcast a message $\mathsf{M}$ to $k$ users, each of which has a subset of the sources and wants to decode another subset of sources. In this section, we consider a generalization of index coding that encompasses the settings in Sections \ref{subsec:butterfly}, \ref{subsec:dis_butterfly} and \ref{subsec:retrieve}. We allow $\mathsf{X}_{1},\ldots,\mathsf{X}_{n}$ to be arbitrary hyperconfusions (not necessarily independent). Assume that user $i\in[k]$ knows $\mathsf{X}_{a}$ for $a\in\mathcal{A}_{i}\subseteq[n]$ and wants to decode $\mathsf{X}_{b}$ for $b\in\mathcal{B}_{i}\subseteq[n]$ if event $F_{i}$ occurs, and the server wants every user to be successful. Using the same arguments as previous sections, the requirement is
\[
\mathsf{M}\rightarrow\bigcap_{i=1}^{k}\left((\lnot\mathsf{E}_{i}\cap\mathsf{X}_{\mathcal{A}_{i}})\to\mathsf{X}_{\mathcal{B}_{i}}\right),
\]
which must evaluate to $2^{\Omega}$ for all users to be successful, where $\lnot\mathsf{E}_{i}=2^{F_{i}}$ represents the event $F_{i}$, and $\mathsf{X}_{\mathcal{A}}:=\bigcap_{j\in\mathcal{A}}\mathsf{X}_{j}$ for $\mathcal{A}\subseteq[n]$. Hence, the most ambiguous $\mathsf{M}$ is given by the right-hand side of the above formula, and we can use the unconfusing lemma to find the minimal entropy of $\mathsf{M}$ if it is restricted to be an ordinary information within a logarithmic gap, similar to Theorem \ref{thm:butterfly}.

We can further generalize this setting to the scenario where the goal is to ensure that the set of successful users satisfies a certain condition. More precisely, we ensure that $\{i\in[k]:\,\text{user \ensuremath{i} is successful}\}\in\mathcal{D}$ for a given $\mathcal{D}\subseteq2^{[k]}$ that is upward closed (if $\mathcal{S}\in\mathcal{D}$, then $\mathcal{T}\in\mathcal{D}$ for every $\mathcal{T}\supseteq\mathcal{S}$). For example, if we only need any one user to be successful, then $\mathcal{D}=\{\mathcal{S}\subseteq[n]:\,|\mathcal{S}|\ge1\}$. The requirement is now 
\begin{equation}
\mathsf{M}\rightarrow\bigcup_{\mathcal{S}\in\mathcal{D}}\bigcap_{i\in\mathcal{S}}\left((\lnot\mathsf{E}_{i}\cap\mathsf{X}_{\mathcal{A}_{i}})\to\mathsf{X}_{\mathcal{B}_{i}}\right).\label{eq:index_or}
\end{equation}

The situation is more complicated when the encoder does not know all information, or when there are more than one encoded messages. These settings will be discussed in Sections \ref{subsec:multiple}, \ref{subsec:bounded}.

\smallskip{}

\subsection{Source Coding with Error}\label{subsec:source_error}

All previous settings are ``zero-error'', which does not allow any positive decoding error probability. We can use logical formula to study error events as well. Consider a source coding setting where $\mathsf{X}$ is compressed to $\mathsf{M}$, and a decoder can recover $\mathsf{X}$ using $\mathsf{M}$ with success probability at least $q$. Let $\mathsf{E}$ be an event hyperconfusion that corresponds to the success event (i.e., $\mathsf{E}=2^{E}$ where $E=\mathrm{supp}(\mathsf{E})$ is the event of successful decoding). The requirement ``if $\lnot\lnot\mathsf{E}$ occurs, then $\mathsf{M}$ can decode $\mathsf{X}$'' is
\begin{align}
\lnot\lnot\mathsf{E}\rightarrow(\mathsf{M}\rightarrow\mathsf{X})\, & =\,(\lnot\lnot\mathsf{E}\cap\mathsf{M})\rightarrow\mathsf{X}\nonumber \\
 & =\,\mathsf{M}\rightarrow(\lnot\lnot\mathsf{E}\rightarrow\mathsf{X})\label{eq:source_error}
\end{align}
due to currying. This must evaluate to $2^{\Omega}$ for the requirement to be satisfied. We use $\lnot\lnot\mathsf{E}$ instead of $\mathsf{E}$ since $\lnot\lnot\mathsf{E}$ is guaranteed to be an event hyperconfusion. The smallest $H(\mathsf{M})$ is therefore 
\begin{equation}
\min_{\mathsf{E}\in\mathrm{Hyps}(\Omega):\,H_{\infty}(\lnot\lnot\mathsf{E})\le-\log q}H(\lnot\lnot\mathsf{E}\rightarrow\mathsf{X}),\label{eq:source_error_E}
\end{equation}
where the constraint on $H_{\infty}(\lnot\lnot\mathsf{E})$ is due to (\ref{eq:Hinfty_event}). 

Since there are two hyperconfusions $\mathsf{E},\mathsf{M}$ to maximize, the answer is no longer given by a single logical formula. Nevertheless, if we are fixing one of $\mathsf{E},\mathsf{M}$ and maximizing the ambiguity of the other, it can still be performed easily. Since $\mathsf{M}\rightarrow(\lnot\lnot\mathsf{E}\rightarrow\mathsf{X})$, the optimal $\mathsf{M}$ for a fixed $\mathsf{E}$ is $\lnot\lnot\mathsf{E}\rightarrow\mathsf{X}$. We also have $\lnot\lnot\mathsf{E}\rightarrow(\mathsf{M}\rightarrow\mathsf{X})$, so a maximal $\mathsf{E}^{*}$ must satisfy $\mathrm{supp}(\mathsf{E}^{*})\in\mathrm{maxs}(\mathsf{M}\rightarrow\mathsf{X})$, i.e., the success event is one of the maximal confusable sets in $\mathsf{M}\rightarrow\mathsf{X}$. To maximize the success probability for a fixed $\mathsf{M}$, we choose $\mathrm{supp}(\mathsf{E}^{*})=\mathrm{argmax}_{A\in\mathsf{M}\rightarrow\mathsf{X}}p(A)$. Interestingly, the largest $p(A)$ is $2^{-H_{\infty}(\mathsf{M}\rightarrow\mathsf{X})}$. Hence, the optimization (\ref{eq:source_error_E}) can be written as 
\begin{equation}
\tilde{H}_{q}(\mathsf{X}):=\min_{\mathsf{M}\in\mathrm{Hyps}(\Omega):\,H_{\infty}(\mathsf{M}\rightarrow\mathsf{X})\le-\log q}H(\mathsf{M}),\label{eq:H_tilde}
\end{equation}
and the problem is a trade-off between $H(\mathsf{M})$ (the compression size) and $H_{\infty}(\mathsf{M}\rightarrow\mathsf{X})$ (negative log success probability). This connection between $H_{\infty}$ and success probability is a vast generalization of the fact that for an ordinary random variable $X$, the success probability of guessing its value without any information is $2^{-H_{\infty}(X)}$. 

This connection can be applied to other settings. For example, in the butterfly network (Section \ref{subsec:butterfly}), if we only require both users to be successful with probability at least $q$, then the problem becomes a trade-off between $H(\mathsf{M})$ (the compression size) and $H_{\infty}(\mathsf{M}\rightarrow((\mathsf{X}\rightarrow\mathsf{Y})\cap(\mathsf{Y}\rightarrow\mathsf{X})))$ (negative log success probability). The answer is simply given by $\tilde{H}_{q}((\mathsf{X}\rightarrow\mathsf{Y})\cap(\mathsf{Y}\rightarrow\mathsf{X}))$, where $\tilde{H}_{q}$ is defined in (\ref{eq:H_tilde}). Generally, we can allow a small error probability by not strictly requiring that the logical formula of the requirement (e.g., (\ref{eq:butterfly_biconditional}), (\ref{eq:dis_butterfly_1}), (\ref{eq:index_or})) must evaluate to  $2^{\Omega}$, but rather requiring it to have a small $H_{\infty}$.

\smallskip{}

\subsection{Multiple Messages}\label{subsec:multiple}

We now study settings with multiple messages. In an erasure code \cite{reed1960polynomial}, the source $\mathsf{X}$ is encoded into the parts $\mathsf{M}_{1},\ldots,\mathsf{M}_{n}$, and the decoder can recover $\mathsf{X}$ as long as at least $k$ parts are available, where $1\le k\le n$. The requirement is
\[
\bigcap_{\mathcal{S}\subseteq[n]:\,|\mathcal{S}|=k}\Big(\Big(\bigcap_{i\in\mathcal{S}}\mathsf{M}_{i}\Big)\rightarrow\mathsf{X}\Big),
\]
which must evaluate to $2^{\Omega}$. The requirement on $\mathsf{M}_{i}$ in terms of $(\mathsf{M}_{j})_{j\neq i}$ can be written as
\begin{equation}
\mathsf{M}_{i}\rightarrow\bigcap_{\mathcal{S}\subseteq[n]\backslash\{i\}:\,|\mathcal{S}|=k-1}\Big(\Big(\bigcap_{j\in\mathcal{S}}\mathsf{M}_{j}\Big)\rightarrow\mathsf{X}\Big).\label{eq:erasure}
\end{equation}
Hence, the right-hand side of (\ref{eq:erasure}) is the most ambiguous $\mathsf{M}_{i}$ when $(\mathsf{M}_{j})_{j\neq i}$ are fixed. Given an initial guess on $\mathsf{M}_{1},\ldots,\mathsf{M}_{n}$, we can update $\mathsf{M}_{i}$ to be the right-hand side of (\ref{eq:erasure}) iteratively for $i=1,\ldots,n$ in order to arrive at a Pareto-optimal $\mathsf{M}_{1},\ldots,\mathsf{M}_{n}$, i.e., none of them can be strictly enlarged while still satisfying the requirement.

For another example, in the Gray-Wyner network \cite{gray1974source}, the sources $\mathsf{X},\mathsf{Y}$ are encoded into the parts $\mathsf{M}_{0},\mathsf{M}_{1},\mathsf{M}_{2}$. Decoder 1 observes $\mathsf{M}_{0},\mathsf{M}_{1}$ and recovers $\mathsf{X}$. Decoder 2 observes $\mathsf{M}_{0},\mathsf{M}_{2}$ and recovers $\mathsf{Y}$. The requirement is
\begin{align}
 & ((\mathsf{M}_{0}\cap\mathsf{M}_{1})\rightarrow\mathsf{X})\cap((\mathsf{M}_{0}\cap\mathsf{M}_{2})\rightarrow\mathsf{Y})\nonumber \\
 & =\,\mathsf{M}_{0}\rightarrow((\mathsf{M}_{1}\rightarrow\mathsf{X})\cap(\mathsf{M}_{2}\rightarrow\mathsf{Y}))\label{eq:graywyner_1}\\
 & =\,(\mathsf{M}_{1}\rightarrow(\mathsf{M}_{0}\rightarrow\mathsf{X}))\cap(\mathsf{M}_{2}\rightarrow(\mathsf{M}_{0}\rightarrow\mathsf{Y})).\label{eq:graywyner_2}
\end{align}
Hence, the optimal $\mathsf{M}_{0}$ for fixed $\mathsf{M}_{1},\mathsf{M}_{2}$ is the right-hand side of (\ref{eq:graywyner_1}), and the optimal $\mathsf{M}_{1},\mathsf{M}_{2}$ for fixed $\mathsf{M}_{0}$ are given in (\ref{eq:graywyner_2}). This allows us to find a Pareto-optimal $\mathsf{M}_{0},\mathsf{M}_{1},\mathsf{M}_{2}$. \smallskip{}

\subsection{Slepian-Wolf Coding and Bounded Information}\label{subsec:bounded}

We now study a special case of Slepian-Wolf coding \cite{slepian1973noiseless}, where an encoder knows the source $\mathsf{X}$ and a decoder knows the side information $\mathsf{Y}$. The encoder sends an information $\mathsf{M}$ to allow the decoder to decode $\mathsf{X}$. The goal is to find the least informative $\mathsf{M}$ to accomplish this task. This setting generalizes zero-error source coding with side information \cite{witsenhausen1976zero,alon1996source,koulgi2003zero} since we use hyperconfusions instead of random variables.

The requirements can be translated to $\mathsf{M}\subseteq(\mathsf{Y}\rightarrow\mathsf{X})$ (decoder obtains $\mathsf{X}$ from $\mathsf{Y}\cap\mathsf{M}$) and $\mathsf{M}\supseteq\mathsf{X}$ ($\mathsf{M}$ is produced by the encoder who knows $\mathsf{X}$). These requirements can be written as a single logical formula:
\[
(\mathsf{X}\rightarrow\mathsf{M})\cap(\mathsf{M}\rightarrow(\mathsf{Y}\rightarrow\mathsf{X})),
\]
which evaluates to $2^{\Omega}$ if and only if $\mathsf{M}$ satisfies the requirements. The most ambiguous $\mathsf{M}$ is $\mathsf{M}^{*}=\mathsf{Y}\rightarrow\mathsf{X}$, which automatically satisfies $\mathsf{X}\rightarrow\mathsf{M}^{*}$.

It might appear that the requirement $\mathsf{X}\rightarrow\mathsf{M}$ does not matter, and the communication cost would be the same if the encoder has both $\mathsf{X}$ and $\mathsf{Y}$, which we know is untrue in non-asymptotic Slepian-Wolf coding. The reason of this discrepancy is that $H(\mathsf{M}^{*})$ is not the correct measure of the communication cost in this setting. If we invoke the unconfusing lemma to convert $\mathsf{M}^{*}$ to an ordinary information $\hat{\mathsf{M}}^{*}\subseteq\mathsf{M}^{*}$, then we may not have $\mathsf{X}\subseteq\hat{\mathsf{M}}^{*}$, so the encoder may not be able to output $\hat{\mathsf{M}}^{*}$. To measure the entropy of $\mathsf{M}^{*}$ with respect to the lower bound $\mathsf{X}$, we require a more general measure $H(\mathsf{M}^{*}\searrow\mathsf{X})$ called \emph{coarse entropy}, to be discussed in Section \ref{sec:coarse}. Other problems such as network coding \cite{ahlswede2000network,li2003linear} can also be analyzed similarly.

\smallskip{}

\subsection{The Logic of Coding}\label{subsec:logic_coding}

In Sections \ref{subsec:intuitionistic_task} and \ref{subsec:modus_ponens}, we have seen that we can use intuitionistic logic to prove that a coding task is trivial, i.e., it can always be accomplished without prior information, regardless of the choices of the hyperconfusion variables (e.g., $\mathsf{X},\mathsf{Y}$ in (\ref{eq:modusponens})). One may raise the question: \emph{what are the logical formulae that correspond to trivial tasks?} We have the following inclusion:
\[
\begin{array}{c}
\text{Theorems of}\\
\text{intuitionistic logic}
\end{array}\subseteq\begin{array}{c}
\text{Trivial}\\
\text{tasks}
\end{array}\subseteq\begin{array}{c}
\text{Theorems of}\\
\text{classical logic}
\end{array}.
\]
The first inclusion is because hyperconfusion forms a Heyting algebra, so formulae that are theorems of intuitionistic logic must evaluate to the greatest element \cite{heyting1930formalen}. The second inclusion ``if a task is trivial, then the formula holds in classical logic'' can be seen from simply taking each hyperconfusion variable to be either $\bot=2^{\emptyset}$ or $\top=2^{\Omega}$, and noting that the Heyting algebra $\mathrm{Hyps}(\Omega)$ has a subalgebra $\{2^{\emptyset},2^{\Omega}\}$ that is a Boolean algebra. If a formula evaluates to $\top$ regardless of how we assign $\bot$ or $\top$ to the values of the variables, then the formula holds in classical logic.

Regarding the precise characterization of the formulae corresponding to trivial tasks, one can check that this set of formulae is the theorems of Medvedev logic \cite{medvedev1962finiteEN} if we consider only finite $\Omega$, or the logic of infinite problems \cite{skvortsov1979logic} if we consider countable $\Omega$. This is mostly similar to the connection between generalized inquisitive logic and Medvedev logic shown in \cite{ciardelli2011inquisitive}. We include a proof in Appendix \ref{subsec:pf_medvedev} for the sake of completeness.

\smallskip{}

\begin{thm}
\label{thm:medvedev}For a formula involving the proposition atoms $\mathsf{X}_{1},\ldots,\mathsf{X}_{n}$, operations $\cup,\cap,\rightarrow$ and constants $\bot,\top$, this formula always evaluates to $\top$ for every combination of $\mathsf{X}_{1},\ldots,\mathsf{X}_{n}\in\mathrm{Hyps}(\Omega)$ (with $\bot=2^{\emptyset}$, $\top=2^{\Omega}$) and every finite set $\Omega$ if and only if the formula holds in Medvedev logic when regarded as a logical formula. It always evaluates to $\top$ for every finite or countable $\Omega$ if and only if the formula holds in the logic of infinite problems.
\end{thm}
\smallskip{}

There are formulae which hold in Medvedev logic (and hence in classical logic) but not in intuitionistic logic. One example is the Kreisel-Putnam formula \cite{kreisel1957eine} (also noted in \cite{ciardelli2011inquisitive})
\[
\big(\lnot\mathsf{E}\rightarrow(\mathsf{A}\cup\mathsf{B})\big)\rightarrow\big((\lnot\mathsf{E}\rightarrow\mathsf{A})\cup(\lnot\mathsf{E}\rightarrow\mathsf{B})\big).
\]
Recall that $\lnot\mathsf{E}$ is an event hyperconfusion. The corresponding coding task is ``given an original coder which perform task $\mathsf{A}$ or $\mathsf{B}$ upon event $\lnot\mathsf{E}$, output either a new coder for ``do task $\mathsf{A}$ upon event $\lnot\mathsf{E}$'', or a new coder for ``do task $\mathsf{B}$ upon event $\lnot\mathsf{E}$''{}''. This task is possible because the original coder has effectively no prior information (knowing event $\lnot\mathsf{E}$ occurs gives no information if the alternative is ignored),\footnote{The original coder has input $\lnot\mathsf{E}$ (let it be $\lnot\mathsf{E}=2^{F}$, $F\subseteq\Omega$), which means that it knows $\omega\in F$ (or an arbitrary subset of $F$) if $F$ occurs, or is provided omniscience if $F$ does not occur (impossible). Hence, it only has to concern the case where $F$ occurs, where its input information is always $F\ni\omega$.} so its operation, as a function of the prior information, is a constant function. Hence, this ``constant coder'' always achieving ``$\mathsf{A}$ or $\mathsf{B}$'' upon $\lnot\mathsf{E}$ implies it always achieves $\mathsf{A}$ or always achieves $\mathsf{B}$ upon $\lnot\mathsf{E}$. This reveals an interesting difference between the coding-logic correspondence and the Curry-Howard correspondence, as such a task is impossible in Curry-Howard since there is no concept of events there (the only type that represents ``no information'' is $\top$).

There are other formulae that hold in Medvedev logic, e.g., \cite{miglioli1989some,chen2025intermediate}. The meaning of these formulae in coding are left for future studies.

\smallskip{}

\section{Hyperconfusions Generated by one Hyperconfusion}\label{sec:gen}

\begin{figure}
\begin{centering}
\includegraphics[scale=1.13]{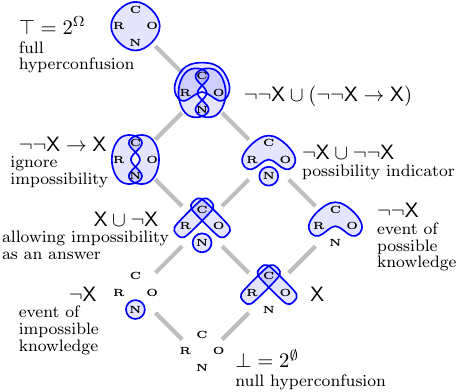}
\par\end{centering}
\caption{Hasse diagram of the hyperconfusions generated by $\mathsf{X}=\{\emptyset,\{\mathrm{C}\},\{\mathrm{R}\},\{\mathrm{O}\},\{\mathrm{C},\mathrm{R}\},\{\mathrm{C},\mathrm{O}\}\}$ over $\Omega=\{\mathrm{C},\mathrm{R},\mathrm{O},\mathrm{N}\}$.}\label{fig:lattice}
\end{figure}

Given a hyperconfusion $\mathsf{X}$, how many different hyperconfusions can be formed by applying the operations $\cup,\cap,\rightarrow$ on $\mathsf{X},\bot=2^{\emptyset},\top=2^{\Omega}$ repeatedly, i.e., how many hyperconfusions are generated by $\mathsf{X}$? There are at most $9$ different hyperconfusions generated by $\mathsf{X}$, as shown in Figure \ref{fig:lattice}. This is a sublattice of the Rieger-Nishimura lattice \cite{nishimura1960formulas}, which only contains $9$ elements because the Scott formula holds in Medvedev logic \cite{sorbi2008intermediate}. We now explain the meaning of the hyperconfusions other than $\mathsf{X},2^{\emptyset},2^{\Omega}$.  

We will use the example in the introduction where $\Omega=\{\mathrm{C},\mathrm{R},\mathrm{O},\mathrm{N}\}$, representing whether Alice has a convertible car, has a car with a fixed roof, has an open-air car without a roof, or has no car, respectively. Consider $\mathsf{X}=\{\emptyset,\{\mathrm{C}\},\{\mathrm{R}\},\{\mathrm{O}\},\{\mathrm{C},\mathrm{R}\},\{\mathrm{C},\mathrm{O}\}\}$, which is the knowledge one would obtain by seeing Alice's car and observing whether it has a roof. If one observes a roof, the confusion set is $\{\mathrm{C},\mathrm{R}\}$ (either convertible or fixed roof). If one observes a car without roof, the confusion set is $\{\mathrm{C},\mathrm{O}\}$.

\begin{itemize}
\item $\lnot\mathsf{X}=2^{\Omega\backslash\mathrm{supp}(\mathsf{X})}$ is the \emph{event of impossible knowledge}. Knowing $\lnot\mathsf{X}$ means knowing that it is impossible to further know $\mathsf{X}$, i.e., it will lead to a contradiction (or omniscience) if we further know $\mathsf{X}$. This is because $\mathsf{X}\cap\lnot\mathsf{X}=2^{\emptyset}$.  For the car example, knowing $\lnot\mathsf{X}$ is knowing that Alice does not have a car, so it is impossible to observe whether the car has a roof.
\item $\lnot\lnot\mathsf{X}=2^{\mathrm{supp}(\mathsf{X})}$ is the \emph{event of possible knowledge}. Knowing $\lnot\lnot\mathsf{X}$ means knowing that it is possible to further know $\mathsf{X}$.  For the car example, knowing $\lnot\lnot\mathsf{X}$ is knowing that Alice has a car.
\item $\mathsf{X}\cup\lnot\mathsf{X}$ is \emph{allowing impossibility as an answer}. Knowing it means that we either know $\mathsf{X}$ or know that it is impossible to know $\mathsf{X}$. This turns $\mathsf{X}$ into a question that always has an answer (i.e., $\mathrm{supp}(\mathsf{X}\cup\lnot\mathsf{X})=\Omega$), and is often the meaning of solving a problem in math. For example, if $\mathsf{X}$ is the knowledge of whether there exists a cardinal number strictly between $\aleph_{0}$ and $2^{\aleph_{0}}$ (the continuum hypothesis), then it is impossible to know $\mathsf{X}$, but we know $\mathsf{X}\cup\lnot\mathsf{X}$ since we know $\lnot\mathsf{X}$ (the continuum hypothesis cannot be proved or disproved in ZFC) \cite{cohen1966set}, so the problem is still considered to be solved. For the car example, knowing $\mathsf{X}\cup\lnot\mathsf{X}$ is knowing the answer to ``does Alice's car have a roof?'' if ``Alice has no car'' is a valid answer.
\item $\lnot\mathsf{X}\cup\lnot\lnot\mathsf{X}$ is the\emph{ possibility indicator}, an ordinary binary random variable indicating whether it is possible to know $\mathsf{X}$.  For the car example, knowing $\lnot\mathsf{X}\cup\lnot\lnot\mathsf{X}$ is knowing whether Alice has a car.
\item $\lnot\lnot\mathsf{X}\rightarrow\mathsf{X}$ is \emph{ignoring impossibility}. Knowing it means that we only need to know $\mathsf{X}$ if it is possible to know $\mathsf{X}$. This is similar to a ``don't-care term'' in digital logic, as a decoder knowing $\lnot\lnot\mathsf{X}\rightarrow\mathsf{X}$ can output arbitrary values when $\omega\notin\mathrm{supp}(\mathsf{X})$.  For the car example, knowing $\lnot\lnot\mathsf{X}\rightarrow\mathsf{X}$ is knowing an answer to ``does Alice's car have a roof?'' if any answer is acceptable when Alice has no car.
\item $\lnot\lnot\mathsf{X}\cup(\lnot\lnot\mathsf{X}\rightarrow\mathsf{X})$ is \emph{refuting or ignoring impossibility}. Knowing it means that we know that it is possible to know $\mathsf{X}$, or know $\mathsf{X}$ if it is possible to know $\mathsf{X}$. 
\end{itemize}

\section{Coarse Entropy}\label{sec:coarse}

An ordinary random variable can sometimes be defined over a coarser $\sigma$-algebra than the original $\sigma$-algebra of the probability space, and its entropy does not depend on the underlying $\sigma$-algebra. Nevertheless, if we change the underlying $\sigma$-algebra so that it induces a partition that is not finer than the partition induced by the random variable (e.g., if $\Omega=\{0,1,2,3\}$, $X(\omega)=\omega\;\mathrm{mod}\;2$, but the $\sigma$-algebra induces the partition $\{0,1\},\{2,3\}$ which confuses $0$ and $1$), then the random variable is no longer defined, or we may say that its entropy becomes infinite.

For hyperconfusion, we can also consider the entropy of a hyperconfusion $\mathsf{X}$ when the underlying $\sigma$-algebra is ``coarsened'' to a hyperconfusion $\mathsf{Y}$. The motivation is that we want to encode $\mathsf{X}$ when the encoder does not have full knowledge on $\omega$, but only knows $\mathsf{Y}$ (e.g., Section \ref{subsec:bounded}). Similar to ordinary information, if $\mathsf{Y}\nsubseteq\mathsf{X}$, then the entropy becomes infinite. Unlike ordinary information, the entropy can depend on $\mathsf{Y}$ even when $\mathsf{Y}\subseteq\mathsf{X}$. This ``coarsened'' entropy is defined below.

\smallskip{}

\begin{defn}
\label{def:entropy-2}For hyperconfusions $\mathsf{X},\mathsf{Y}$ over a probability space $(\Omega,p)$ where $\Omega=\mathrm{supp}(p)$,\footnote{If $\Omega\neq\mathrm{supp}(p)$, then we consider the restricted space $(\mathrm{supp}(p),p)$ instead, and restrict $\mathsf{X},\mathsf{Y}$ to the new space before computing $H(\mathsf{X}\searrow\mathsf{Y})$.} the \emph{coarse entropy} of $\mathsf{X}$ with respect to $\mathsf{Y}$ is 
\[
H(\mathsf{X}\searrow\mathsf{Y}):=\begin{cases}
\max_{p_{B}}\min_{p_{A|B}:\,B\subseteq A\in\mathsf{X}}I(B;A) & \mathrm{if}\;\tilde{\mathsf{Y}}\subseteq\mathsf{X},\\
\infty & \mathrm{if}\;\tilde{\mathsf{Y}}\nsubseteq\mathsf{X},
\end{cases}
\]
where $\tilde{\mathsf{Y}}:=\mathsf{Y}\cup\mathrm{sing}(\Omega)$, the maximization is over distributions $p_{B}$ over $\tilde{\mathsf{Y}}$ satisfying $p(S)\le\sum_{b\in\tilde{\mathsf{Y}}:\,S\cap b\neq\emptyset}p_{B}(b)$ for all $S\subseteq\Omega$, or equivalently, there exists a conditional distribution $p_{B|Z}$ from $\Omega$ to $\tilde{\mathsf{Y}}$ such that $p_{B}$ is the marginal distribution of $B$ when $Z\sim p$ and $B|Z\sim p_{B|Z}$, and $Z\in B$ almost surely. The minimization is over conditional distributions $p_{A|B}$ from $\tilde{\mathsf{Y}}$ to $\mathsf{X}$ satisfying $B\subseteq A$ almost surely. 
\end{defn}
\smallskip{}

Note that $H(\mathsf{X}\searrow\mathsf{Y})$ is decreasing in $\mathsf{X}$ but increasing in $\mathsf{Y}$. It should not be regarded as a ``distance'' between $\mathsf{X}$ and $\mathsf{Y}$ since $H(\mathsf{X}\searrow\mathsf{X})\ge H(\mathsf{X})$. The following proposition gives some properties of coarse entropy. In particular, the generalized unconfusing lemma shows that if $\mathsf{Y}$ is an ordinary information, then we can convert $\mathsf{X}$ to an ordinary information that is more ambiguous than $\mathsf{Y}$, with entropy approximately $H(\mathsf{X}\searrow\mathsf{Y})$. This makes the coarse entropy useful for communication tasks where there is a lower bound on the ambiguity of the message (e.g., Section \ref{subsec:bounded}). The proof is in Appendix \ref{subsec:pf_coarse}.

\smallskip{}

\begin{prop}
\label{prop:coarse}$H(\mathsf{X}\searrow\mathsf{Y})$ satisfies the following properties:
\begin{itemize}
\item $H(\mathsf{X}\searrow\mathsf{Y})\le H(\mathsf{X}'\searrow\mathsf{Y}')$ if $\mathsf{X}\supseteq\mathsf{X}'$ and $\mathsf{Y}\subseteq\mathsf{Y}'$.
\item $H(\mathsf{X})=H(\mathsf{X}\searrow2^{\emptyset})=H(\mathsf{X}\searrow\mathrm{sing}(\Omega))$.
\item If $\mathsf{X}\in\mathrm{OIs}(\Omega)$ and $\mathsf{Y}\subseteq\mathsf{X}$, then $H(\mathsf{X}\searrow\mathsf{Y})=H(\mathsf{X})$.
\item (Unconfusing lemma) If $\mathsf{Y}\in\mathrm{OIs}(\Omega)$ and $\mathsf{Y}\subseteq\mathsf{X}$, then there exists $\hat{\mathsf{X}}\in\mathrm{OIs}(\Omega)$ such that $\mathsf{Y}\subseteq\hat{\mathsf{X}}\subseteq\mathsf{X}$ and
\[
H(\hat{\mathsf{X}})\le H(\mathsf{X}\searrow\mathsf{Y})+\log(H(\mathsf{X}\searrow\mathsf{Y})+3.4)+1.
\]
\end{itemize}
\end{prop}
\smallskip{}

\section{Hyperconfusable Set-Valued Functions and Random Variables}\label{sec:setvalued}

Hyperconfusion is a generalization of partitions of the sample space (i.e., $\sigma$-algebras over finite sample spaces). While one can regard information as a partition of the sample space, another essentially equivalent  way is to treat information as random variables. One can convert a random variable into a partition, by noting that the random variable $X:\Omega\to\mathcal{X}$ induces the partition $\{X^{-1}(x):x\in\mathcal{X}\}$ of the sample space $\Omega$. 

It is natural to raise the question whether we can define a hyperconfusion generalization of the concept of random variables. Recall that a measurable space is a set with a $\sigma$-algebra, and a probability space is a measurable space with a probability measure. We now define analogous concepts with hyperconfusions in place of $\sigma$-algebras. We say that $(\Omega,\mathsf{X})$ is a \emph{hyperconfusable space} if $\mathsf{X}$ is a hyperconfusion over a nonempty sample space $\Omega$. Moreover, if $(\Omega,p)$ is a finite probability space with probability mass function $p:\Omega\to[0,1]$, then we call $(\Omega,p,\mathsf{X})$ a \emph{hyperconfusable probability space}.

We recall some basic concepts of set-valued functions \cite{aubin2009set}. A \emph{set-valued function} from $\Omega$ to $\Gamma$ is a function $\beta:\Omega\to2^{\Gamma}$. For a set $A\in\Omega$, its \emph{image} is $\beta(A):=\bigcup_{x\in A}\beta(x)$. We say that $\beta$ is \emph{surjective} if $\beta(\Omega)=\Gamma$. The \emph{inverse} of $\beta$ is $\beta^{-1}:\Gamma\to2^{\Omega}$, $\beta^{-1}(\gamma):=\{\omega\in\Omega:\,\gamma\in\beta(\omega)\}$. An ordinary function $f:\Omega\to\Gamma$ can be regarded as a set-valued function $\omega\mapsto\{f(\omega)\}$. 

We now define the concept of hyperconfusable set-valued functions, which correspond to ``realistic'' mappings that cannot reduce ambiguity. Intuitively, local processing should not be able to clear up confusion, so for a realistic $\beta$, when $A$ is a confusable set, then $\beta(A)$ should be confusable as well.

\smallskip{}

\begin{defn}
Given set-valued function $\beta:\Omega\to2^{\Gamma}$ and a hyperconfusion $\mathsf{X}$ over $\Omega$, the \emph{image hyperconfusion} is $\beta(\mathsf{X}):=\bigcup_{A\in\mathsf{X}}2^{\beta(A)}$, which is a hyperconfusion over $\Gamma$. Given hyperconfusable spaces $(\Omega,\mathsf{X})$, $(\Gamma,\mathsf{Y})$ and set-valued function $\beta:\Omega\to2^{\Gamma}$, we say that $\beta$ is \emph{hyperconfusable} if $\beta(\mathsf{X})\subseteq\mathsf{Y}$.\footnote{If $\beta$ is restricted to be a function, then $\beta$ is a simplicial map \cite{munkres2018elements}.} In this case, its \emph{preimage hyperconfusion} is $\beta^{-1}(\mathsf{Y})$ (the image hyperconfusion of $\mathsf{Y}$ through $\beta^{-1}$), which must satisfy $\beta^{-1}(\mathsf{Y})\supseteq\mathsf{X}$.
\end{defn}
\smallskip{}

We can see that for a function $f:\Omega\to\Gamma$, the ordinary information $\mathrm{oi}(f)$ is the preimage hyperconfusion of $f$ as a hyperconfusable set-valued function from $(\Omega,2^{\emptyset})$ to $(\Gamma,\mathrm{sing}(\Gamma))$. The set of ordinary informations $\mathrm{OIs}(\Omega)$ is precisely the set of preimage hyperconfusions of hyperconfusable $\beta$'s from $(\Omega,2^{\emptyset})$ to $(\Gamma,\mathrm{sing}(\Gamma))$ where $\beta$ must be a partial function (i.e., $|\beta(\omega)|\le1$ for $\omega\in\Omega$). We now extend this definition to probability spaces.

\smallskip{}

\begin{defn}
Given hyperconfusable probability spaces $(\Omega,p,\mathsf{X})$, $(\Gamma,q,\mathsf{Y})$ and set-valued function $\beta:\Omega\to2^{\Gamma}$, we say that $\beta$ is \emph{hyper-expanding} if $\beta$ is hyperconfusable, and $q(\beta(A))\ge p(A)$ for all $A\subseteq\Omega$. Moreover, we say that $\beta$ is a \emph{hyperconfusable random variable} if $\beta$ is a function (i.e., $|\beta(\omega)|=1$ for $\omega\in\Omega$) and is hyper-expanding (which implies measure-preserving). In this case, the entropy of $\beta$ is
\[
H(\beta):=H(\mathsf{Y})=H(\beta^{-1}(\mathsf{Y})).
\]
\end{defn}
\smallskip{}

A hyper-expanding mapping does not increase entropy. The proof is in Appendix \ref{subsec:pf_entropy_decrease}.

\smallskip{}

\begin{prop}
\label{prop:entropy_decrease}Given a hyper-expanding set-valued function $\beta$ from $(\Omega,p,\mathsf{X})$ to $(\Gamma,q,\mathsf{Y})$, we have
\[
H(\mathsf{X})\ge H(\mathsf{Y}).
\]
\end{prop}
\smallskip{}

We can check that hyperconfusable spaces form a category with hyperconfusable set-valued functions as morphisms. Also, hyperconfusable probability spaces form a category with hyper-expanding set-valued functions as morphisms. Properties of these categories are left for future studies.

\smallskip{}

\section{Zero-Error Joint-Source-Channel Coding}\label{sec:joint}

We can raise the question whether there exists a hyperconfusable set-valued function $\beta$ from $(\Omega,\mathsf{X})$ to $(\Gamma,\mathsf{Y})$. This question is trivial since we can always set $\beta(\omega)=\emptyset$. Hence, we require $\beta$ to be surjective.

\smallskip{}

\begin{defn}
Given hyperconfusable spaces $(\Omega,\mathsf{X})$ and $(\Gamma,\mathsf{Y})$, we say that $(\Omega,\mathsf{X})$ \emph{confuses onto} $(\Gamma,\mathsf{Y})$ (or simply $\mathsf{X}$ confuses onto $\mathsf{Y}$), denoted as $\mathsf{X}\sqsubseteq\mathsf{Y}$, if there exists a hyperconfusable surjective set-valued function $\beta$ from $(\Omega,\mathsf{X})$ to $(\Gamma,\mathsf{Y})$. The \emph{confusion ratio} between $\mathsf{X}$ and $\mathsf{Y}$ is
\[
\mathrm{CR}(\mathsf{X}/\mathsf{Y}):=\sup\Big\{\frac{m}{n}:\,\mathsf{X}^{\otimes n}\sqsubseteq\mathsf{Y}^{\otimes m}\Big\},
\]
where $\mathsf{X}^{\otimes n}$ is the $n$-fold product hyperconfusion over $\Omega^{n}$ (Definition \ref{def:product}).
\end{defn}
\smallskip{}

A simple bound can be given as a corollary of Proposition \ref{prop:entropy_decrease}. The proof is in Appendix \ref{subsec:pf_capacity}. This can be seen as a hypergraph extension of the result in \cite{korner2001relative}.

\smallskip{}

\begin{prop}
\label{prop:capacity}If $\mathsf{X}\sqsubseteq\mathsf{Y}$, then $C(\mathsf{X})\ge C(\mathsf{Y})$, where
\[
C(\mathsf{X}):=\max_{p}H(\mathsf{X})
\]
is the \emph{capacity} of $\mathsf{X}$ (we maximize over probability distributions $p$ over $\Omega$). As a result, 
\[
\mathrm{CR}(\mathsf{X}/\mathsf{Y})\le C(\mathsf{X})/C(\mathsf{Y}).
\]
\end{prop}
\smallskip{}

We now explain the meaning of $\mathsf{X}\sqsubseteq\mathsf{Y}$ via a hyperconfusion compression problem, which is a generalization of the setting in \cite{wang2017graph} to hypergraphs. Given $\gamma\in\Gamma$, the encoder compresses it into $\omega=f(\gamma)\in\Omega$. Nevertheless, the decoder does not observe $\omega$ precisely, but can only know the hyperconfusion $\mathsf{X}$, i.e., it is given $A\in\mathsf{X}$ with $A\ni\omega$. How should we design $f$ such that the decoder can know the hyperconfusion $\mathsf{Y}$ using $\mathsf{X}$, i.e., there exists a decoding function $g:\mathsf{X}\to\mathsf{Y}$ such that we must have $g(A)\ni\gamma$ for every $\gamma\in\Gamma$, $\omega=f(\gamma)$ and $A\in\mathsf{X}$ with $A\ni\omega$?\footnote{The direction of encoding is from $\mathsf{Y}$ to $\mathsf{X}$, which is different from the direction of the set-valued function $\beta$ from $\mathsf{X}$ to $\mathsf{Y}$. This is because it is the decoding function $g:\mathsf{X}\to\mathsf{Y}$ that cannot reduce ambiguity, so $\beta$ plays the role of the decoder, not the encoder.}

We now show that such a code $(f,g)$ exists if and only if $\mathsf{X}\sqsubseteq\mathsf{Y}$. If such a code exists, then taking $\beta(\omega)=\{\gamma:\,f(\gamma)=\omega\}$, for every $A\in\mathsf{X}$, we have $\beta(A)=\{\gamma:\,f(\gamma)\in A\}\subseteq g(A)\in\mathsf{Y}$, so $\beta$ is hyperconfusable. For the reverse direction, if a hyperconfusable surjective $\beta$ exists, then we can take $f:\Gamma\to\Omega$ to be any function with $f(\gamma)\in\beta^{-1}(\{\gamma\})$, and $g(A)=\beta(A)$, which satisfies $\gamma\in\beta(f(\gamma))\subseteq\beta(A)=g(A)$ for every $\gamma\in\Gamma$ and $A\in\mathsf{X}$ with $A\ni f(\gamma)$.

We can also consider a more concrete zero-error joint-source-channel coding setting. The encoder compresses a random source $U\in\Gamma$, $U\sim p_{U}$ (assume $\mathrm{supp}(p_{U})=\Gamma$) into $X=f(U)\in\Omega$. Then $X$ is passed through a noisy channel $p_{Z|X}$. The decoder observes the channel output $Z\in\mathcal{Z}$ and recovers $\hat{U}=\hat{g}(Z)\in\hat{\Gamma}$. The goal is to ensure $(U,\hat{U})\in\mathcal{S}$ almost surely, where $\mathcal{S}\subseteq\Gamma\times\hat{\Gamma}$ is the set of allowed source-reconstruction pairs. Let $\rho(x):=\{z\in\mathcal{Z}:\,p_{Z|X}(z|x)>0\}$ be a hyperconfusable set-valued function from $(\Omega,2^{\emptyset})$ to $(\mathcal{Z},\mathrm{sing}(\mathcal{Z}))$, and $\sigma(u):=\{\hat{u}\in\hat{\Gamma}:\,(u,\hat{u})\in\mathcal{S}\}$ be a hyperconfusable set-valued function from $(\Gamma,2^{\emptyset})$ to $(\hat{\Gamma},\mathrm{sing}(\hat{\Gamma}))$. Taking 
\begin{align}
\mathsf{X} & :=\rho^{-1}(\mathrm{sing}(\mathcal{Z}))\nonumber \\
 & =\{A\subseteq\Omega:\,\exists z\in\mathcal{Z}.\,\forall x\in A.\,p_{Z|X}(z|x)>0\},\label{eq:zeroerror_X}\\
\mathsf{Y} & :=\sigma^{-1}(\mathrm{sing}(\hat{\Gamma}))\nonumber \\
 & =\{B\subseteq\Gamma:\,\exists\hat{u}\in\hat{\Gamma}.\,\forall u\in B.\,(u,\hat{u})\in\mathcal{S}\}\label{eq:zeroerror_Y}
\end{align}
to be the preimage hyperconfusions, we can check that this problem is equivalent to the aforementioned hyperconfusion compression problem.

This setting generalizes  zero-error channel coding \cite{shannon1956zero}. The zero-error capacity of a channel with confusion graph $G=(\Omega,E)$, where $E=\{\{u,v\}:\,u,v\text{ confusable}\}\subseteq2^{\Omega}$, can be given as $\mathrm{CR}(\mathsf{X}/\mathrm{sing}([2]))$, where $\mathsf{X}=\mathrm{sing}(\Omega)\cup E$.

We can also study the problem of zero-error channel coding with list decoding \cite{elias1988zero,korner1990capacity} using hyperconfusions. The encoder compresses $M\in[k]$ into $X=f(M)\in\Omega$, which is passed through a noisy channel $p_{Z|X}$. The decoder observes $Z\in\mathcal{Z}$ and outputs a list $\hat{\mathcal{M}}\subseteq[k]$ with $|\hat{\mathcal{M}}|\le\ell$ which satisfies $M\in\hat{\mathcal{M}}$ almost surely. We can see that this is possible if and only if $\mathsf{X}\sqsubseteq\mathsf{Y}$, where $\mathsf{X}$ is given in \ref{eq:zeroerror_X}, and $\mathsf{Y}=\{B\subseteq[k]:\,|B|\le\ell\}$. 

The confusion ratio is a generalization of the information ratio between confusion graphs studied in \cite{wang2017graph}. For two confusion graphs $(\Omega,E)$ and $(\Gamma,F)$, a function $f:\Gamma\to\Omega$ is called non-adjacency preserving \cite{wang2017graph} if for every $\gamma,\gamma'\in\Gamma$ that are non-adjacent in $(\Gamma,F)$, $f(\gamma),f(\gamma')$ are also non-adjacent in $(\Omega,E)$. Letting $\mathsf{X}=\mathrm{sing}(\Omega)\cup E$, $\mathsf{Y}=\mathrm{sing}(\Gamma)\cup F$, we can see that a non-adjacency preserving mapping exists if and only if $\mathsf{X}\sqsubseteq\mathsf{Y}$. Therefore, the information ratio \cite{wang2017graph} is given as $\mathrm{CR}(\mathsf{X}/\mathsf{Y})$. Also refer to \cite{korner2001relative} for a related problem.

\smallskip{}

\section{Algorithms for Hyperconfusions}\label{sec:algorithm}

We now describe some algorithms for computations about hyperconfusions. To store a hyperconfusion $\mathsf{X}$, it suffices to store the maximal sets $\mathrm{maxs}(\mathsf{X}):=\{A\in\mathsf{X}:\,\forall B\in\mathsf{X}:\,B\supseteq A\,\to\,B=A\}$, since $\mathsf{X}=\bigcup_{A\in\mathrm{maxs}(\mathsf{X})}2^{A}$ is the downward closure of $\mathrm{maxs}(\mathsf{X})$. This reduces the storage cost. For example, the full hyperconfusion $2^{\Omega}$ can be stored as $\mathrm{maxs}(2^{\Omega})=\{\Omega\}$. To compute the disjunction, we have to compute $\mathrm{maxs}(\mathsf{X}\cup\mathsf{Y})$, which can be performed by considering $\mathcal{S}=\mathrm{maxs}(\mathsf{X})\cup\mathrm{maxs}(\mathsf{Y})$, and then taking $\mathrm{maxs}(\mathsf{X}\cup\mathsf{Y})=\mathrm{maxs}(\mathcal{S})$. To compute the conjunction, we have $\mathrm{maxs}(\mathsf{X}\cap\mathsf{Y})=\mathrm{maxs}(\{A\cap B:\,A\in\mathrm{maxs}(\mathsf{X}),\,B\in\mathrm{maxs}(\mathsf{Y})\})$. Nevertheless, for implication, there does not appear to be a simple algorithm that directly computes $\mathrm{maxs}(\mathsf{X}\rightarrow\mathsf{Y})$ efficiently other than exhausting all $2^{|\Omega|}$ possible sets.

To find the entropy $H(\mathsf{X})=\min_{p_{A|Z}:\,Z\in A\in\mathsf{X}\;\mathrm{a.s.}}I(Z;A)$ numerically, it suffices to consider $A\in\mathrm{maxs}(\mathsf{X})$.\footnote{We can construct $\tilde{A}$ by adding elements in $\Omega$ to $A$ one by one in a fixed order as long as $\tilde{A}\in\mathsf{X}$. At the end, we have $\tilde{A}\in\mathrm{maxs}(\mathsf{X})$ and $H(\tilde{A}|A)$, so $I(Z;A)\ge I(Z;\tilde{A})$.} We can use the Blahut-Arimoto algorithm \cite{blahut1972computation,arimoto1972algorithm}, which iteratively computes for $k=0,1,2,\ldots$, and $a\in\mathrm{maxs}(\mathsf{X})$, $z\in a$:
\[
p_{A|Z}^{(k+1)}(a|z):=\frac{\mathbf{1}\{z\in a\}p_{A}^{(k)}(a)}{\sum_{a'\in\mathrm{maxs}(\mathsf{X}),a'\ni z}p_{A}^{(k)}(a')},
\]
where $p_{A}^{(k)}(a):=\sum_{z\in a}p(z)p_{A|Z}^{(k)}(a|z)$. We may initialize $p_{A|Z}^{(0)}(a|z)=\mathbf{1}\{z\in a\}/|\{a'\in\mathrm{maxs}(\mathsf{X}),a'\ni z\}|$ for $a\in\mathrm{maxs}(\mathsf{X})$, $z\in a$.

One may also be interested in symbolic algorithms for the automated proofs of linear inequalities involving entropies of hyperconfusions (e.g., $H(\mathsf{X}\rightarrow\mathsf{Y})\ge H(\mathsf{X}\cap\mathsf{Y})-H(\mathsf{Y})$), in a manner similar to the linear programming framework for entropy inequalities for random variables \cite{yeung1997framework,yeung2021machine,li2023automated}. Nevertheless, while there are only $2^{n}-1$ different joint entropies that can be formed by random variables $X_{1},\ldots,X_{n}$, making the linear program in \cite{yeung1997framework} tractable, the number of hyperconfusions that can be formed by $\mathsf{X}_{1},\ldots,\mathsf{X}_{n}$ is significantly larger (there are $7$ different non-null non-full hyperconfusions that can be formed by just $\mathsf{X}_{1}$, as shown in Section \ref{sec:gen}). Developing an efficient symbolic representation of the hyperconfusions generated by $\mathsf{X}_{1},\ldots,\mathsf{X}_{n}$ is left for future studies.

\section{From Heyting Algebra to Hyperconfusion}\label{sec:Heyting_to}

We have seen how the set of hyperconfusions $\mathrm{Hyps}(\Omega)$ forms a Heyting algebra. In this section, we consider the reverse direction, and study whether an arbitrary Heyting algebra $\mathcal{H}$ can be embedded in the set of hyperconfusions $\mathrm{Hyps}(\Omega)$ for some $\Omega$. If this is possible, then we can generalize the concept of information to any Heyting algebra. 

Recall that the negation is defined as $\lnot x:=(x\rightarrow\bot)$. The \emph{regularization} of $\mathcal{H}$ is \cite{grilletti2024esakia}
\[
\mathcal{H}_{\lnot}:=\{x\in\mathcal{H}:\,x=\lnot\lnot x\}.
\]
Note that $\mathcal{H}_{\lnot}$ is a Boolean algebra with meet $\wedge$ (same as $\mathcal{H}$) and join $x\dot{\vee}y:=\lnot(\lnot x\wedge\lnot y)$ \cite{grilletti2024esakia}. By Stone's representation theorem \cite{stone1936theory}, $\mathcal{H}_{\lnot}$ can be embedded into a power set. Let $\psi:\mathcal{H}_{\lnot}\to2^{\mathrm{U}(\mathcal{H}_{\lnot})}$ be the Stone embedding of $\mathcal{H}_{\lnot}$,\footnote{$\mathrm{U}(\mathcal{H}_{\lnot})\subseteq2^{\mathcal{H}_{\lnot}}$ is the set of ultrafilters of $\mathcal{H}_{\lnot}$, and $\psi(x):=\{u\in\mathrm{U}(\mathcal{H}_{\lnot}):\,x\in u\}$. We will not require these details in this paper.} which is injective, and preserves and reflects order, join, meet and negation (i.e., $x\le y$ if and only if $\psi(x)\subseteq\psi(y)$, $\psi(x\wedge y)=\psi(x)\cap\psi(y)$, $\psi(x\dot{\vee}y)=\psi(x)\cup\psi(y)$, $\psi(\lnot x)=\mathrm{U}(\mathcal{H}_{\lnot})\backslash\psi(x)$, $\psi(\bot)=\emptyset$).

If $\mathcal{H}=\mathrm{Hyps}(\Omega)$, then $\mathcal{H}_{\lnot}=\{2^{E}:\,E\subseteq\Omega\}$ are the event hyperconfusions (Section \ref{subsec:negation}), $\mathrm{U}(\mathcal{H}_{\lnot})=\Omega$, and $\psi(2^{E})=E$. Hence, $\mathcal{H}_{\lnot}$ and $\psi$ allows us to extract the sample space and events from the structure of $\mathcal{H}$, even if $\mathcal{H}$ may not be $\mathrm{Hyps}(\Omega)$ for some $\Omega$. This inspires the following general embedding of any Heyting algebra to a set of hyperconfusions.

\smallskip{}

\begin{defn}
Given a Heyting algebra $\mathcal{H}$, the \emph{hyperconfusion pseudo-embedding} is the function $\zeta:\mathcal{H}\to\mathrm{Hyps}(\Omega)$, where $\Omega=\mathrm{U}(\mathcal{H}_{\lnot})$,
\[
\zeta(x):=\bigcup_{z\in\mathcal{H}_{\lnot}:\,z\le x}2^{\psi(z)}.
\]
\end{defn}
\smallskip{}

The hyperconfusion pseudo-embedding preserves order and meet, but does not necessarily preserve implication and join.

\smallskip{}

\begin{thm}
\label{thm:pseudoembedding}For all $x,y\in\mathcal{H}$, we have
\begin{itemize}
\item $\zeta(x\wedge y)=\zeta(x)\cap\zeta(y)$, $\zeta(\bot)=2^{\emptyset}$ and $\zeta(\top)=2^{\Omega}$.
\item $x\le y$ implies $\zeta(x)\subseteq\zeta(y)$. Moreover, $\zeta(x)\subseteq\zeta(y)$ implies $x\le y$ if $\mathcal{H}_{\lnot}$ is join-dense in $\mathcal{H}$.\footnote{$\mathcal{H}_{\lnot}$ is join-dense in $\mathcal{H}$ if for all $w\in\mathcal{H}$, $w=\bigvee_{t\in\mathcal{H}_{\lnot}:\,t\le w}t$ \cite{halavs2022join}.}
\item $\zeta(x\rightarrow y)\subseteq(\zeta(x)\rightarrow\zeta(y))$ and $\zeta(\lnot x)\subseteq\lnot\zeta(x)$. Equality holds if $\mathcal{H}_{\lnot}$ is complete and atomic,\footnote{$\mathcal{H}_{\lnot}$ is complete and atomic if it is isomorphic to $2^{\mathrm{U}(\mathcal{H}_{\lnot})}$ \cite{halmos2018lectures}.} and is join-dense in $\mathcal{H}$.
\item $\zeta(x\vee y)\supseteq(\zeta(x)\cup\zeta(y))$. Equality holds if all elements in $\mathcal{H}_{\lnot}$ are join-prime in $\mathcal{H}$.\footnote{$z$ is join-prime if for every $x,y\in\mathcal{H}$, $z\le x\vee y$ implies $z\le x$ or $z\le y$ \cite{gaskill1981join}.}
\end{itemize}
\end{thm}
\smallskip{}

\begin{IEEEproof}
To prove $\zeta(\bot)=2^{\emptyset}$, for $A\subseteq\mathrm{U}(\mathcal{H}_{\lnot})$, we have $A\in\zeta(\bot)$ iff $A\subseteq\psi(z)$ for some $z\in\mathcal{H}_{\lnot}$, $z\le\bot$, iff $A\subseteq\psi(\bot)$, iff $A=\emptyset$. To prove $\zeta(\top)=2^{\Omega}$, for $A\subseteq\mathrm{U}(\mathcal{H}_{\lnot})$, we have $A\in\zeta(\top)$ iff $A\subseteq\psi(z)$ for some $z\in\mathcal{H}_{\lnot}$, $z\le\top$, iff $A\subseteq\psi(\top)=\mathrm{U}(\mathcal{H}_{\lnot})=\Omega$. 

To prove $\zeta(x\wedge y)=\zeta(x)\cap\zeta(y)$, for $A\subseteq\mathrm{U}(\mathcal{H}_{\lnot})$, we have $A\in\zeta(x\wedge y)$ iff $A\subseteq\psi(z)$ for some $z\in\mathcal{H}_{\lnot}$, $z\le x\wedge y$, iff $A\subseteq\psi(z_{1})\cap\psi(z_{2})=\psi(z_{1}\wedge z_{2})$ for some $z_{1},z_{2}\in\mathcal{H}_{\lnot}$, $z_{1}\le x$, $z_{2}\le y$, iff $A\in\zeta(x)\cap\zeta(y)$. 

$x\le y$ implies $\zeta(x)\subseteq\zeta(y)$ follows directly from the definition. To prove $\zeta(x)\subseteq\zeta(y)$ implies $x\le y$ if $\mathcal{H}_{\lnot}$ is join-dense, assume $\zeta(x)\subseteq\zeta(y)$. Then for every $z\in\mathcal{H}_{\lnot}$ with $z\le x$, we have $\psi(z)\subseteq\psi(z')$ (equivalently, $z\le z'$) for some $z'\in\mathcal{H}_{\lnot}$ with $z'\le y$, and hence $z\le\bigvee_{z'\in\mathcal{H}_{\lnot}:z'\le y}z'=y$ by join-density. Hence, by join-density, $x=\bigvee_{z\in\mathcal{H}_{\lnot}:z\le x}z\le y$.

To prove $\zeta(x\rightarrow y)\subseteq(\zeta(x)\rightarrow\zeta(y))$, consider $A\in\zeta(x\rightarrow y)$. Then $A\subseteq\psi(z)$ for some $z\in\mathcal{H}_{\lnot}$, $z\le(x\rightarrow y)$, so $(z\wedge x)\le y$. For any $B\subseteq A$ with $B\in\zeta(x)$, we have $B\subseteq\psi(z')$ for some $z'\in\mathcal{H}_{\lnot}$, $z'\le x$, and hence $B\subseteq\psi(z)\cap\psi(z')=\psi(z\wedge z')$ and $z\wedge z'\le z\wedge x\le y$. Therefore, $\zeta(x)\cap2^{A}\subseteq\zeta(y)$, and $A\in(\zeta(x)\rightarrow\zeta(y))$. 

To prove $(\zeta(x)\rightarrow\zeta(y))\subseteq\zeta(x\rightarrow y)$ if $\mathcal{H}_{\lnot}$ is complete, atomic and join-dense, consider $A\in(\zeta(x)\rightarrow\zeta(y))$, and assume $A=\psi(a)$ for some $a\in\mathcal{H}_{\lnot}$. Then $\zeta(x)\cap2^{\psi(a)}\subseteq\zeta(y)$. For every $z\in\mathcal{H}_{\lnot}$ with $z\le x$, we have $\psi(z)\cap\psi(a)\subseteq\psi(z')$ (equivalently, $z\wedge a\le z'$) for some $z'\in\mathcal{H}_{\lnot}$ with $z'\le y$, and hence $z\wedge a\le\bigvee_{z'\in\mathcal{H}_{\lnot}:z'\le y}z'=y$ by join-density, and $z\le(a\rightarrow y)$. Hence, by join-density, $x=\bigvee_{z\in\mathcal{H}_{\lnot}:z\le x}z\le(a\rightarrow y)$. We have $x\wedge a\le y$, $a\le(x\rightarrow y)$, and $A=\psi(a)\in\zeta(x\rightarrow y)$.

$\zeta(x\vee y)\supseteq(\zeta(x)\cup\zeta(y))$ follows directly from the monotonicity of $\zeta$. To prove $\zeta(x\vee y)\subseteq(\zeta(x)\cup\zeta(y))$ for join-prime $\mathcal{H}_{\lnot}$, consider $A\in\zeta(x\vee y)$. Then $A\subseteq\psi(z)$ for some $z\in\mathcal{H}_{\lnot}$, $z\le x\vee y$. Since $z$ is join-prime, we have $z\le x$ or $z\le y$, so $A\in\zeta(x)$ or $A\in\zeta(y)$.

\end{IEEEproof}
\smallskip{}

Given Theorem \ref{thm:pseudoembedding}, we can define a class of Heyting algebras where the hyperconfusion pseudo-embedding is indeed an embedding (in fact an isomorphism) that preserves all operations.

\smallskip{}

\begin{defn}
We say that a Heyting algebra $\mathcal{H}$ is \emph{strictly hyperconfusable} if $\mathcal{H}_{\lnot}$ is complete and atomic, is join-dense in $\mathcal{H}$, and all elements in $\mathcal{H}_{\lnot}$ are join-prime in $\mathcal{H}$.
\end{defn}
\smallskip{}

Note that $\mathrm{Hyps}(\Omega)$ is a strictly hyperconfusable Heyting algebra. We remark that a complete atomic Boolean algebra is not necessarily strictly hyperconfusable since it may fail the join-prime condition. Strictly hyperconfusable Heyting algebra is a rather stringent condition. Nevertheless, it is not necessarily for interpreting a Heyting algebra as a model of information. We can generalize the concept of hyperconfusion to be an element of \emph{any} Heyting algebra. The operations of hyperconfusion are conjunction (meet), disjunction (join) and implication, which are defined over any Heyting algebra. It is left to generalize the concept of entropy to an element of an arbitrary finite Heyting algebra, which is simply taken to be the entropy of the pseudo-embedding.

\smallskip{}

\begin{defn}
Given a finite Heyting algebra $\mathcal{H}$, and a probability mass function $p:\mathrm{U}(\mathcal{H}_{\lnot})\to[0,1]$, the \emph{entropy} of $x\in\mathcal{H}$ is 
\[
H(x):=H(\zeta(x)),
\]
which is the entropy of the pseudo-embedding $\zeta(x)$, which is a hyperconfusion over the probability space $(\mathrm{U}(\mathcal{H}_{\lnot}),p)$.
\end{defn}
\smallskip{}

Theorem \ref{thm:pseudoembedding} implies that the entropy over an arbitrary Heyting algebra retains many desirable properties of the entropy of hyperconfusions. For example, we still have $H(x\wedge y)\le H(x)+H(y)$ and $H(x\rightarrow y)\ge H(x\wedge y)-H(y)$. We can even define entropy in an abstract manner to be any function satisfying the basic properties.

\smallskip{}

\begin{defn}
Given a Heyting algebra $\mathcal{H}$, an \emph{abstract entropy} over $\mathcal{H}$ is a function $H:\mathcal{H}\to[0,\infty]$ satisfying $H(\top)=0$, $H$ is non-increasing ($x\le y$ implies $H(x)\ge H(y)$), and $H(x\wedge y)\le H(x)+H(y)$ for all $x,y\in\mathcal{H}$.
\end{defn}
\smallskip{}

A more abstract investigation of the concept of information via Heyting algebra is left for future studies.

\section{Acknowledgement}

This work was partially supported by two grants from the Research Grants Council of the Hong Kong Special Administrative Region, China {[}Project No.s: CUHK 24205621 (ECS), CUHK 14209823 (GRF){]}. The author would like to thank Raymond W. Yeung, Chandra Nair, Amin Gohari, Peter Harremo\"{e}s, Kenneth W. Shum and Abdellatif Zaidi for the insightful discussions.

\smallskip{}

\appendix{}

\subsection{Proof of Proposition \ref{prop:convex}}\label{subsec:pf_convex}

To prove $H(\mathsf{X})\ge\min_{v\in\mathcal{C}}\sum_{\omega}p(\omega)\log(1/v(\omega))$, consider any $p_{A|Z}$ with $Z\in A\in\mathsf{X}$ almost surely when $Z\sim p$, $A|Z\sim p_{A|Z}$. Let $\mathcal{S}_{\omega}:=\{A\in\mathsf{X}:\,A\ni\omega\}$. By the data processing inequality for KL divergence, 
\begin{align*}
I(Z;A) & =\mathbb{E}[D(p_{A|Z}(\cdot|Z)\Vert p_{A})]\\
 & \ge\mathbb{E}[D(\mathrm{Bern}(p_{A|Z}(\mathcal{S}_{Z}|Z))\Vert\mathrm{Bern}(p_{A}(\mathcal{S}_{Z})))]\\
 & =\mathbb{E}[D(\mathrm{Bern}(1)\Vert\mathrm{Bern}(p_{A}(\mathcal{S}_{Z})))]\\
 & =\mathbb{E}[-\log p_{A}(\mathcal{S}_{Z})]\\
 & =\mathbb{E}\Big[-\log\sum_{A\in\mathsf{X}}p_{A}(A)\mathbf{1}_{A}(Z)\Big]\\
 & \ge\min_{v\in\mathcal{C}}\mathbb{E}[-\log v(Z)].
\end{align*}
To prove $H(\mathsf{X})\le\min_{v\in\mathcal{C}}\sum_{\omega}p(\omega)\log(1/v(\omega))$, for any $v\in\mathcal{C}$, let $v=\sum_{A\in\mathsf{X}}q(A)\mathbf{1}_{A}$ for a probability mass function $q:\mathsf{X}\to[0,1]$. Take $p_{A|Z}(a|z)=\mathbf{1}_{A}(z)q(a)/q(\mathcal{S}_{z})$, we have 
\begin{align*}
I(Z;A) & \le\mathbb{E}[D(p_{A|Z}(\cdot|Z)\Vert q)]\\
 & =\mathbb{E}[-\log q(\mathcal{S}_{Z})]\;=\;\mathbb{E}[-\log v(Z)].
\end{align*}

\subsection{Proof of Proposition \ref{prop:properties}}\label{subsec:pf_properties}

To prove $H_{\infty}(\mathsf{X})\le H(\mathsf{X})$ using Proposition \ref{prop:convex}, letting $v(\Omega):=\sum_{\omega\in\Omega}v(\omega)$,
\begin{align*}
H(\mathsf{X}) & =\min_{v\in\mathcal{C}}\sum_{\omega\in\Omega}p(\omega)\log\frac{1}{v(\omega)}\\
 & =\min_{v\in\mathcal{C}}\Big(\sum_{\omega\in\Omega}p(\omega)\log\frac{1}{v(\omega)/v(\Omega)}-\log v(\Omega)\Big)\\
 & \ge\min_{v\in\mathcal{C}}(-\log v(\Omega))\;=\;H_{\infty}(\mathsf{X}).
\end{align*}
To prove $H(\mathsf{X})\le H_{\epsilon}(\mathsf{X})$, consider any fractional covering $\mu$. Let $\mu(\mathsf{X}):=\sum_{A\in\mathsf{X}}\mu(A)$. Take $v(\omega)=\sum_{A\in\mathsf{X}}(\mu(A)/\mu(\mathsf{X}))\mathbf{1}_{A}(\omega)$. We have $v(\omega)\ge1/\mu(\mathsf{X})$ for every $\omega\in\mathrm{supp}(p)$. By Proposition \ref{prop:convex}, $H(\mathsf{X})\le\sum_{\omega\in\Omega}p(\omega)\log(1/v(\omega))\le\log\mu(\mathsf{X})$.

$H_{\epsilon}(\mathsf{X})\le H_{0}(\mathsf{X})$ and $H_{\infty}(\mathsf{X})\ge0$ follow from the definition. If $2^{\mathrm{supp}(p)}\subseteq\mathsf{X}$, then $H_{0}(\mathsf{X})=0$ by taking $\mathcal{S}=\{\mathrm{supp}(p)\}$, and hence all entropies are zero. If any of the entropies is zero, then $H_{\infty}(\mathsf{X})=0$, and there exists $A\in\mathsf{X}$ such that $p(A)$. Monotonicity follows from the definitions. If $\mathsf{X}=\mathrm{oi}(\tilde{X})$ for an ordinary random variable $\tilde{X}$, then $H_{\infty}(\mathsf{X})=H_{\infty}(\tilde{X})$ by definition, $H(\mathsf{X})=H(\tilde{X})$ since we have a Markov chain $Z\to A\to\tilde{X}^{-1}(\tilde{X}(Z))$ (where $\tilde{X}^{-1}(\tilde{X}(z))=\{\omega\in\Omega:\,\tilde{X}(\omega)=\tilde{X}(z)\}$), so $I(Z;A)\le I(Z;\tilde{X}^{-1}(\tilde{X}(Z)))=H(\tilde{X})$ with equality if $A=\tilde{X}^{-1}(\tilde{X}(Z))$, and $H_{0}(\mathsf{X})=H_{\epsilon}(\mathsf{X})=H_{0}(\tilde{X})$ since the (integral/fractional) covering number of a partition is the number of parts in the partition.

\smallskip{}

\subsection{Proof of Lemmas \ref{lem:unconfuse} and \ref{lem:unconfuse2}}\label{subsec:pf_unconfuse}

For lemma \ref{lem:unconfuse}, since both sides are infinite when $H(\mathsf{X})=\infty$, we focus on the case $H(\mathsf{X})<\infty$. Assume $\mathrm{supp}(p)=\Omega$ without loss of generality. Consider any $p_{A|Z}$ with $Z\in A\in\mathsf{X}$ almost surely. By the strong functional representation lemma \cite{sfrl_trans,li2024channel,li2025discrete}, there exists a random variable $W\in\mathcal{W}$ independent of $Z$ and a function $f:\Omega\times\mathcal{W}\to\mathsf{X}$ such that $A=f(Z,W)$ and $H(A|W)\le I(Z;A)+\log(I(Z;A)+3.4)+1$. Hence, there exists a constant $w\in\mathcal{W}$ such that $H(A|W=w)\le I(Z;A)+\log(I(Z;A)+3.4)+1$. Take $\mathsf{Y}=\mathrm{oi}(\omega\mapsto f(\omega,w))$. We have $\mathsf{Y}\subseteq\mathsf{X}$ since $\omega\in A$ whenever $f(\omega,w)=A$. We have $H(\mathsf{Y})=H(f(Z,w))=H(A|W=w)$ satisfying the upper bound. The result follows from minimizing over $p_{A|Z}$.

For lemma \ref{lem:unconfuse2} with $H_{0}$, we assume $\mathrm{supp}(p)=\Omega$ without loss of generality. If $\mathcal{S}\subseteq\mathsf{X}$ is an integral cover, there exists $f:\mathcal{S}\to2^{\Omega}$ such that $f(A)\subseteq A$ and $\{f(A)\}_{A\in\mathcal{S}}$ forms a partition of $\Omega$. Taking $\mathsf{Y}=\bigcup_{A\in\mathcal{S}}2^{f(A)}$ completes the proof. For $H_{\infty}$, let $H_{\infty}(\mathsf{X})=-\log p(A^{*})$ for $A^{*}\in\mathsf{X}$. The result follows from taking $\mathsf{Y}$ to be induced by an arbitrary partition of $\bigcup_{A\in\mathsf{X}}A$ where each part is a subset of a set in $\mathsf{X}$, and one of the parts is $A^{*}$.

\smallskip{}

\subsection{Proof of Proposition \ref{prop:conjunction}}\label{subsec:pf_conjunction}

To prove $H(\mathsf{X}\cap\mathsf{Y})\le H(\mathsf{X})+H(\mathsf{Y})$, consider any $p_{A|Z}$ and $p_{B|Z}$ with $Z\in A\in\mathsf{X}$ and $Z\in B\in\mathsf{Y}$ almost surely. Generate $Z\sim p$, and $A|Z\sim p_{A|Z}$, $B|Z\sim p_{B|Z}$ conditionally independently given $Z$. Let $C=A\cap B\in\mathsf{X}\cap\mathsf{Y}$ since $\mathsf{X}$, $\mathsf{Y}$ are downward closed. We have $H(\mathsf{X}\cap\mathsf{Y})\le I(Z;C)\le I(Z;A,B)\le I(Z;A)+I(Z;B)$ since $I(A;B|Z)=0$. The result follows from minimizing over $p_{A|Z},p_{B|Z}$.

To prove $H(\mathsf{X}\cap\mathsf{Y})=H(\mathsf{X})+H(\mathsf{Y})$ when $\mathsf{X},\mathsf{Y}$ are independent, note that $\mathsf{X},\mathsf{Y}$ being independent means that the $\sigma$-algebra generated by $\mathrm{maxs}(\mathsf{X})$ is independent of the $\sigma$-algebra generated by $\mathrm{maxs}(\mathsf{Y})$. Assume the $\sigma$-algebra generated by $\mathrm{maxs}(\mathsf{X})$ is also generated by the function $\phi:\Omega\to\mathbb{N}$, and the $\sigma$-algebra generated by $\mathrm{maxs}(\mathsf{Y})$ is also generated by the function $\psi:\Omega\to\mathbb{N}$. Consider any $C\in\mathrm{maxs}(\mathsf{X}\cap\mathsf{Y})$ (to evaluate $H(\mathsf{X}\cap\mathsf{Y})$, it suffices to consider maximal $C$). Let $A\in\mathrm{maxs}(\mathsf{X})$ be such that $C\subseteq A$ (if multiple $A$'s are possible, select the lexicographically smallest). Similarly, let $B\in\mathrm{maxs}(\mathsf{Y})$ be such that $C\subseteq B$. If there exists $\omega\in(A\cap B)\backslash C$, then $C\cup\{\omega\}\in\mathsf{X}\cap\mathsf{Y}$, contradicting $C\in\mathrm{maxs}(\mathsf{X}\cap\mathsf{Y})$. Hence, $C=A\cap B$. This implies that the $\sigma$-algebra generated by $\mathrm{maxs}(\mathsf{X}\cap\mathsf{Y})$ is also generated by the functions $\phi$ and $\psi$. 

Consider any $p_{C|Z}$ satisfying $Z\in C\in\mathrm{maxs}(\mathsf{X}\cap\mathsf{Y})$ almost surely. Let $U=\phi(Z)$, $V=\psi(Z)$ be independent. Since the $\sigma$-algebra generated by $\mathrm{maxs}(\mathsf{X}\cap\mathsf{Y})$ is also generated by $\phi$ and $\psi$, whether $Z\in C$ for $C\in\mathrm{maxs}(\mathsf{X}\cap\mathsf{Y})$ only depends on $(U,V)$. Hence, it suffices to assume $Z\to(U,V)\to C$ forms a Markov chain, and consider $p_{C|U,V}$. We have $I(Z;C)=I(U,V;C)\ge I(U;C)+I(V;C)$ since $I(U;V)=0$. Since the $\sigma$-algebra generated by $\mathrm{maxs}(\mathsf{X})$ is also generated by $\phi$, we have $I(U;C)\ge I(U;A)\ge H(\mathsf{X})$. Hence, $I(Z;C)\ge H(\mathsf{X})+H(\mathsf{Y})$. The result follows from minimizing over $p_{C|Z}$.

\smallskip{}

\subsection{Proof of Theorem \ref{thm:disjunctive}}\label{subsec:pf_disjunctive}

We have $H(\bigcup_{i=1}^{n}\mathsf{X}_{i})\ge H_{\infty}(\bigcup_{i=1}^{n}\mathsf{X}_{i})=H_{\infty}(\mathsf{X})$. It is left to show $\lim_{n}H(\bigcup_{i=1}^{n}\mathsf{X}_{i})\le H_{\infty}(\mathsf{X})$. Consider an i.i.d. sequence $Z_{1},\ldots,Z_{n}\in\Omega$ following $p$. Write $Z^{n}=(Z_{1},\ldots,Z_{n})$. Our goal is to design random variables $K\in[n]$, $A\in\mathsf{X}$ such that $Z_{K}\in A$ almost surely, and $I(Z^{n};K,A)\to H_{\infty}(\mathsf{X})$ as $n\to\infty$. Let $H_{\infty}(\mathsf{X})=-\log p(A^{*})$ for $A^{*}\in\mathsf{X}$. Fix $\epsilon>0$. Let $S:=\{i\in[n]:Z_{i}\in A^{*}\}$, $E$ be the event ``$||S|/n-p(A^{*})|<\epsilon$'', and $\delta:=\mathbb{P}(E^{\mathrm{c}})$. We take $K\sim\mathrm{Unif}(S)$ and $A=A^{*}$ if $E$ occurs, or $K=1$ and $A=\{Z_{1}\}$ if $E$ does not occur. We have
\begin{align*}
 & I(Z^{n};K,A)\\
 & \le I(Z^{n};K,A|E)+\delta I(Z^{n};K,A|E^{\mathrm{c}})+H_{b}(\delta)\\
 & \le I(S;K|E)+\delta\log|\Omega|+H_{b}(\delta)\\
 & =H(K|E)-H(K|S,E)+\delta\log|\Omega|+H_{b}(\delta)\\
 & =\log n-\mathbb{E}[\log|S|\,|\,E]+\delta\log|\Omega|+H_{b}(\delta)\\
 & \le\log n-\log(np(A^{*})-n\epsilon)+\delta\log|\Omega|+H_{b}(\delta)\\
 & \to-\log(p(A^{*})-\epsilon)
\end{align*}
as $n\to\infty$ since $\delta\to0$ due to the law of large numbers, where $H_{b}$ is the binary entropy function. Taking $\epsilon\to0$ completes the proof.

\smallskip{}

\subsection{Proof of Proposition \ref{prop:condition}}\label{subsec:pf_condition}

 Assume $\mathrm{supp}(p)=\Omega$ without loss of generality. $H(\mathsf{X}|2^{\Omega})=H(\mathsf{X})$ and $H(\mathsf{X}|2^{\emptyset})=0$ follow directly from the definition. To prove $H(\mathsf{X}|\mathsf{Y})+H(\mathsf{Y})=H(\mathsf{X}\cap\mathsf{Y})$, if $\mathrm{supp}(\mathsf{Y})=\Omega$, then $H(\mathsf{X}|\mathsf{Y})=H(\mathsf{X}\cap\mathsf{Y})-H(\mathsf{Y})$. Otherwise if $\mathrm{supp}(\mathsf{Y})\neq\Omega$, then $H(\mathsf{Y})=H(\mathsf{X}\cap\mathsf{Y})=\infty$, so the result is also true. 

To prove $H(\mathsf{X}|\mathsf{Y})\le H(\mathsf{X})$, letting $E=\mathrm{supp}(\mathsf{Y})$, we have $H(\mathsf{X}|\mathsf{Y})=p(E)(H(\mathsf{X}\cap\mathsf{Y}|E)-H(\mathsf{Y}|E))\le p(E)H(\mathsf{X}|E)$ by Proposition \ref{prop:conjunction}. For any $p_{A|Z}$ such that $Z\in A\in\mathsf{X}$ almost surely for $Z\sim p$, we also have $Z\in A\cap E\in\mathsf{X}\cap2^{E}$ almost surely conditional on the event $Z\in E$, so $I(Z;A)\ge I(Z;A\cap E|\mathbf{1}_{E})\ge p(E)I(Z;A\cap E|E)\ge p(E)H(\mathsf{X}|E)$. Minimizing over $p_{A|Z}$ gives $H(\mathsf{X})\ge p(E)H(\mathsf{X}|E)$, completing the proof.

We now prove that if $\mathsf{Y}\cap\mathsf{Z}=2^{\emptyset}$, then $H(\mathsf{X}|\mathsf{Y}\cup\mathsf{Z})=H(\mathsf{X}|\mathsf{Y})+H(\mathsf{X}|\mathsf{Z})$. Let $E=\mathrm{supp}(\mathsf{Y})$, $F=\mathrm{supp}(\mathsf{Z})$, $E\cap F=\emptyset$. Consider $p_{A|Z}$ such that $Z\in A\in\mathsf{X}\cap(\mathsf{Y}\cup\mathsf{Z})$ almost surely for $Z\sim p(\cdot|E\cup F)$. We have
\[
I(Z;A)=H_{b}\Big(\frac{p(E)}{p(E\cup F)}\Big)+\frac{p(E)I(Z;A|E)+p(F)I(Z;A|F)}{p(E\cup F)},
\]
where $H_{b}$ is the binary entropy function. We can see that the minimization of $I(Z;A)$ can be separated into the minimization of $I(Z;A|A\in\mathsf{Y})$ and of $I(Z;A|A\in\mathsf{Z})$, with minimums $\min_{p_{A|Z,E}:\,Z\in A\in\mathsf{X}\cap\mathsf{Y}}I(Z;A|E)=H(\mathsf{X}\cap\mathsf{Y}|E)$ and $H(\mathsf{X}\cap\mathsf{Z}|F)$, respectively. Hence, 
\begin{align}
 & H(\mathsf{X}\cap(\mathsf{Y}\cup\mathsf{Z})|E\cup F)\nonumber \\
 & =H_{b}\Big(\frac{p(E)}{p(E\cup F)}\Big)+\frac{p(E)H(\mathsf{X}\cap\mathsf{Y}|E)+p(F)H(\mathsf{X}\cap\mathsf{Z}|F)}{p(E\cup F)}\label{eq:disjoint_condition}
\end{align}
In particular, substituting $\mathsf{X}=2^{\Omega}$, $H(\mathsf{Y}\cup\mathsf{Z}|E\cup F)=H_{b}(\cdots)+\frac{p(E)H(\mathsf{Y}|E)+p(F)H(\mathsf{Z}|F)}{p(E\cup F)}$. Subtracting this from (\ref{eq:disjoint_condition}) gives the desired result.

 We then prove $H(\mathsf{X}|\mathsf{Y})=\sum_{F\in\mathcal{S}}p(F)H(\mathsf{X}|F)$ when $\mathcal{S}\subseteq2^{\Omega}$ is a set of disjoint events such that $\mathsf{Y}=\bigcup_{F\in\mathcal{S}}2^{F}$. Let $E=\mathrm{supp}(\mathsf{Y})=\bigcup_{F\in\mathcal{S}}F$. Consider $p_{A|Z}$ such that $Z\in A\in\mathsf{X}\cap\mathsf{Y}$ almost surely for $Z\sim p(\cdot|E)$. Let $C\in\mathcal{S}$ be a random set such that $Z\in C$, so $H(C|Z)=0$. Also, $A\subseteq C$, so $H(C|A)=0$. Hence, 
\begin{align*}
I(Z;A) & =H(C)+I(Z;A|C)\\
 & =H(\mathsf{Y}|E)+\sum_{c\in\mathcal{S}}p(c|E)I(Z;A|C=c).
\end{align*}
We can see that the minimization of $I(Z;A)$ can be separated into the minimization of $I(Z;A|C=c)$ for each $c\in\mathcal{S}$, and each has a minimum $\min_{p_{A|Z,C=c}:\,Z\in A\in\mathsf{X}\cap2^{c}}I(Z;A|C=c)=H(\mathsf{X}|C=c)$. Hence, $H(\mathsf{X}\cap\mathsf{Y}|E)=H(\mathsf{Y}|E)+\sum_{c\in\mathcal{S}}p(c|E)H(\mathsf{X}|C=c)$. Rearranging the terms gives the desired result.

\smallskip{}

\subsection{Proof of Proposition \ref{prop:implication}}\label{subsec:pf_implication}

Assume $\mathrm{supp}(p)=\Omega$ without loss of generality. Let $E=\mathrm{supp}(\mathsf{X})$. By Proposition \ref{prop:condition},
\begin{align*}
H(\mathsf{X}\rightarrow\mathsf{Y}) & \ge H(\mathsf{X}\rightarrow\mathsf{Y}|2^{E})\\
 & =p(E)H(\mathsf{X}\rightarrow\mathsf{Y}|E)\\
 & \stackrel{(a)}{\ge}p(E)(H(\mathsf{X}\cap\mathsf{Y}|E)-H(\mathsf{X}|E))\\
 & =H(\mathsf{Y}|\mathsf{X}),
\end{align*}
where (a) is due to $\mathsf{X}\cap(\mathsf{X}\rightarrow\mathsf{Y})\subseteq\mathsf{X}\cap\mathsf{Y}$ and Proposition \ref{prop:conjunction}.

If $\mathsf{X}\in\mathrm{OIs}(\Omega)$, we let $\mathcal{S}\subseteq2^{\Omega}$ be a set of disjoint events such that $\mathsf{X}=\bigcup_{F\in\mathcal{S}}2^{F}$. Let $E=\mathrm{supp}(\mathsf{X})=\bigcup_{F\in\mathcal{S}}F$. For $F\in\mathcal{S}$, consider any $p_{A|Z}^{(F)}$ such that $Z\in A\in\mathsf{Y}\cap2^{F}$ almost surely if $Z\sim p(\cdot|F)$, $A|Z\sim p_{A|Z}^{(F)}$. Let $Z\sim p$. Generate $A^{(F)}\in\mathsf{Y}\cap2^{F}$ for $F\in\mathcal{S}$ randomly, where $A^{(F)}|Z\sim p_{A|Z}^{(F)}$ if $Z\in F$, and $A^{(F)}\sim p_{A}^{(F)}$ if $Z\notin F$ where $p_{A}^{(F)}$ is the $A$-marginal of $p_{A|Z}^{(F)}p(\cdot|E)$. Let $A^{(\emptyset)}:=\Omega\backslash E$ and $B:=A^{(\emptyset)}\cup\bigcup_{F\in\mathcal{S}}A^{(F)}$. We have 
\begin{align*}
\mathsf{X}\cap2^{B} & =\bigcup_{F\in\mathcal{S}}2^{F}\cap2^{B}\\
 & =\bigcup_{F\in\mathcal{S}}2^{A^{(F)}}\\
 & \subseteq\mathsf{Y},
\end{align*}
so $B\in(\mathsf{X}\rightarrow\mathsf{Y})$. Let $C\in\mathcal{S}\cup2^{\emptyset}$ be such that $C\ni Z$ if $Z\in E$, or $C=\emptyset$ if $Z\notin E$. Note that $I(B;C)=0$ since $(A^{(F)})_{F}\sim\prod_{F}p_{A}^{(F)}$ conditional on any value of $C$. We have
\begin{align*}
 & \sum_{F\in\mathcal{S}}p(F)I(Z;A^{(F)}|Z\in F)\\
 & =I(Z;A^{(C)}|C)\\
 & =I(Z;B|C)\\
 & =I(Z;B)-I(B;C)\\
 & =I(Z;B)\\
 & \ge H(\mathsf{X}\rightarrow\mathsf{Y}).
\end{align*}
Minimizing over $p_{A|Z}^{(F)}$, we have $\sum_{F\in\mathcal{S}}p(F)H(\mathsf{Y}|C=F)\ge H(\mathsf{X}\rightarrow\mathsf{Y})$. By Proposition \ref{prop:condition}, $H(\mathsf{Y}|\mathsf{X})\ge H(\mathsf{X}\rightarrow\mathsf{Y})$.

\smallskip{}

\subsection{Proof of Theorem \ref{thm:medvedev}}\label{subsec:pf_medvedev}

Medvedev logic \cite{medvedev1962finiteEN} is the logic defined by the Kripke frames $\langle2^{[n]}\backslash\emptyset,\supseteq\rangle$ (where $[n]:=\{1,\ldots,n\}$) for $n\in\mathbb{Z}_{+}$, where each element in $2^{[n]}\backslash\emptyset$ is regarded as a world which holds the truth values of each logical formula. A satisfaction relation $\Vdash$ is a relation between $2^{[n]}\backslash\emptyset$ and logical formulae, where $w\Vdash\phi$ means ``$\phi$ is true in world $w$''. It must satisfy the following conditions for all $w,w'\in2^{[n]}\backslash\emptyset$ and formulae $\phi,\psi$: 1) $w'\subseteq w$ and $w\Vdash\phi$ implies $w'\Vdash\phi$, 2) not $w\Vdash\bot$, 3) $w\Vdash\phi\wedge\psi$ iff $w\Vdash\phi$ and $w\Vdash\psi$, 4) $w\Vdash\phi\vee\psi$ iff $w\Vdash\phi$ or $w\Vdash\psi$, and 5) $w\Vdash\phi\rightarrow\psi$ iff ($w'\Vdash\phi$ implies $w'\Vdash\psi$) for every $w'\subseteq w$ \cite{simpson1994proof}. Recall that $\top=\lnot\bot$ and $\lnot\phi=(\phi\rightarrow\bot)$. A formula $\phi$ holds in Medvedev logic if $w\Vdash\phi$ for every $n\in\mathbb{Z}_{+}$, satisfaction relation $\Vdash$, and $w\in2^{[n]}\backslash\emptyset$.

Theorem \ref{thm:medvedev} follows from the fact that the dual Heyting algebra of $\langle2^{[n]}\backslash\emptyset,\supseteq\rangle$ is the hypergraph Heyting algebra $\mathrm{Hyps}([n])$. We now spell out the arguments. For a formula $\phi$, let $\mathrm{hyp}_{\Vdash}(\phi):=\{w\in2^{[n]}\backslash\emptyset:\,w\Vdash\phi\}\cup\{\emptyset\}\in\mathrm{Hyps}([n])$ (which depends on $n$ and $\Vdash$). It is straightforward to check that $\mathrm{hyp}_{\Vdash}(\phi\wedge\psi)=\mathrm{hyp}_{\Vdash}(\phi)\cap\mathrm{hyp}_{\Vdash}(\psi)$, $\mathrm{hyp}_{\Vdash}(\phi\vee\psi)=\mathrm{hyp}_{\Vdash}(\phi)\cup\mathrm{hyp}_{\Vdash}(\psi)$, and $\mathrm{hyp}_{\Vdash}(\phi\rightarrow\psi)=\mathrm{hyp}_{\Vdash}(\phi)\rightarrow\mathrm{hyp}_{\Vdash}(\psi)$, i.e., $\mathrm{hyp}_{\Vdash}$ is a homomorphism. For the forward direction, if a formula $\phi$ always evaluate to $2^{[n]}$ regardless of the choice of hyperconfusions for the atoms $\mathsf{X}_{1},\ldots,\mathsf{X}_{k}$, then for every $n$ and $\Vdash$, we have $\mathrm{hyp}_{\Vdash}(\phi)=2^{[n]}$, so $\phi$ holds in Medvedev logic. For the reverse direction, if $\phi$ holds in Medvedev logic, then for every combinations of hyperconfusions for the atoms $\mathsf{X}_{1},\ldots,\mathsf{X}_{k}\in\mathrm{Hyps}([n])$, we take $\Vdash$ to be satisfying $w\Vdash\mathsf{X}_{i}$ iff $w\in\mathsf{X}_{i}$ (as a hyperconfusion), so the formula evaluates to $\mathrm{hyp}_{\Vdash}(\phi)=2^{[n]}$ since $\phi$ holds in Medvedev logic and $\mathrm{hyp}_{\Vdash}$ is a homomorphism.

The logic of infinite problems \cite{skvortsov1979logic} concerns the Kripke frame $\langle2^{\mathbb{N}}\backslash\emptyset,\supseteq\rangle$. The same arguments hold except that we now consider $\mathrm{Hyps}(\mathbb{N})$.

\subsection{Proof of Proposition \ref{prop:coarse}}\label{subsec:pf_coarse}

The first two properties follows from the definition. Assume $\mathsf{X}\in\mathrm{OIs}(\Omega)$ and $\mathsf{Y}\subseteq\mathsf{X}$. Assume $\mathrm{supp}(p)=\Omega$ without loss of generality. If $\mathrm{supp}(\mathsf{X})\neq\Omega$, then both sides are infinite. Hence, we assume $\mathrm{supp}(\mathsf{X})=\Omega$. Let $\mathsf{X}=\mathrm{oi}(X)$ for a random variable $X$. We have $\min_{p_{A|B}:\,B\subseteq A\in\mathsf{X}}I(B;A)=H(X)$ regardless of $p_{B}$ since we can deduce $X$ from $B$ or $A$ as the unique value of $X(B)=X(A)$. Hence, $H(\mathsf{X}\searrow\mathsf{Y})=H(X)$.

For the unconfusing lemma, we assume $\tilde{\mathsf{Y}}\subseteq\mathsf{X}$ where $\tilde{\mathsf{Y}}=\mathsf{Y}\cup\mathrm{sing}(\Omega)$ (otherwise the right-hand side is infinite). Let $\tilde{\mathsf{Y}}=\mathrm{oi}(\tilde{Y})$. Let $Z\sim p$, and $B=\tilde{Y}^{-1}(\tilde{Y}(Z))\in\tilde{\mathsf{Y}}$. Consider any $p_{A|B}$ with $B\subseteq A\in\mathsf{X}$ almost surely. By the strong functional representation lemma \cite{sfrl_trans,li2024channel,li2025discrete}, there exists a random variable $W\in\mathcal{W}$ independent of $Z$ and a function $f:\tilde{\mathsf{Y}}\times\mathcal{W}\to\mathsf{X}$ such that $A=f(B,W)$ and $H(A|W)\le I(B;A)+\log(I(B;A)+3.4)+1$. Hence, there exists a constant $w\in\mathcal{W}$ such that $H(A|W=w)\le I(B;A)+\log(I(B;A)+3.4)+1$. Take $\hat{\mathsf{X}}=\mathrm{oi}(\omega\mapsto f(\tilde{Y}^{-1}(\tilde{Y}(\omega)),w))$. We have $\mathsf{Y}\subseteq\tilde{\mathsf{Y}}\subseteq\hat{\mathsf{X}}$, and $\hat{\mathsf{X}}\subseteq\mathsf{X}$ since $\omega\in\tilde{Y}^{-1}(\tilde{Y}(\omega))\subseteq A$ whenever $f(\tilde{Y}^{-1}(\tilde{Y}(\omega)),w)=A$. We have $H(\hat{\mathsf{X}})=H(A|W=w)$ satisfying the upper bound. The result follows from minimizing over $p_{A|B}$.

\subsection{Proof of Proposition \ref{prop:entropy_decrease}}\label{subsec:pf_entropy_decrease}

By max-flow min-cut theorem, $\beta$ is hyper-expanding if and only if there exists a joint distribution $\kappa:\Omega\times\Gamma\to[0,1]$ which is a coupling of $p$ and $q$, and $\kappa(\omega,\gamma)>0$ implies $\gamma\in\beta(\omega)$. Let $(Z,W)\sim\kappa$. Consider any $p_{A|Z}$ such that $Z\in A\in\mathsf{X}$ almost surely when $A|Z\sim p_{A|Z}$ (assume $W-Z-A$ forms a Markov chain). Let $B=\beta(A)\in\mathsf{Y}$. We also have $W\in\beta(Z)\subseteq B$ almost surely, and by data processing inequality, $I(Z;A)\ge I(W;B)\ge H(\mathsf{Y})$. The result follows from minimizing over $p_{A|Z}$.

\subsection{Proof of Proposition \ref{prop:capacity}}\label{subsec:pf_capacity}

Consider a hyperconfusable surjective set-valued function $\beta$ from $(\Omega,\mathsf{X})$ to $(\Gamma,\mathsf{Y})$. Fix any distribution $q$ over $\Gamma$. Since $\beta$ is surjective, we can find a joint distribution $\kappa:\Omega\times\Gamma\to[0,1]$ which has a $\gamma$-marginal $q$, and $\kappa(\omega,\gamma)>0$ implies $\gamma\in\beta(\omega)$. Let $p$ be the $\omega$-marginal of $\kappa$. By Proposition \ref{prop:entropy_decrease}, $H(\mathsf{X})\ge H(\mathsf{Y})$, where $H(\mathsf{X})$ is evaluated using $p$, and $H(\mathsf{Y})$ is evaluated using $q$. Hence, $C(\mathsf{X})\ge H(\mathsf{Y})$. The result follows from maximizing over $q$.

Next, we show that the capacity tensorizes, i.e., $C(\mathsf{X}\otimes\mathsf{Y})=C(\mathsf{X})+C(\mathsf{Y})$. For any distribution $p$ over $\Omega$ and distribution $q$ over $\Gamma$, consider the product distribution $p\times q$ over $\Omega\times\Gamma$, by Proposition \ref{prop:conjunction}, $H(\mathsf{X})+H(\mathsf{Y})=H(\mathsf{X}\otimes\mathsf{Y})\le C(\mathsf{X}\otimes\mathsf{Y})$. Hence, $C(\mathsf{X}\otimes\mathsf{Y})\ge C(\mathsf{X})+C(\mathsf{Y})$. For the other direction, given any joint distribution $\kappa:\Omega\times\Gamma\to[0,1]$ which has marginal distributions $p,q$. By Proposition \ref{prop:conjunction}, $H(\mathsf{X}\otimes\mathsf{Y})\le H(\mathsf{X})+H(\mathsf{Y})\le C(\mathsf{X})+C(\mathsf{Y})$. Hence, $C(\mathsf{X}\otimes\mathsf{Y})\le C(\mathsf{X})+C(\mathsf{Y})$.

To show $\mathrm{CR}(\mathsf{X}/\mathsf{Y})\le C(\mathsf{X})/C(\mathsf{Y})$, if $\mathsf{X}^{\otimes n}\sqsubseteq\mathsf{Y}^{\otimes m}$, then $C(\mathsf{X}^{\otimes n})=nC(\mathsf{X})\ge C(\mathsf{Y}^{\otimes m})\ge mC(\mathsf{Y})$, so $m/n\le C(\mathsf{X})/C(\mathsf{Y})$. 

\smallskip{}

\bibliographystyle{IEEEtran}
\bibliography{ref}

\end{document}